\documentclass[twocolumn,preprint]{aastex631}

\usepackage[normalem]{ulem}

\usepackage{xcolor}
\usepackage{amsmath,amssymb}
\usepackage{orcidlink}
\usepackage{hyperref}

\begin{document}

\title{Are all Binary Black Holes Detected by LIGO-Virgo-KAGRA Following the Universal Time-Delay Distributions? Probably Not}

\author{Samsuzzaman Afroz \orcidlink{0009-0004-4459-2981}}
\email{samsuzzaman.afroz@tifr.res.in}
\affiliation{Department of Astronomy and Astrophysics, Tata Institute of Fundamental Research, Mumbai 400005, India}
\author{Navdha \orcidlink{0009-0002-3202-1335}}
\affiliation{Indian Institute of Technology Bombay, Mumbai-400076, India}
\author{Suvodip Mukherjee \orcidlink{0000-0002-3373-5236}}
\email{suvodip@tifr.res.in}
\affiliation{Department of Astronomy and Astrophysics, Tata Institute of Fundamental Research, Mumbai 400005, India}

\begin{abstract}
The delay time distribution (DTD) of binary black hole (BBH) mergers encodes the evolutionary link between stellar formation history and gravitational-wave (GW) emission. We present a non-parametric reconstruction of the DTD using BBHs from the GWTC-4 catalog, employing a grid-based framework that avoids restrictive power-law assumptions. We divide the BBH population into two mass bins, $20$--$40\,M_\odot$ and $40$--$100\,M_\odot$, and reconstruct the DTD independently for each bin using grid-based trajectories, under three different star formation rate density (SFRD) models. We find that both mass bins favor an extended plateau of intermediate-to-long delay times rather than a single characteristic timescale, and that in both bins the reconstructed merger rate $\mathcal{R}(z)$ rises from its local value to a peak at intermediate redshift before declining. The two bins differ in degree rather than kind: the $20$--$40\,M_\odot$ bin is broadly constrained, with a local rate of $R_0 \approx 18$--$23\,\mathrm{Gpc^{-3}\,yr^{-1}}$ and a merger rate peaking at $z \sim 0.8$--$1.05$, while the $40$--$100\,M_\odot$ bin prefers a more sharply defined intermediate delay range of $\sim$2--4 Gyr with a steeper suppression of long delay times, within a noticeably tighter set of high-evidence trajectories, a lower local rate of $R_0 \approx 9$--$13\,\mathrm{Gpc^{-3}\,yr^{-1}}$, and a peak shifted to slightly higher redshift, $z \sim 1.0$--$1.05$. The results indicate that a single power-law DTD may not be sufficient to simultaneously describe both mass bins, pointing toward possible mass-dependent binary evolution pathways that merit further investigation with larger GW catalogs.
\end{abstract}

\section{Introduction}
\label{sec:intro}

The delay time distribution (DTD) of binary black hole (BBH) mergers represents one of the most fundamental quantities connecting stellar evolution to gravitational-wave (GW) observations. It encodes the probability distribution of times between the formation of massive stellar progenitors and their eventual coalescence as compact binaries, directly linking the cosmic star formation history to the observable merger rate across redshift. Understanding these evolutionary timescales is crucial for constraining binary evolution physics, supernova dynamics, and the efficiency of compact object formation channels \citep{Bailyn:1997xt,Dominik:2012kk,Barrett:2017fcw,Mapelli:2021taw,Franciolini:2021tla,Bouffanais:2021wcr,Antonelli:2023gpu,Cheng:2023ddt,Ye:2024ypm,Afroz:2024fzp,Afroz:2025efn,Afroz:2025ikg,Afroz:2025xpp}.

Gravitational-wave astronomy has opened unprecedented opportunities to constrain DTDs through direct observations of BBH mergers across cosmic time \citep{LIGOScientific:2016aoc,LIGOScientific:2018mvr,LIGOScientific:2020kqk,KAGRA:2021duu,LIGOScientific:2025pvj}. The relationship between stellar formation and merger observations is fundamentally described by a convolution integral that relates the present-day merger rate to the historical star formation rate weighted by the DTD. This connection enables inference of evolutionary timescales from the mass and redshift distribution of detected GW events \citep{Dominik:2013tma,Mapelli:2016vca,Mukherjee:2021rtw,Bailes:2021tot,Arimoto:2021cwc,Iorio:2022sgz,Sicilia:2021gtu,Schiebelbein-Zwack:2024roj,Rinaldi:2023bbd,Rinaldi:2025emt,Gennari:2025nho,Lalleman:2025xcs}.

Most of the previous approaches to DTD inference from GWTC-3 have not found any evidence of departure from a simple power-law form \citep{Fishbach:2018edt,Safarzadeh:2020qru,Mukherjee:2021qam,Fishbach:2021mhp,Karathanasis:2022rtr,Smith:2024awx}. While the simple power-law forms have provided valuable constraints on characteristic timescales, they inherently restrict the range of evolutionary behaviors that can be discovered. More flexible or semi-agnostic approaches have also been developed \citep{Edelman:2022ydv,Callister:2023tgi,Stevenson:2022djs,Heinzel:2023hlb}. Currently there are limitations both in theoretical modeling as well as in data analysis techniques to capture complex DTDs arising from multiple formation channels, environmental dependencies, or mass-dependent evolutionary pathways.

The mass dependence of DTDs is particularly important from both theoretical and observational perspectives. Massive BBH progenitors experience different stellar evolution pathways, including altered wind mass loss rates, distinct supernova dynamics, and potentially different formation environments. Systems forming through isolated binary evolution may follow different delay time characteristics compared to those assembled through dynamical interactions in dense stellar environments. Additionally, the efficiency of common-envelope evolution and the survival probability of wide binaries may depend sensitively on component masses \citep{Belczynski:2016jno,Rodriguez:2016avt,Mandel:2018hfr}.

In this work, we introduce a novel grid-based approach for non-parametric inference of DTD that addresses the limitations of traditional parametric methods, and allows us to infer mass-dependent DTD and merger rate in a non-parametric way from GW observations. Our framework divides the delay time-probability space into a structured grid and systematically explores all physically allowed evolutionary trajectories. The key innovation is the incorporation of causality constraints: evolutionary trajectories can progress forward in delay time with varying merger probabilities but cannot violate the fundamental requirement that delay times increase monotonically.

This grid-based methodology provides several advantages over conventional approaches. First, it enables discovery of complex, non-monotonic DTD shapes without assuming specific functional forms. Second, it naturally incorporates physical constraints while maintaining maximum flexibility for capturing unexpected features. Third, it provides a systematic framework for model comparison through Bayesian evidence calculations, enabling objective selection among competing evolutionary scenarios.

We apply this framework to the population of BBH mergers reported in GWTC-4.0 \citep{LIGOScientific:2025slb}, which is the cumulative Gravitational-Wave Transient Catalog compiled by the LIGO \citep{LIGOScientific:2016dsl}, Virgo \citep{VIRGO:2014yos}, and KAGRA \citep{KAGRA:2020tym} (LVK) collaboration. GWTC-4.0 combines the new candidates identified during the first part of the fourth observing run (O4a) with the events previously reported in GWTC-3.0 \citep{KAGRA:2021vkt}, which in turn already incorporates the earlier GWTC-2.1 \citep{LIGOScientific:2021usb} events, so that the events from all three catalogs are contained within the single GWTC-4.0 release. We select the subset of these events with a false alarm rate (FAR) below $1\,\mathrm{yr}^{-1}$ in at least one search pipeline.

The analysis is performed under three star formation rate density (SFRD) assumptions, the standard Madau-Dickinson model, which is metallicity-averaged and treats all cosmic star formation as equally capable of producing BBH progenitors; the high-metallicity effective SFRD ($Z > 0.1\,Z_\odot$), and the low-metallicity effective SFRD ($Z < 0.1\,Z_\odot$) from \citet{Chruslinska2019MNRAS}. The two effective SFRDs are motivated by the  expectation that BBH formation is suppressed at high metallicity due to enhanced stellar wind mass loss, so restricting the star formation history to a metallicity range above or below $0.1\,Z_\odot$ provides a first order approximation to the star formation environments preferentially associated with lower-mass and higher-mass BBH progenitors, respectively \citep{2010ApJ...714.1217B,Mapelli:2016vca,Giacobbo:2017qhh,vanSon:2024bxr}. This choice allows us to assess the sensitivity of the inferred DTD to the assumed star formation history. Across all three assumptions, we find that both mass bins favor an extended plateau of intermediate-to-long delay times rather than a single characteristic timescale, and that the reconstructed merger rate $\mathcal{R}(z)$ rises from its local value to a peak at intermediate redshift in both bins.

The two bins differ in degree rather than kind: lower-mass binaries ($20$--$40\,M_\odot$) are broadly constrained, with a wide range of allowed delay time distributions, a local merger rate of $R_0 \approx 18$--$23\,\mathrm{Gpc^{-3}\,yr^{-1}}$, and a merger rate peaking at $z \sim 0.8$--$1.05$, whereas higher-mass binaries ($40$--$100\,M_\odot$) prefer a more sharply defined intermediate delay range of $\sim$2--4 Gyr, with a lower local rate of $R_0 \approx 9$--$13\,\mathrm{Gpc^{-3}\,yr^{-1}}$, a peak shifted to slightly higher redshift $z \sim 1.0$--$1.05$, and a tighter envelope of high-evidence trajectories. In both bins the peak-to-local rate ratio is largest under the low-metallicity SFRD. These differences suggest that higher-mass BBHs may merge on somewhat more characteristic timescales than lower-mass systems, and that a single power-law DTD may not simultaneously describe both populations. These findings illustrate the potential of non-parametric grid-based approaches for extracting mass-dependent evolutionary information from current and future GW observations.

The remainder of this paper is organized as follows. Section~\ref{sec:Framework} presents the theoretical framework, including the metallicity-dependent star formation histories, the grid-based non-parametric DTD construction, and the Bayesian inference framework. Section~\ref{sec:data} describes the GW catalog and analysis setup used in this work. Section~\ref{sec:result} describes the GW catalog and analysis setup, and presents our results for the reconstructed DTDs, merger rate evolution, and Bayesian evidence in both mass bins. Section~\ref{sec:conclusions} summarizes our findings and conclusions. Appendix~\ref{app:6x5} presents a robustness check using an alternative delay-time grid.

\section{Theoretical Framework and Grid Construction}
\label{sec:Framework}

The theoretical foundation for merger rate density inference rests on the fundamental relationship between BBH formation, evolution, and the observable GW merger rates across cosmic time. For a population of BBHs with component masses $m_1$ and $m_2$, the differential merger rate density at redshift $z$ can be expressed as \citep{2019PASA...36...10T,Mandel:2018mve,Loredo:2004nn}
\begin{equation}\label{eq:dngw}
\frac{dN_{\rm GW}}{dz}(z,m) = T_{\rm obs} \frac{dV_c}{dz}
\frac{R(z,m)}{(1+z)} \cdot p(m|z) \cdot \mathcal{S}(m,z),
\end{equation}
where $R(z,m)$ represents the intrinsic merger rate density we seek to infer, $p(m|z)$ is the normalized mass distribution at redshift $z$, $T_{\rm obs}$ is the observation time, $dV_c/dz$ is the comoving volume element, $(1+z)^{-1}$ accounts for cosmological time dilation, and $\mathcal{S}(m,z)$ accounts for detector selection effects derived from injection campaigns.

The merger rate density can be linked to the cosmic star formation history through a convolution with the DTD. In normalized form, it can be written as \citep{Dominik:2014yma,Karathanasis:2022hrb}
\begin{equation}
R(z,m) = R_0(m)
\frac{ \displaystyle \int_z^{\infty} p_t(t_d(m))\, R_{\rm SFR}(z_f)\, 
\frac{dt}{dz_f}\, dz_f}
{\displaystyle \int_0^{\infty} p_t(t_d(m))\, R_{\rm SFR}(z_f)\, 
\frac{dt}{dz_f}\, dz_f},
\label{eq:mergrate}
\end{equation}
where $R_{\rm SFR}(z_f)$ denotes the cosmic star formation rate density at formation redshift $z_f$, $p_t(t_d(m))$ is the DTD (potentially mass dependent), and $t_d(m) = t(z) - t(z_f(m))$ is the delay between formation and merger. The denominator provides normalization, ensuring that $R(z,m)$ has a proper probabilistic interpretation.

The choice of $R_{\rm SFR}(z_f)$ in Equation~\ref{eq:mergrate} is a key ingredient of the inference. In the usual approach, one adopts a single metallicity-averaged SFRD to describe all BBH progenitors regardless of their mass \citep{Fishbach:2021mhp,Karathanasis:2022rtr,Turbang:2023tjk}. A widely used parameterization of this kind is the Madau-Dickinson model \citep{Madau:2014bja}:
\begin{equation}
R_{\rm SFR}(z) = \frac{0.015 (1+z)^{2.7}}{1 + \left[ 
\tfrac{1+z}{2.9} \right]^{5.6}}
\quad \text{M}_\odot \, \text{Mpc}^{-3}\,\text{yr}^{-1}.
\end{equation}
This model provides a convenient, well-tested baseline and is used as our primary SFRD assumption. However, the Madau-Dickinson model does not provide a metallicity dependent SFR. Since the efficiency of BBH formation depends strongly on metallicity, treating all star formation as equally contributing to BBH production regardless of component mass may limit the accuracy of the inferred DTD. To assess the magnitude of this effect, we additionally consider metallicity-dependent effective SFRDs as described below.

\subsection{Metallicity-Dependent Star Formation Rate Density}
\label{sec:metallicity_sfrd}

The formation of massive BBHs is strongly suppressed at high metallicities due to enhanced stellar wind mass loss \citep{2010ApJ...714.1217B,Mapelli:2016vca,2018IAUS..338...40L,
Mukherjee:2021rtw,2019MNRAS.490.3740N}. This physical expectation motivates supplementing the metallicity-averaged Madau-Dickinson SFRD with effective SFRDs that integrate only over a restricted metallicity range.

The physical origin of this metallicity dependence lies in line-driven stellar winds, whose mass-loss rate scales with metallicity \citep{2000A&A...362..295V}: at high metallicity, massive stars lose more mass before collapse and form lower mass black holes, so the most massive BBHs form preferentially from low metallicity progenitors \citep{2010ApJ...714.1217B,Giacobbo:2017qhh,Spera:2017fyx}. This wind-driven mass$-$metallicity relation is a feature of the isolated binary evolution channel; black holes assembled dynamically in dense stellar environments, in AGN disks, or through hierarchical mergers can instead attain high masses through repeated mergers or stellar collisions, and need not follow the same metallicity dependence \citep{Mckernan:2017ssq,Gerosa:2021mno,Rodriguez:2016avt}. Our metallicity-restricted effective SFRDs should therefore be understood as an approximation appropriate for the isolated channel contribution to the BBH population, adopted to test the sensitivity of the inferred DTD to the assumed star formation history rather than as a channel-independent prescription.

Physically, high-mass BBH progenitors ($40$--$100\,M_\odot$) are expected to form preferentially in low-metallicity environments ($Z \lesssim 0.1\,Z_\odot$), while lower-mass progenitors ($20$--$40\,M_\odot$) can also form at moderate metallicities ($Z \gtrsim 0.1\,Z_\odot$). Since the cosmic metallicity increases with time, the fraction of star formation falling below or above a given metallicity threshold evolves strongly with redshift, so the high- and low-metallicity effective SFRDs have markedly different redshift dependence. Rather than assigning a single metallicity range to each mass bin a priori, we apply both effective SFRDs, together with the metallicity-averaged Madau-Dickinson model, to both mass bins, and let the data determine which delay-time distributions are preferred under each assumption.

\begin{figure}[t]
    \centering
    \includegraphics[width=0.43\textwidth]{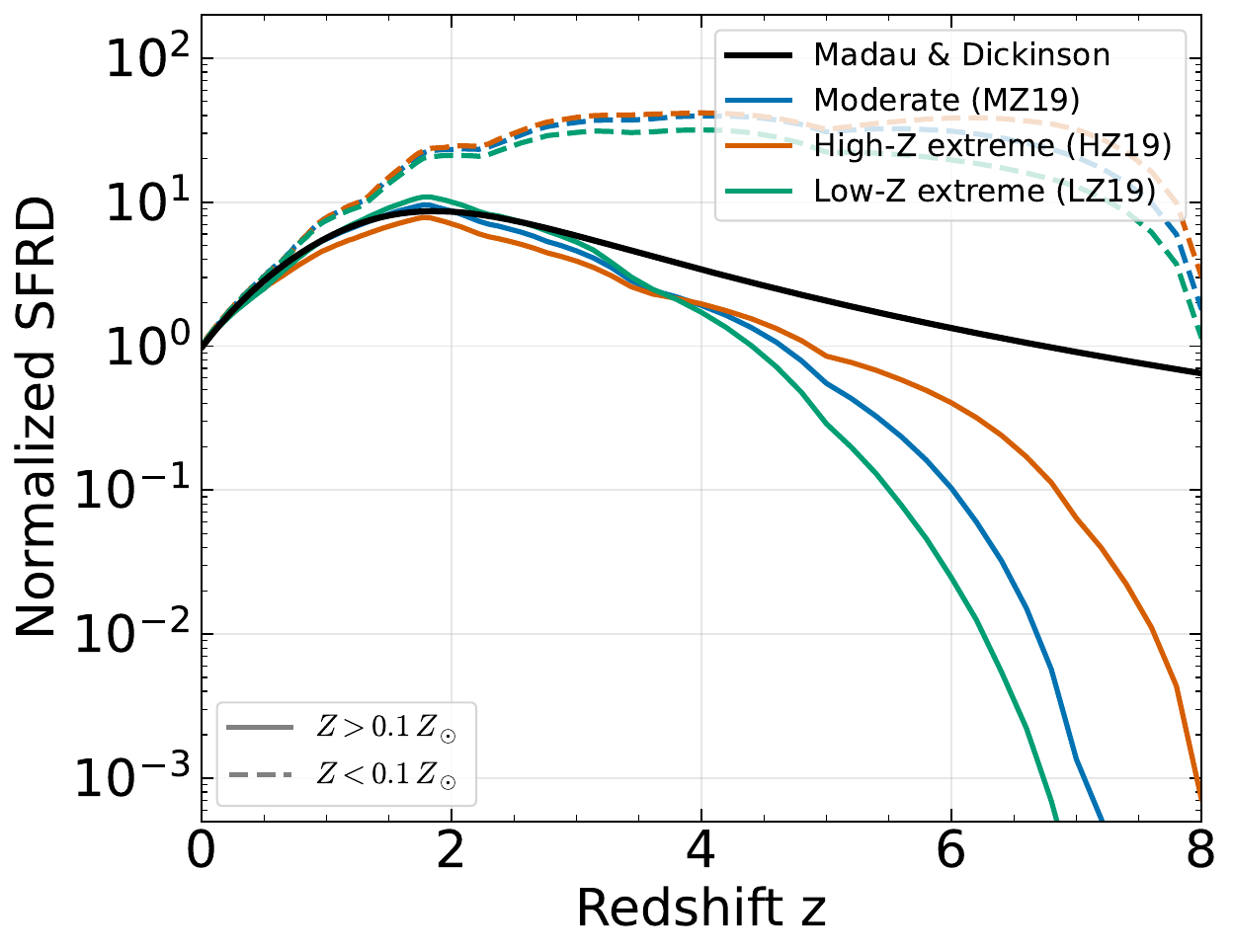}
    \caption{ Normalized effective star formation rate density (SFRD) as a function of redshift, compared with the standard Madau \& Dickinson (2014) model (black). Left panel: high-metallicity effective SFRD ($Z > 0.1\,Z_\odot$), which closely tracks the Madau-Dickinson shape since most star formation occurs above this metallicity threshold at all redshifts. Right panel: low-metallicity effective SFRD ($Z < 0.1\,Z_\odot$), which rises significantly more steeply with redshift, reflecting the increasing fraction of low-metallicity star formation at earlier cosmic epochs. Each curve is normalized to its value at $z \approx 0$. The three line styles correspond to the moderate (MZ19), high-$Z$ extreme (HZ19), and low-$Z$ extreme (LZ19) model 
    variations from \citet{Chruslinska2019MNRAS}; we use the moderate model in our analysis and show the extreme variations for reference. Both effective SFRDs, together with the Madau-Dickinson model, are applied to both mass bins. The close agreement between the $Z > 0.1\,Z_\odot$ SFRD and the Madau-Dickinson model means that the DTD inference is expected to be insensitive to the choice between them, whereas the steeper rise of the $Z < 0.1\,Z_\odot$ SFRD means that the DTD inference can differ more significantly under that assumption.}
    \label{fig:normalized_sfrd}
\end{figure}

To quantify this effect, we incorporate the observation-based metallicity-dependent SFRD $\dot{\rho}_\star(Z_{\rm O/H}, z)$ from \citet{Chruslinska2019MNRAS}, which provides the distribution of cosmic star formation over both metallicity and redshift\footnote{The data file is publicly available at 
\url{https://ftp.science.ru.nl/astro/mchruslinska/}.}. This framework combines three empirical ingredients the galaxy stellar mass function (GSMF), the star formation rate stellar mass relation (SFMR), and the mass metallicity relation (MZR) to construct the two-dimensional distribution of SFRD over gas-phase oxygen abundance $Z_{\rm O/H} \equiv 12 + \log_{10}(\mathrm{O/H})$ and redshift. The metallicity axis spans $Z_{\rm O/H} \in [5.3, 9.7]$ in 200 bins, with solar metallicity at $Z_{\rm O/H,\odot} = 8.83$ corresponding to $Z_\odot = 0.017$ \citep{GrevesseSauval1998}. \citet{Chruslinska2019MNRAS} provide three model variations moderate (MZ19), high-metallicity extreme (HZ19), and low-metallicity extreme (LZ19) which bracket the observationally allowed uncertainty in the metallicity evolution. In this work we use the moderate (MZ19) model as our metallicity-dependent SFRD, and the two extreme variations are shown in Figure~\ref{fig:normalized_sfrd} for reference.

We define two effective SFRDs by integrating over a restricted metallicity range. Following the metallicity threshold at $Z = 0.1\,Z_\odot$ (corresponding to $Z_{\rm O/H} = 7.83$), the high-metallicity effective SFRD is
\begin{equation}
\dot{\rho}_{\star,\mathrm{eff}}^{>}(z) = \int_{Z_{\rm O/H} > 7.83} 
\dot{\rho}_\star(Z_{\rm O/H}, z) \, dZ_{\rm O/H},
\label{eq:sfrd_eff_high}
\end{equation}
which collects star formation at $Z > 0.1\,Z_\odot$, and the low-metallicity effective SFRD is
\begin{equation}
\dot{\rho}_{\star,\mathrm{eff}}^{<}(z) = \int_{Z_{\rm O/H} \leq 7.83} 
\dot{\rho}_\star(Z_{\rm O/H}, z) \, dZ_{\rm O/H},
\label{eq:sfrd_eff_low}
\end{equation}
which collects star formation at $Z < 0.1\,Z_\odot$.

In practice, these integrals are computed as discrete sums over the 200 metallicity bins provided by 
\citet{Chruslinska2019MNRAS}:
\begin{equation}
\dot{\rho}_{\star,\mathrm{eff}}(z_i) = \sum_{j \in \mathcal{M}} 
\dot{\rho}_\star(Z_{\rm O/H}^{(j)}, z_i) \, \Delta Z_{\rm O/H},
\end{equation}
where $\mathcal{M}$ denotes the set of metallicity bins satisfying the relevant threshold condition and $\Delta Z_{\rm O/H} = 0.022$ dex is the bin width.

The choice of the $Z = 0.1\,Z_\odot$ threshold sets how strongly the two effective SFRDs differ from the metallicity-averaged Madau-Dickinson model, and therefore affects the two mass bins differently. Because most cosmic star formation occurs above this threshold at the low-to-intermediate redshifts that dominate the current detectable population, the high-metallicity effective SFRD $\dot{\rho}_{\star,\mathrm{eff}}^{>}(z)$ closely tracks the Madau-Dickinson shape, so the inferred DTD under this assumption is not expected to depend strongly on the precise threshold value. The low-metallicity effective SFRD $\dot{\rho}_{\star,\mathrm{eff}}^{<}(z)$ is more sensitive to this choice: a lower threshold shifts a larger fraction of its star formation to high redshift and steepens its redshift dependence, which concentrates the inferred DTD support at intermediate delay times and amplifies the intermediate-redshift peak of $\mathcal{R}(z)$ (Section~\ref{sec:result}).

The merger rate density in Equation~\ref{eq:mergrate} then generalizes to
\begin{equation}
R(z,m) = R_0(m) \frac{\displaystyle \int_z^{\infty} 
p_t(t_d(m))\, \dot{\rho}_{\star,\mathrm{eff}}(z_f)\, 
\frac{dt}{dz_f}\, dz_f}{\displaystyle \int_0^{\infty} 
p_t(t_d(m))\, \dot{\rho}_{\star,\mathrm{eff}}(z_f)\, 
\frac{dt}{dz_f}\, dz_f},
\label{eq:mergrate_metallicity}
\end{equation}
where $\dot{\rho}_{\star,\mathrm{eff}}(z_f)$ denotes either the high-metallicity effective SFRD of Equation~\ref{eq:sfrd_eff_high} or the low-metallicity effective SFRD of Equation~\ref{eq:sfrd_eff_low}. We carry out the inference for each mass bin under all three SFRD assumptions (Madau-Dickinson, high-$Z$ effective, and low-$Z$ effective), so that the same set of star formation histories is applied uniformly to both bins.

Figure~\ref{fig:normalized_sfrd} shows the normalized effective SFRDs compared with the standard Madau-Dickinson model. The high-metallicity effective SFRD ($Z > 0.1\,Z_\odot$) closely tracks the standard Madau-Dickinson shape, since most star formation at all redshifts occurs above this metallicity threshold; consequently, for any mass bin the DTD inferred under the Madau-Dickinson assumption and under the high-metallicity effective SFRD are expected to be broadly similar. In contrast, the low-metallicity effective SFRD ($Z < 0.1\,Z_\odot$) rises much more steeply with redshift than the Madau-Dickinson model, reflecting the increasing fraction of low-metallicity star formation at earlier cosmic epochs; the DTD inferred under this assumption can therefore differ more significantly from the Madau-Dickinson result, as the two SFRDs weight different redshift epochs differently. Because the high-metallicity SFRD is the physically motivated choice for the $20$--$40\,M_\odot$ progenitors and the low-metallicity SFRD for the $40$--$100\,M_\odot$ progenitors, comparing the DTD inference across all three SFRDs for each mass bin allows us to disentangle the contribution of metallicity-driven selection effects from intrinsic delay-time preferences in the data.

In summary, we perform the DTD inference under three SFRD assumptions for each mass bin: (i) the standard Madau-Dickinson model, a metallicity-averaged baseline; (ii) the high-metallicity effective SFRD ($Z > 0.1\,Z_\odot$), physically motivated for the lower-mass ($20$--$40\,M_\odot$) progenitors; and (iii) the low-metallicity effective SFRD ($Z < 0.1\,Z_\odot$), physically motivated for the higher-mass ($40$--$100\,M_\odot$) progenitors. All three are applied to both mass bins. By comparing the DTD inference across all three assumptions for each mass bin, we can assess the degree to which the inferred DTD shapes and local merger rates are driven by the assumed star formation history versus intrinsic features of the data.

\subsection{Grid-Based Delay Time Distribution Construction}

The key innovation of our approach lies in the non-parametric construction of DTDs through a structured grid methodology. Rather than assuming specific power-law forms for $p_t(t_d)$, we discretize the delay time-probability space and systematically explore all physically allowed evolutionary trajectories. We construct a structured grid in the $(t_d, \log p_t(t_d))$ plane, where $t_d$ represents delay time and $p_t(t_d)$ denotes the probability density of the DTD. The logarithmic transformation of the probability axis ensures numerical stability while capturing distributions spanning multiple orders of magnitude.

The delay time axis is divided from $t_{\rm d,min}$ to $t_{\rm d,max}$ using $N_t$ equally spaced grid points in logarithmic space:
\begin{equation}
t_d^{(i)} = t_{\rm d,min} \left(\frac{t_{\rm d,max}}{t_{\rm d,min}}\right)^{(i-1)/(N_t-1)}, \quad i = 1, 2, \ldots, N_t.
\end{equation}

The probability density values are divided using $N_p$ grid points in logarithmic space:
\begin{equation}
\log p_t^{(j)} = \log p_{\rm t,min} + (j-1) \Delta \log p_t, \quad j = 1, 2, \ldots, N_p,
\end{equation}
where $\Delta \log p_t = (\log p_{\rm t,max} - \log p_{\rm t,min})/(N_p-1)$ provides adequate dynamic range for diverse DTD shapes.

The fundamental physical constraint governing our grid exploration is causality: delay times cannot decrease as we move forward through the evolutionary sequence. Each trajectory begins from a single starting point in the grid and can only progress to delay times that are greater than the current position, while the probability values can move freely upward or downward to any accessible grid point. Any valid trajectory through the grid must satisfy:
\begin{equation}
\mathcal{T} = \{(t_d^{(i_1)}, \log p_t^{(j_1)}), (t_d^{(i_2)}, \log p_t^{(j_2)}), \ldots, (t_d^{(i_{N_t})}, \log p_t^{(j_{N_t})})\},
\end{equation}
with the causality constraint:
\begin{equation}
i_1 \leq i_2 \leq i_3 \leq \ldots \leq i_{N_t} \quad \text{(no backward time evolution)}.
\end{equation}

For each valid trajectory $\mathcal{T}$, we construct a continuous DTD by interpolating between grid points using monotone cubic (PCHIP) splines:
\begin{equation}
p_t^{\rm raw}(t_d|\mathcal{T}) = \text{PchipInterpolator}\left(\{t_d^{(i_k)}, \log p_t^{(j_k)}\}_{k=1}^{N_t}\right).
\end{equation}

The interpolated function is then normalized to ensure proper probability density normalization:
\begin{equation}
p_t(t_d|\mathcal{T}) = \frac{p_t^{\rm raw}(t_d|\mathcal{T})}{\int_0^{\infty} p_t^{\rm raw}(t_d|\mathcal{T}) dt_d}.
\end{equation}

This procedure converts each discrete trajectory into a continuous, normalized DTD suitable for convolution with the cosmic star formation history.

\begin{figure*}
    \centering
    \includegraphics[width=0.90\textwidth, height=8.3cm]{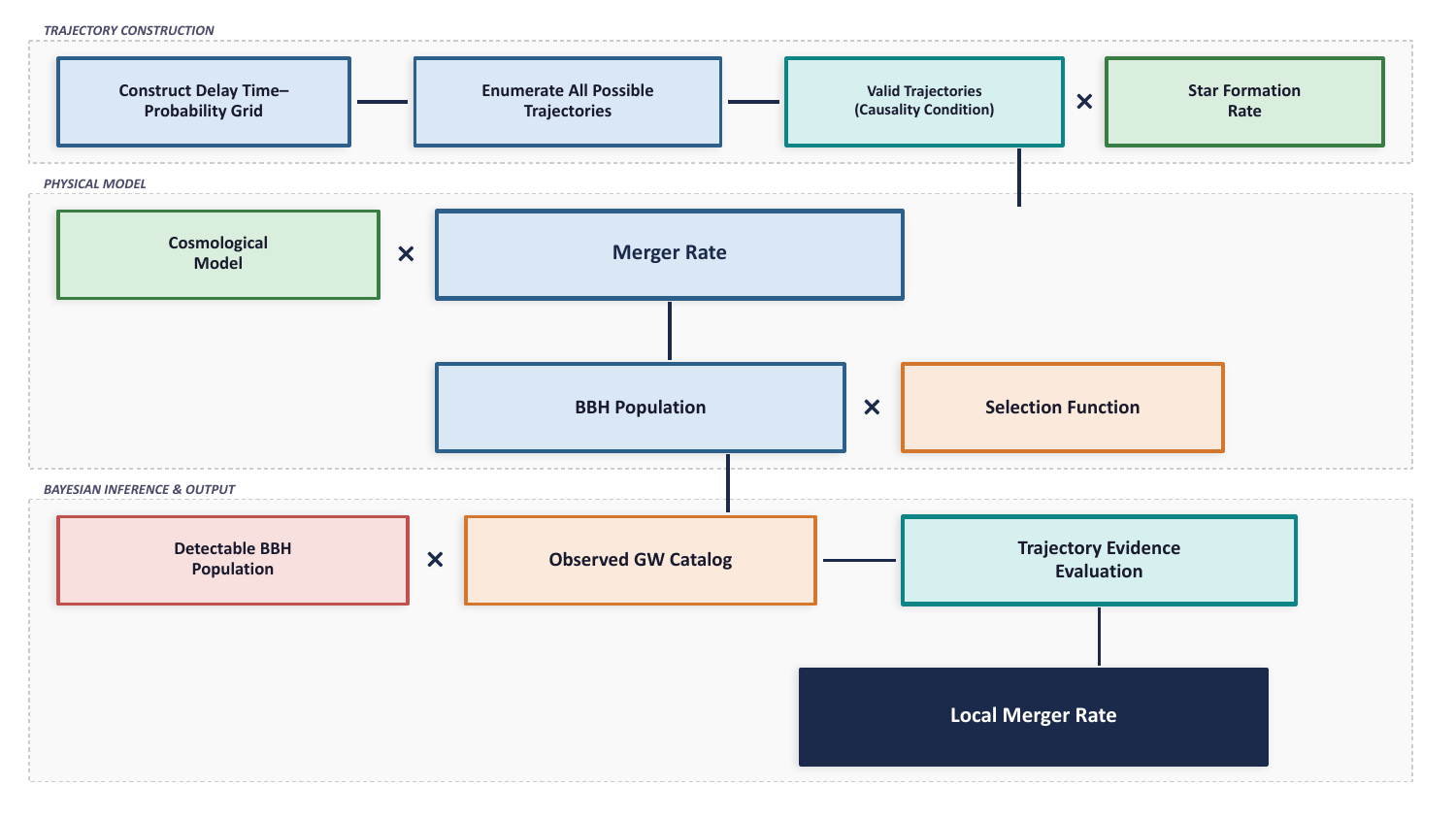}
    \caption{Flowchart of the non-parametric grid-based framework for BBH DTD and merger rate inference. A delay time-probability grid is constructed, and all causality-allowed trajectories are enumerated. Trajectories are convolved with the cosmic star-formation rate and selection effects to obtain the detectable population, which is compared with GW observations. Bayesian evidence then yields the DTD shape and local merger rate.}
    \label{fig:Flowchart}
\end{figure*}

\subsection{Bayesian Inference Framework}
\label{sec:bayesian}

The inference is performed independently for each mass bin $m_k$. For each mass bin we consider three SFRD assumptions: the standard Madau--Dickinson model, the high-metallicity effective SFRD ($Z > 0.1\,Z_\odot$), and the low-metallicity effective SFRD ($Z < 0.1\,Z_\odot$).

\subsubsection{The fixed population mass model}
\label{sec:massmodel}

Throughout the inference we adopt a \emph{fixed} parametric model for the binary mass distribution and hold its hyperparameters constant; we do not fit them. We use the \textsc{Broken Power Law + 2 Peaks} model, the fiducial BBH
mass model of the GWTC-4.0 population analysis \citep{LIGOScientific:2025pvj}, applied to the same GWTC-4.0 event sample analyzed here.

The goal of this work is to infer the delay time distribution  associated with the redshift evolution of the merger-rate of the BBH without invoking any explicit redshift evolution of the mass distribution of BBHs or dependence of delay time $t_d$ on BBHs masses. As a result, the BBH merger rate and mass distributions (for both the component masses) are separable in Eq. \eqref{eq:dngw}. However, the formalism developed in this work can be extended to mass-dependent delay time distribution as well.  We therefore make the assumption of fixing the mass model to the fiducial GWTC-4.0 \textsc{Broken Power Law + 2 Peaks} result, holding its hyperparameters (Table~\ref{tab:massparams}) constant rather than fitting them jointly with the DTD. This assumption is valid for our purpose for two reasons. First, the mass model enters our analysis in an essentially redshift-independent way, as a fixed weighting of the events that is common to every trajectory, whereas the delay-time and merger-rate information is carried by the redshift dependence of the merger rate. The reconstructed DTD is therefore sensitive to the mass model only through the assignment of events to a primary-mass bin and their relative weighting within it. Second, those hyperparameters are already tightly constrained by the full GWTC-4.0 population analysis \citep{LIGOScientific:2025pvj}, which uses the entire catalog; adopting its posterior medians thus provides a more precise, data-driven mass model than could be obtained from the present DTD-focused analysis alone. Opening up the mass hyperparameters is difficult here for a statistical and a methodological reason. Statistically, the current catalog of confident BBH events, split across the two mass bins, is insufficient to constrain the full set of mass hyperparameters simultaneously with the delay-time distribution and rate evolution. Methodologically, our framework enumerates all causality-allowed trajectories and evaluates each one's evidence by nested sampling; treating the mass hyperparameters as free would require recomputing the weights and marginalizing over them inside every trajectory's evidence integral, a high-dimensional marginalization repeated over the full trajectory ensemble that is computationally prohibitive. Fixing the mass model to its well-measured values therefore isolates the DTD and merger-rate inference as the target of this work.

Working in primary mass and mass ratio $(m_1, q)$ with $q = m_2/m_1$, the joint population density factorizes as

\begin{equation}
    p_{\rm pop}(m_1, q) = p(m_1)\, p(q \mid m_1).
    \label{eq:poppop}
\end{equation}

The primary-mass distribution is a mixture of a broken power-law continuum and two left-truncated Gaussian peaks, with a low-mass taper applied to the whole distribution. The broken power law is

\begin{equation}
    p_{\rm BP}(m_1) \propto
    \begin{cases}
        \left(\dfrac{m_1}{m_{\rm break}}\right)^{-\alpha_1}
        & m_{1,\rm low} \leq m_1 < m_{\rm break}, \\[2.2ex]
        \left(\dfrac{m_1}{m_{\rm break}}\right)^{-\alpha_2}
        & m_{\rm break} \leq m_1 < m_{\rm high},
    \end{cases}
    \label{eq:brokenpl}
\end{equation}

with low- and high-mass spectral indices $\alpha_1$ and $\alpha_2$ and a break at $m_{\rm break}$. The full primary mass density is

\begin{equation}
\begin{split}
    p(m_1) \propto \Big[ &\lambda_0\, p_{\rm BP}(m_1)
    + \lambda_1\, \mathcal{G}_{\rm lt}(m_1 \mid \mu_1, \sigma_1) \\
    &+ (1 - \lambda_0 - \lambda_1)\,
    \mathcal{G}_{\rm lt}(m_1 \mid \mu_2, \sigma_2) \Big]\, \\
    & \times S(m_1 \mid m_{1,\rm low}, \delta_{m,1}),
    \label{eq:bp2p}
\end{split}
\end{equation}

where $\mathcal{G}_{\rm lt}(m_1 \mid \mu, \sigma)$ is a Gaussian of mean $\mu$ and width $\sigma$ left-truncated at $m_{1,\rm low}$, $\lambda_0$ and $\lambda_1$ are the mixture weights of the continuum and the lower-mass peak, and $S(m_1 \mid m_{1,\rm low}, \delta_{m,1})$ is a Planck-taper that smoothly turns the distribution on over an interval $\delta_{m,1}$ above $m_{1,\rm low}$. The two Gaussian components capture the observed over densities near $\sim\!10\,M_\odot$ and $\sim\!35\,M_\odot$. The conditional mass-ratio density is a power law in $q$ with low-mass tapering of the secondary,

\begin{equation}
    p(q \mid m_1) \propto q^{\beta_q}\,
    S\!\left(m_1 q \mid m_{2,\rm low}, \delta_{m,2}\right),
    \label{eq:qmodel}
\end{equation}

normalized numerically at fixed $m_1$. The hyperparameter values, held fixed throughout and set to the GWTC-4.0 \textsc{Broken Power Law + 2 Peaks} posterior medians \citep{LIGOScientific:2025pvj}, are listed in Table~\ref{tab:massparams}.

\begin{table}[t]
    \centering
    \begin{tabular}{l c l}
        \hline\hline
        Parameter & Value & Description \\
        \hline
        $\alpha_1$      & $1.7283$            & Low-mass continuum slope \\
        $\alpha_2$      & $4.5117$            & High-mass continuum slope \\
        $m_{\rm break}$ & $35.6223\,M_\odot$    & Continuum break mass \\
        $\mu_1$         & $9.7637\,M_\odot$   & Lower Gaussian peak location \\
        $\sigma_1$      & $0.6492\,M_\odot$ & Lower peak width \\
        $\mu_2$         & $32.7629\,M_\odot$  & Upper Gaussian peak location \\
        $\sigma_2$      & $3.9181\,M_\odot$   & Upper peak width \\
        $m_{1,\rm low}$ & $5.0586$\,$M_\odot$ & Lower edge of $m_1$ taper \\
        $\delta_{m,1}$  & $4.3207$\,$M_\odot$ & $m_1$ low-mass smoothing scale \\
        $m_{\rm max}$   & $300\,M_\odot$   & Upper mass cutoff \\
        $\lambda_0$     & $0.3610$      & Continuum mixture weight \\
        $\lambda_1$     & $0.5861$      & Lower-peak mixture weight \\
        $\beta_q$       & $1.1710$            & Mass-ratio power-law slope \\
        $m_{2,\rm low}$ & ${3.5511}$\,$M_\odot$ & Lower edge of $m_2$ taper \\
        $\delta_{m,2}$  & $4.9101$\,$M_\odot$ & $m_2$ smoothing scale \\
        \hline\hline
    \end{tabular}
    \caption{Fixed hyperparameters of the \textsc{Broken Power Law + 2 Peaks} primary-mass model $p(m_1)$ and the mass-ratio model $p(q\mid m_1)$, set to the GWTC-4.0 population posterior medians (Data publicly available at \href{https://zenodo.org/records/16911563}{https://zenodo.org/records/16911563}) and held constant during the inference; they are not free parameters.}
    \label{tab:massparams}
\end{table}

\subsubsection{Expected rate and likelihood}

The hierarchical inference framework we adopt is the standard treatment of an inhomogeneous Poisson process in the presence of selection effects \citep{Loredo:2004nn,Mandel:2018mve,Talbot:2018cva,Vitale:2020aaz}, as used throughout GW population analyses. We collect the source-frame mass parameters into $\boldsymbol{m} \equiv (m_1, q)$, with $q = m_2/m_1$, and all mass integrals below run over this two-dimensional space. The analysis bin $\Delta m_k$ is defined by the primary-mass interval, $m_1 \in [m_{k,\rm low}, m_{k,\rm high}]$. The expected intrinsic differential merger rate per unit redshift and unit mass-parameter element is

\begin{equation}
\begin{split}
    \frac{d^2N_{\rm int}}{dz\,d\boldsymbol{m}}(z, \boldsymbol{m} \mid m_k, \mathcal{T}, R_0)
    & = T_{\rm obs}\, R_0(m_k)\,
    \mathcal{R}(z \mid m_k, \mathcal{T})\,
    \\ & \times p_{\rm pop}(\boldsymbol{m})\,
    \frac{dV_c}{dz}\, \frac{1}{1+z},
    \label{eq:Nint}
\end{split}
\end{equation}

where $p_{\rm pop}(\boldsymbol{m}) = p(m_1)\,p(q \mid m_1)$ is the fixed population mass density of Equation~(\ref{eq:poppop}), $T_{\rm obs}$ is the total observation time, and $dV_c/dz$ is the differential comoving volume element. This quantity counts all mergers in the universe regardless of detectability. The total expected number of detectable events in mass bin $m_k$ is

\begin{equation}
    N_{\rm exp}(m_k, \mathcal{T}, R_0) =
    \int_0^{z_{\rm max}} \int_{\Delta m_k}
    \frac{d^2N_{\rm int}}{dz\,d\boldsymbol{m}}\,
    \mathcal{S}(\boldsymbol{m}, z)\, d\boldsymbol{m}\, dz,
    \label{eq:ntot_full}
\end{equation}

where $\mathcal{S}(\boldsymbol{m}, z)$ is the probability that a binary with mass parameters $\boldsymbol{m}$ at redshift $z$ passes our detection criterion, defined as a false-alarm rate below $1\,\mathrm{yr}^{-1}$ and $d\boldsymbol{m} = dm_1\,dq$. The selection function is estimated from the LVK injection campaign.

The hierarchical likelihood for $N_{\rm det}(m_k)$ detected events in mass bin $m_k$ is
\begin{equation}
\begin{split}
    \mathcal{L}_k(\mathcal{T}, R_0) & =
    \exp\!\left[-N_{\rm exp}(m_k, \mathcal{T}, R_0)\right]
    \\ & \times \prod_{i=1}^{N_{\rm det}(m_k)}
    \mathcal{I}_i(m_k, \mathcal{T}, R_0),
    \label{eq:like_full}
\end{split}
\end{equation}
where the per-event likelihood integral is
\begin{equation}
\begin{split}
    \mathcal{I}_i & = \int_0^{z_{\rm max}}
    \int_{\Delta m_k}
    \frac{d^2N_{\rm int}}{dz\,d\boldsymbol{m}}(z, \boldsymbol{m} \mid m_k, \mathcal{T}, R_0)\,
    \\ & \times \frac{p_i(\boldsymbol{m}, z \mid d_i)}
    {\pi(\boldsymbol{m}, z)}\, d\boldsymbol{m}\, dz,
    \label{eq:per_event}
\end{split}
\end{equation}
with $p_i(\boldsymbol{m}, z \mid d_i)$ the single-event posterior given data $d_i$ and $\pi(\boldsymbol{m}, z)$ the parameter-estimation prior under which that posterior was produced.

\subsubsection{Monte Carlo evaluation and the role of the mass model}

Substituting Equation~(\ref{eq:Nint}) into Equation~(\ref{eq:per_event}) and evaluating the integral as a Monte Carlo sum over the $N_s$ posterior samples $\{\boldsymbol{m}_s^{(i)}, z_s^{(i)}\}_{s=1}^{N_s}$ of event $i$, where $\boldsymbol{m}_s^{(i)} = (m_{1,s}^{(i)}, q_s^{(i)})$, gives

\begin{equation}
\begin{split}
    \mathcal{I}_i & = \frac{T_{\rm obs}\, R_0(m_k)}{N_s}
    \sum_{s=1}^{N_s}
    \frac{p_{\rm pop}\!\big(\boldsymbol{m}_s^{(i)}\big)\,
    \mathbf{1}\!\left[m_{1,s}^{(i)} \in \Delta m_k\right]}
    {\pi\!\big(\boldsymbol{m}_s^{(i)}, z_s^{(i)}\big)}
    \\ & \times \mathcal{R}\!\big(z_s^{(i)} \mid m_k, \mathcal{T}\big)\,
    \frac{dV_c}{dz}\bigg|_{z_s^{(i)}}
    \frac{1}{1+z_s^{(i)}}.
    \label{eq:mc_sum}
\end{split}
\end{equation}

The mass distribution enters this expression through the explicit factor $p_{\rm pop}(\boldsymbol{m}_s^{(i)})$. We emphasize that this is an importance reweighting, not a marginalization over an unknown mass model: each posterior sample, originally drawn under the PE prior $\pi$, is reweighted by the ratio $p_{\rm pop}(\boldsymbol{m}_s^{(i)})/\pi(\boldsymbol{m}_s^{(i)}, z_s^{(i)})$ so that the sample population is mapped onto the fixed
\textsc{Broken Power Law + 2 Peaks} model of Equations~(\ref{eq:poppop})--(\ref{eq:qmodel}) with the hyperparameters of Table~\ref{tab:massparams}. The integral over mass is thereby evaluated against this fixed parametric distribution rather than against any free-floating or per-event mass model; no mass hyperparameters are inferred in this work. The indicator $\mathbf{1}[m_{1,s}^{(i)} \in \Delta m_k]$ restricts the sum to samples whose \emph{primary} mass falls in the analysis bin, so that each event contributes to bin $m_k$ in proportion to its fractional posterior support there. Within each primary-mass bin the mass-ratio dependence is integrated over by the Monte Carlo sum, weighted by the conditional density $p(q \mid m_1)$ of Equation~(\ref{eq:qmodel}). Grouping the Monte Carlo sum by redshift bin $j$, Equation~(\ref{eq:mc_sum}) reduces to

\begin{equation}
\begin{split}
    \mathcal{I}_i & = T_{\rm obs}\, R_0(m_k)
    \sum_{j=1}^{N_z}
    \mathcal{R}(z_j \mid m_k, \mathcal{T})\,
     \frac{dV_c}{dz}\bigg|_{z_j}
    \frac{\Delta z}{1+z_j}\, w_{ij}(m_k),
    \label{eq:binned}
\end{split}
\end{equation}

where $\Delta z = 0.1$ is the uniform bin width and the reweighted, mass-model-weighted fractional support of event $i$ in bin $(\Delta m_k, \Delta z_j)$ is

\begin{equation}
    w_{ij}(m_k) = \frac{1}{N_s}\sum_{s=1}^{N_s}
    \frac{p_{\rm pop}\!\big(\boldsymbol{m}_s^{(i)}\big)\,
    \mathbf{1}\!\left[m_{1,s}^{(i)} \in \Delta m_k\right]\,
    \mathbf{1}\!\left[z_s^{(i)} \in \Delta z_j\right]}
    {\pi\!\big(\boldsymbol{m}_s^{(i)}, z_s^{(i)}\big)}.
    \label{eq:weights}
\end{equation}

Equation~(\ref{eq:weights}) makes explicit that the population mass model $p_{\rm pop}$ acts as a per-sample weight: it is fixed, parametric, and applied identically to every event, so that differences between events in $w_{ij}$ reflect only their distinct posterior supports in $(\boldsymbol{m},z)$, not any event-specific mass assumption. Defining the intrinsic expected number of mergers in redshift bin $j$ as

\begin{equation}
\begin{split}
    N_{{\rm int},j}(m_k, \mathcal{T}, R_0) &=
    T_{\rm obs}\, R_0(m_k)\,
    \\& \times \mathcal{R}(z_j \mid m_k, \mathcal{T})\,
    \frac{dV_c}{dz}\bigg|_{z_j}
    \frac{\Delta z}{1+z_j},
    \label{eq:Nintj}
\end{split}
\end{equation}

the per-event integral becomes $\mathcal{I}_i \approx \sum_j N_{{\rm int},j}\, w_{ij}$, and the full likelihood is

\begin{equation}
\begin{split}
    \mathcal{L}_k(\mathcal{T}, R_0)
    &= \exp\!\left[- N_{\rm exp}(m_k, \mathcal{T}, R_0)\right] \\
    &\quad \times \prod_{i=1}^{N_{\rm det}(m_k)}
    \left[ \sum_{j=1}^{N_z}
    N_{{\rm int},j}(m_k, \mathcal{T}, R_0)\, w_{ij}(m_k)
    \right],
\end{split}
\label{eq:like}
\end{equation}

where the product runs over all $N_{\rm det}(m_k)$ detected events in mass bin $m_k$ and the sum runs over $N_z = 15$ uniform bins of width $\Delta z = 0.1$ spanning $0 \leq z \leq 1.5$. The sum over $j$ marginalizes each event's redshift uncertainty against the predicted merger-rate shape, correctly handling events whose posteriors span multiple bins.

\subsubsection{Marginalization over $R_0$ and trajectory evidence}

For each trajectory $\mathcal{T}$, the DTD shape $p_t(t_d \mid \mathcal{T})$ is fully specified by construction, so the only free parameter is the local rate amplitude $R_0(m_k)$. We adopt a uniform prior $\pi(R_0) = \mathcal{U}(1, 200)\,\mathrm{Gpc^{-3}\,yr^{-1}}$ and use the nested sampler \textsc{Dynesty} \citep{speagle2020dynesty} to evaluate the likelihood for each trajectory, marginalizing over $R_0(m_k)$. The Bayesian evidence for each trajectory is

\begin{equation}
    \mathcal{Z}_k(\mathcal{T}) = \int
    \mathcal{L}_k(\mathcal{T}, R_0)\, \pi(R_0)\, dR_0,
\end{equation}

which quantifies how well a given DTD shape explains the observed redshift distribution of BBH mergers in mass bin $m_k$, after marginalizing over $R_0$. Because the local rate amplitude $R_0(m_k)$ is defined as the normalization of $\mathcal{R}(z \mid m_k, \mathcal{T})$ at $z = 0$ over the analysis redshift range $0 \leq z \leq z_{\rm max} = 1.5$, it corresponds to the present-day merger rate of the sub-population whose mergers fall within this range. The best-supported trajectory is

\begin{equation}
    \mathcal{T}_{\rm best}(m_k) =
    \underset{\mathcal{T}}{\arg\max}\;\ln\mathcal{Z}_k(\mathcal{T}).
\end{equation}

The overall workflow of our framework is summarized in Figure~\ref{fig:Flowchart}.

\section{Data Characterization}
\label{sec:data}
\subsection{Mass Binning and Redshift Distribution}

We analyze BBH mergers observed until GWTC-4 \citep{LIGOScientific:2025slb} catalog. For each event we use the released parameter-estimation posterior samples of the source-frame primary mass $m_1$, the mass ratio $q = m_2/m_1$, and the redshift $z$, working throughout in the $(m_1, q)$ parameterization (Section~\ref{sec:bayesian}). Each sample is reweighted from its parameter-estimation prior $\pi_{\rm PE}(m_1, q, z)$ to the population model as described in Section~\ref{sec:bayesian}; we retain events whose secondary source mass exceeds $3\,M_\odot$ to select confident BBHs. We adopt a spatially flat $\Lambda$CDM cosmology with $H_0 = 67.74\,\mathrm{km\,s^{-1}\,Mpc^{-1}}$ and $\Omega_m = 0.3075$ \citep{Planck:2018vyg}, consistently for the event posteriors, the injection set, and all comoving-volume and lookback-time integrals.

We divide the primary-mass distribution into two analysis bins, low-mass systems ($20$--$40\,M_\odot$) and high-mass systems ($40$--$100\,M_\odot$), and infer the DTD and local merger rate independently in each (events contribute to a bin in proportion to their fractional posterior support there; Section~\ref{sec:bayesian}).

Detections in the real catalog are defined by a False Alarm Rate below $1\,\mathrm{yr}^{-1}$ in at least one GW search pipeline, yielding the confident BBH sample of $153$ events, consistent with the selection adopted in \citet{LIGOScientific:2025pvj} and \citet{Tong:2025wpz}. The detection probability as a function of source parameters, $\mathcal{S}(m_1, q, z)$, is estimated LVK injection campaign \citep{Essick:2025zed}; its role in the rate likelihood and its Monte Carlo evaluation are described in Section~\ref{sec:bayesian}.

\subsection{Grid Configuration}

The primary grid configuration used in this work is the asymmetric $7 \times 5$ grid, which provides finer sampling in the delay time axis while maintaining five probability levels. The delay time nodes are logarithmically spaced:
\begin{equation}
\{t_d^{(i)}\}_{i=1}^7 = \{0.50,\, 0.85,\, 1.44,\, 2.45,\, 4.16,\, 7.07,\, 12.0\}~\mathrm{Gyr}.
\end{equation}
The probability density axis spans:
\begin{equation}
\{\log_{10} p_t^{(j)}\}_{j=1}^5 = \{2.0,\, 1.5,\, 1.0,\, 0.5,\, 0.0\}.
\end{equation}

The logarithmic spacing of the delay time axis ensures dense sampling at short delays where the DTD shape has the most physical discriminating power. A complementary analysis using a $6 \times 5$ grid is presented in Appendix~\ref{app:6x5} for robustness verification.

\section{Results}
\label{sec:result}

We restrict to component masses above $20\,M_\odot$ and focus on two broad intervals: $20$--$40\,M_\odot$ and $40$--$100\,M_\odot$. We present results for three complementary assumptions about the underlying star formation history: (i) the standard metallicity-averaged Madau-Dickinson SFRD, (ii) the high-metallicity effective SFRD ($Z > 0.1\,Z_\odot$) from \citet{Chruslinska2019MNRAS}, and (iii) the low-metallicity effective SFRD ($Z < 0.1\,Z_\odot$) from \citet{Chruslinska2019MNRAS}. In all cases we use the $7\times5$ logarithmically spaced grid. Results for an alternative $6\times5$ grid configuration are deferred to Appendix~\ref{app:6x5}.

\begin{figure*}[ht]
    \centering
    \includegraphics[width=0.32\textwidth]{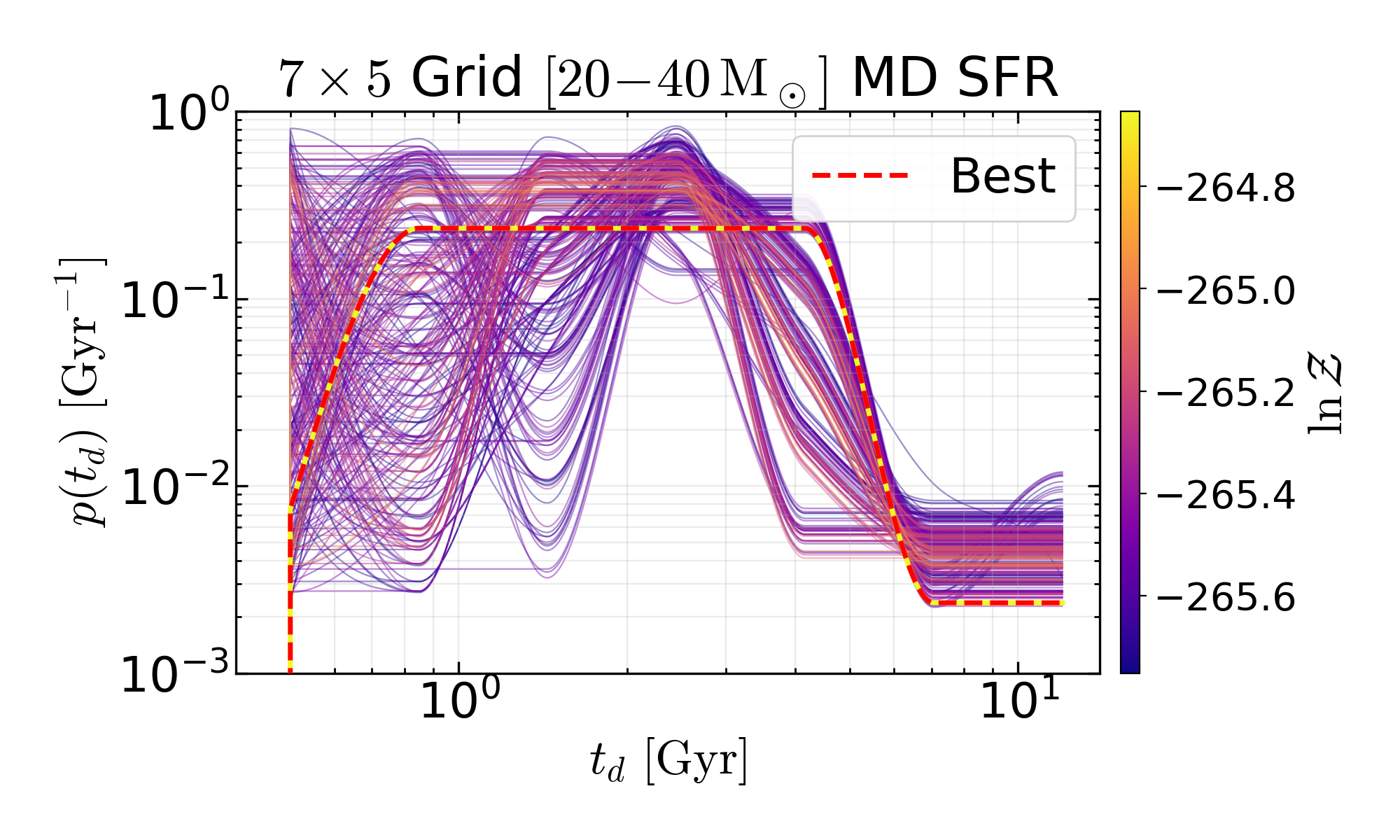}
    \includegraphics[width=0.32\textwidth]{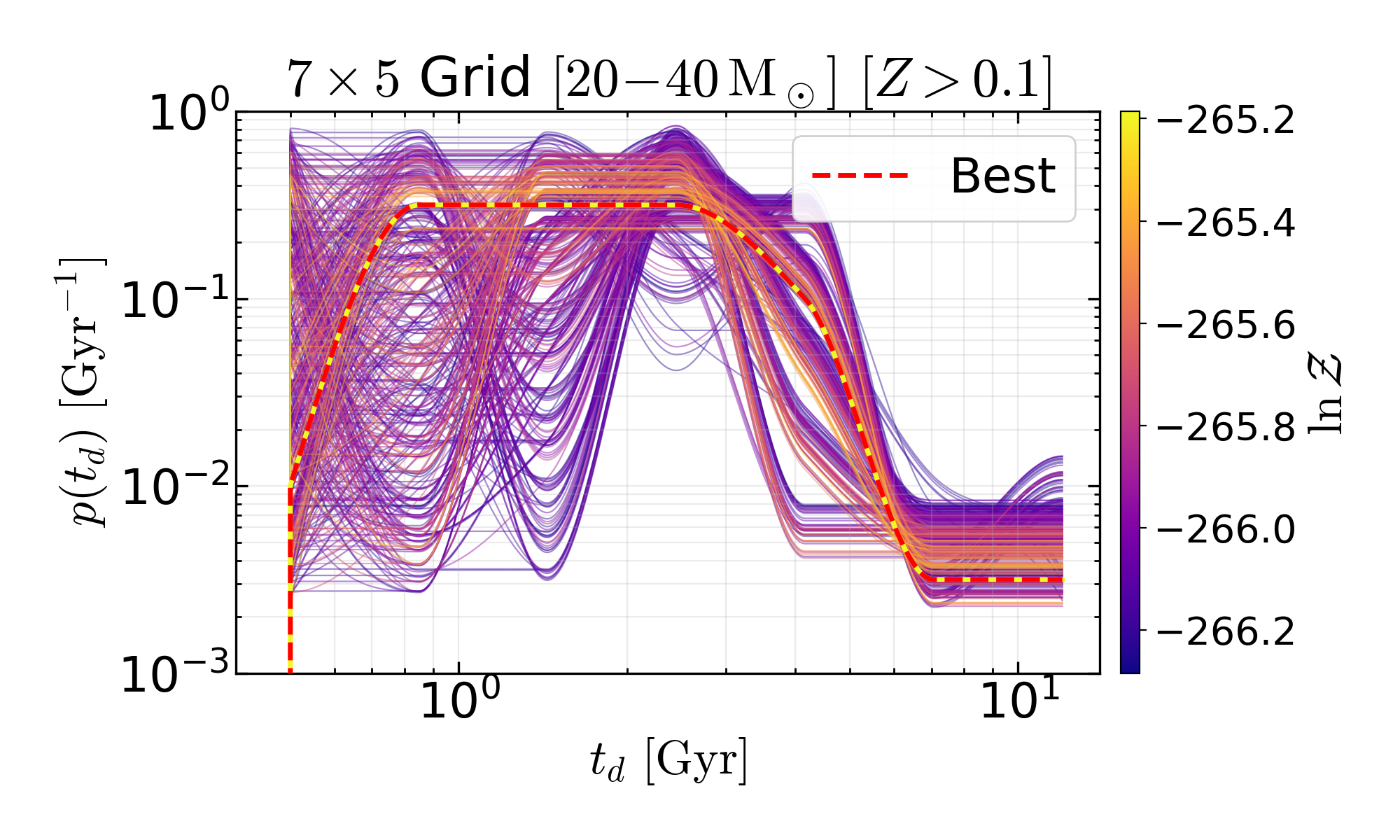}
    \includegraphics[width=0.32\textwidth]{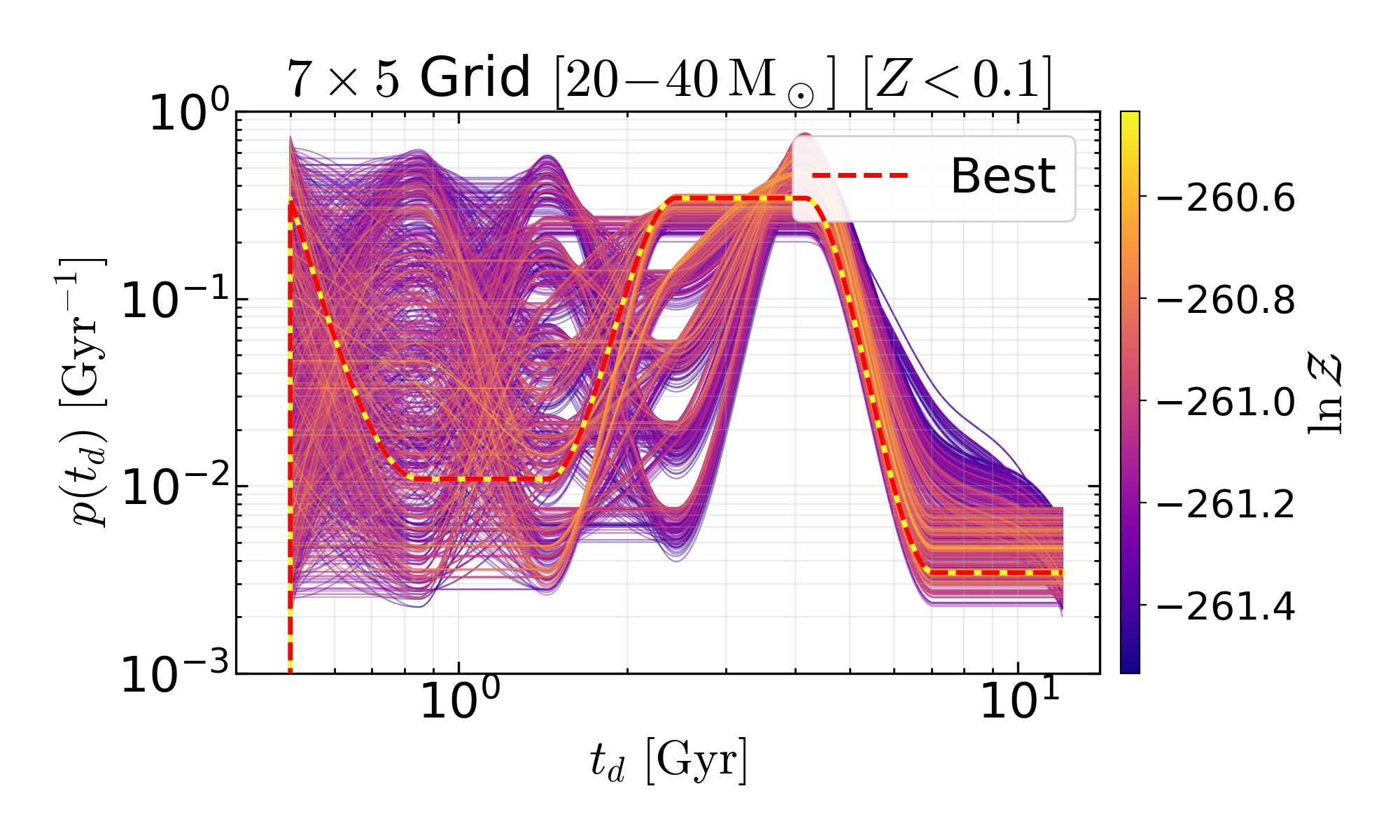}\\[6pt]
    \includegraphics[width=0.32\textwidth]{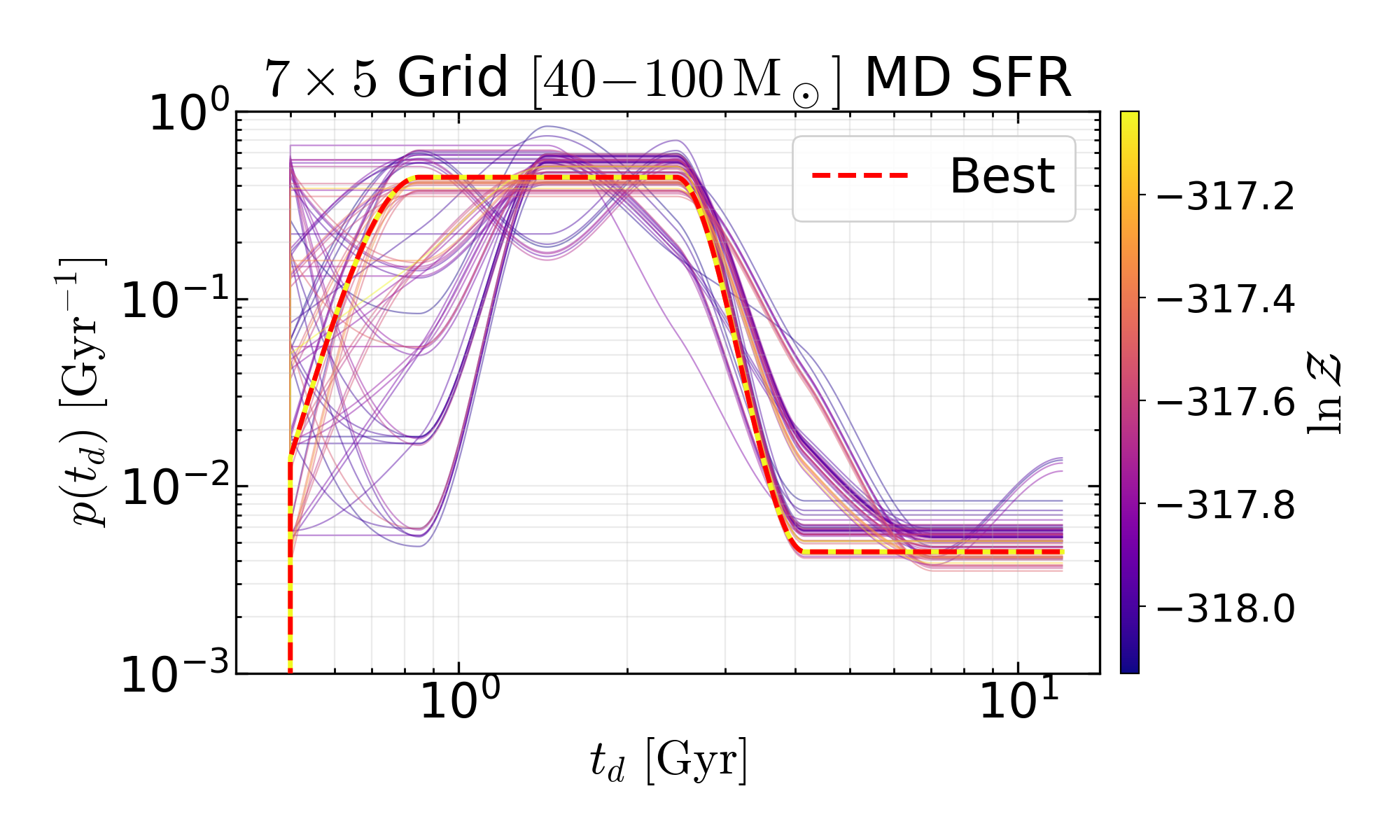}
    \includegraphics[width=0.32\textwidth]{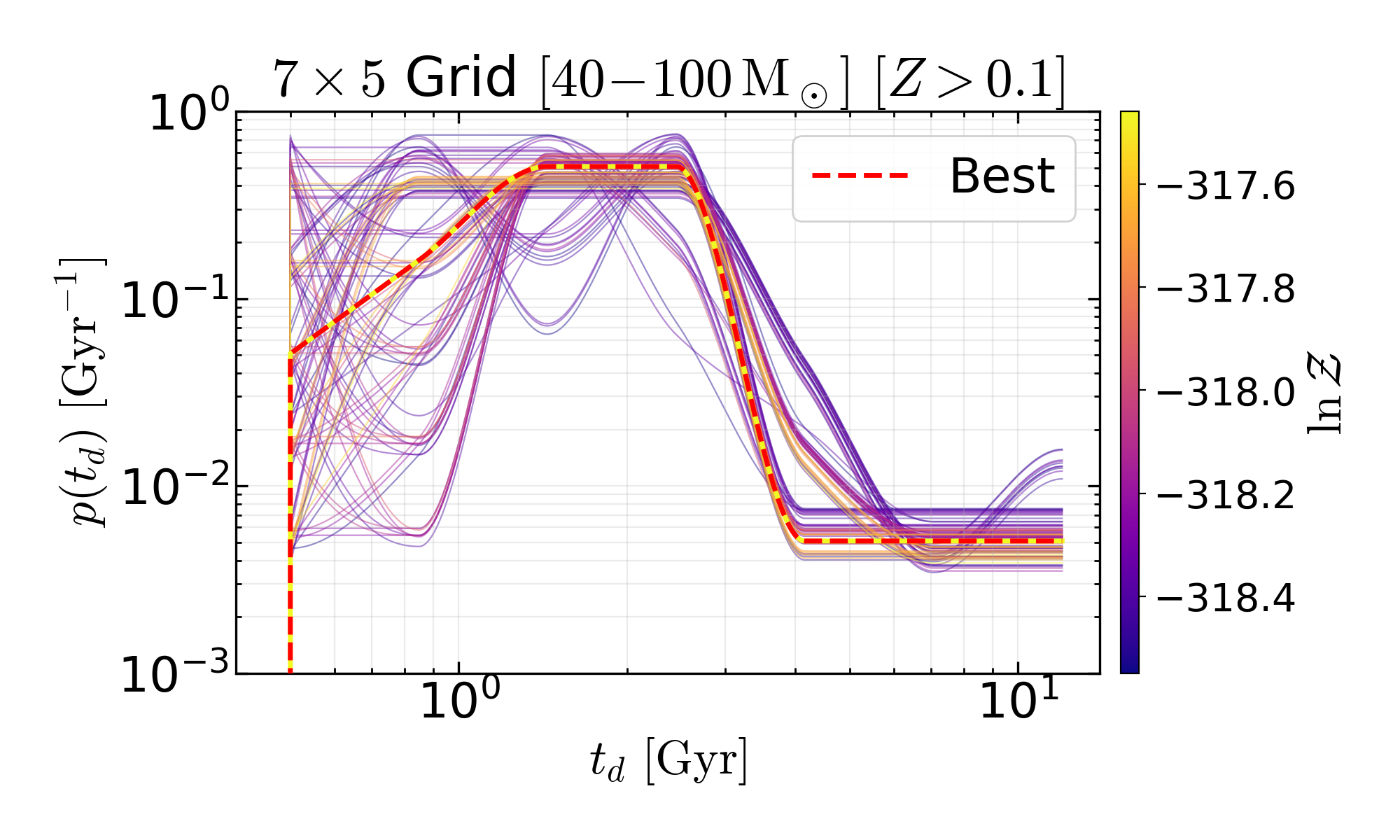}
    \includegraphics[width=0.32\textwidth]{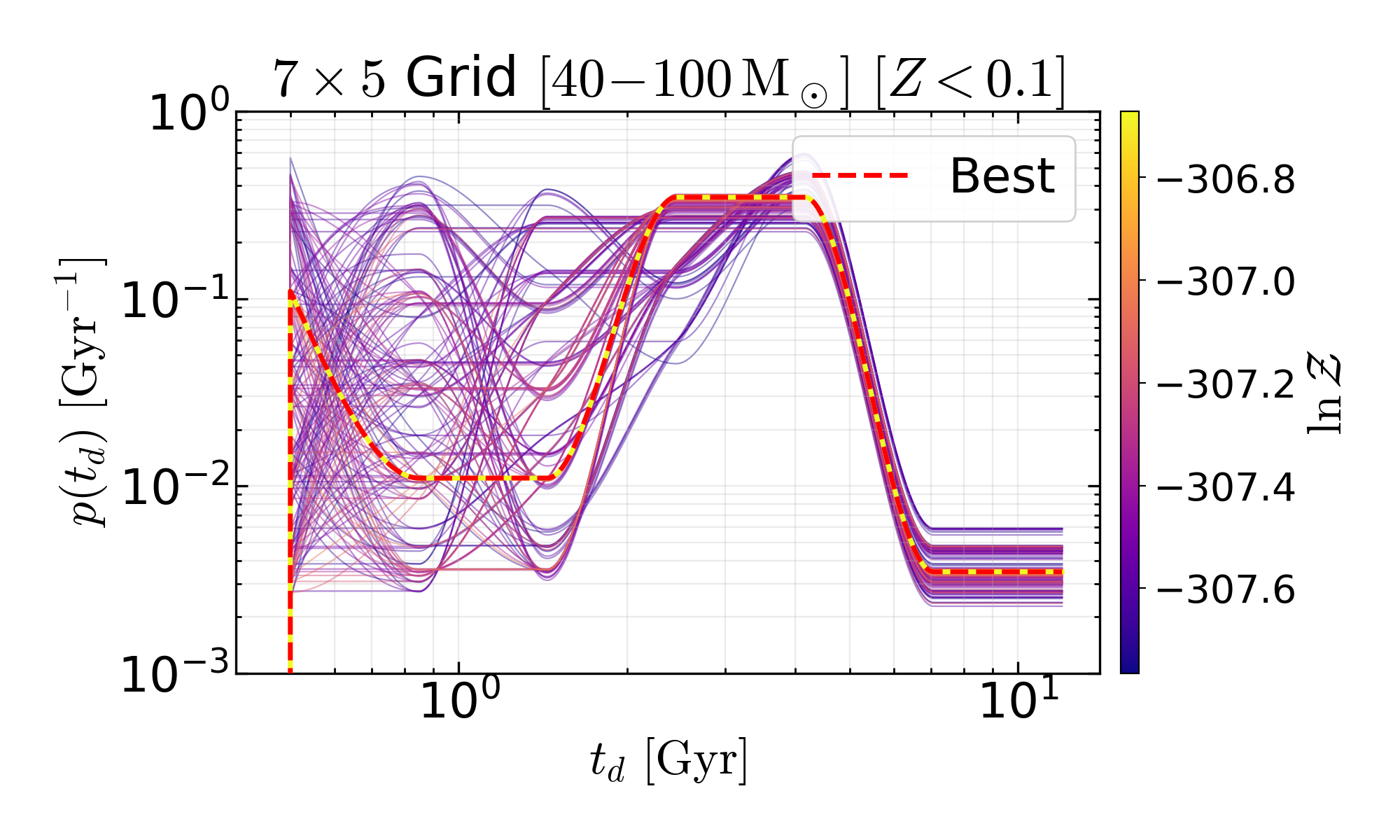}
    \caption{Reconstructed DTD trajectories $p(t_d)$ [Gyr$^{-1}$] within Bayes factor $\mathrm{BF} < 3$ ($\Delta\ln\mathcal{Z} < \ln 3$ relative to the highest-evidence trajectory) of the best-fit solution, for the $7\times5$ logarithmically spaced grid. Top row: $20$--$40\,M_\odot$ bin. Bottom row: $40$--$100\,M_\odot$ bin. Left to right: standard Madau-Dickinson SFRD, high-metallicity effective SFRD ($Z > 0.1\,Z_\odot$), and low-metallicity effective SFRD ($Z < 0.1\,Z_\odot$). Lines are coloured by $\ln\mathcal{Z}$ from purple (near the BF$=3$ boundary) to yellow (near the best-fit evidence); the dashed red curve with yellow outline marks the single highest-evidence trajectory. For both mass bins the highest-evidence trajectories favour an extended plateau of intermediate-to-long delay times followed by a decline at the longest delays, with the low-metallicity SFRD producing the most sharply structured DTD support.}
    \label{fig:dtd_all}
\end{figure*}

We organize the presentation of the results by mass bin. For each bin we discuss, in turn, the reconstructed delay time distribution and the ensemble of high-evidence trajectories (Figure~\ref{fig:dtd_all}), the associated BBH merger rate evolution $\mathcal{R}(z)$ that follows from convolving these trajectories with the star formation history (Figure~\ref{fig:Rz_all}), and the corresponding Bayesian evidence and local merger rate $R_0 = \mathcal{R}(z=0)$ (Figure~\ref{fig:evidence_all}). Because the merger rate and the local rate are derived from the same trajectories as the DTD, we present all three quantities together for each mass bin rather than in separate sections, so that the inferential chain from the reconstructed DTD to the associated $\mathcal{R}(z)$ and $R_0$ is made explicit.

\subsection{Reconstructed DTD, Merger Rate, and Local Rate}
\label{sec:dtd_results}

Figure~\ref{fig:dtd_all} shows all DTD trajectories within a Bayes factor of $\mathrm{BF} < 3$ (i.e., $\Delta\ln\mathcal{Z} < \ln 3$ relative to the highest-evidence trajectory) of the best-fit solution, for both mass bins and all three SFRD assumptions. Each trajectory is coloured by its $\ln\mathcal{Z}$ value, with yellow indicating trajectories closest to the best-fit evidence and purple indicating those near the BF$=3$ boundary. The dashed red curve with yellow outline marks the single highest-evidence trajectory. The associated physical merger rate $\mathcal{R}(z)$ for the same BF$<3$ trajectories is shown in Figure~\ref{fig:Rz_all}, and the distribution of Bayesian log-evidence $\ln\mathcal{Z}$ against the local rate $R_0$ is shown in Figure~\ref{fig:evidence_all}, with the right panel of each subfigure giving the marginalized $R_0$ histogram and the red star marking the highest-evidence trajectory.

Before discussing the individual mass bins, we caution that not all features of the reconstructed DTD are equally well constrained. As is visible in Figure~\ref{fig:dtd_all}, the scatter among the BF$<3$ trajectories is substantially larger at short delay times ($t_d \lesssim 1$~Gyr) than at intermediate and long delays, in both mass bins and under all three SFRD assumptions. This reflects the limited sensitivity of the current data to the short-delay portion of the DTD: because relatively few detectable mergers over the analysis redshift range $0 \le z \le 1.5$ map onto very short delays, the probability density at small $t_d$ is only weakly constrained, and the apparent structure there, including the short-delay suppression or dip seen in several best-fit trajectories, carries little statistical significance and should not be over-interpreted. The feature that is robustly constrained is instead the \emph{slope} of the DTD across intermediate-to-long delays, where the BF$<3$ envelope is narrow: the data consistently favour an elevated probability density at intermediate delays that declines toward the longest delays, and it is this overall trend, rather than the detailed low-$t_d$ behavior, that drives the mass-dependent differences discussed below.

\subsubsection{Low-mass bin: $20$--$40\,M_\odot$}

For the $20$--$40\,M_\odot$ bin, the BF$<3$ DTD envelope is broad under all three SFRD assumptions, confirming that the current data do not strongly discriminate among different delay time distributions for this mass bin. The ensemble of BF$<3$ trajectories spans the full delay time range from $t_d \sim 0.5$ to $\sim$13 Gyr, but the highest-evidence trajectories share a common qualitative feature across all three SFRD assumptions: a clear preference for an extended plateau of elevated probability density at intermediate delay times, followed by a decline toward the longest delays. As noted above, the scatter of the BF$<3$ trajectories is largest at short delays, so for this bin the robustly constrained feature is the negative slope of the DTD from intermediate to long delays rather than any structure at small $t_d$.

The precise structure of this preferred plateau, however, depends on the assumed star formation history in a physically understandable way.

Under the standard Madau-Dickinson SFRD (Figure~\ref{fig:dtd_all}, top left), the best-fit trajectory rises from short delays to a broad plateau of $p(t_d) \approx 0.2$--$0.3\,\mathrm{Gyr^{-1}}$ extending across $t_d \sim 1$--$5$ Gyr, before dropping by roughly an order of magnitude at $t_d \gtrsim 5$--$6$ Gyr. The surrounding BF$<3$ envelope is broad and admits a wide range of shapes, reflecting the relatively smooth and broad redshift dependence of the Madau-Dickinson SFRD, which does not strongly single out any narrow delay time range.

Under the high-metallicity effective SFRD ($Z > 0.1\,Z_\odot$, Figure~\ref{fig:dtd_all}, top middle), which is the physically motivated assumption for this mass bin, the best-fit DTD and the BF$<3$ envelope are closely similar to the MD SFR case, with the same intermediate-delay plateau spanning $t_d \sim 1$--$5$ Gyr and a comparable decline at longer delays. This similarity is physically understandable from the left panel of Figure~\ref{fig:normalized_sfrd}: the $Z > 0.1\,Z_\odot$ SFRD tracks the Madau-Dickinson shape closely at low to intermediate redshifts ($z \lesssim 2$), where most of the detectable merger population originates. Since the two SFRDs have similar shapes over the redshift range that dominates the likelihood, they naturally produce similar constraints on the DTD. This agreement reinforces the stability of the low-mass DTD reconstruction and confirms that, for the $20$--$40\,M_\odot$ bin, the inferred DTD is insensitive to whether one uses the metallicity-averaged or the high-metallicity effective SFRD.

Under the low-metallicity effective SFRD ($Z < 0.1\,Z_\odot$, Figure~\ref{fig:dtd_all}, top right), which is not the physically motivated assumption for the $20$--$40\,M_\odot$ bin but is shown for completeness, the best-fit DTD develops a more sharply structured shape: the highest-evidence trajectory shows a pronounced concentration of support at intermediate delays $t_d \sim 2$--$4$ Gyr, separated from the short-delay region by a dip near $t_d \sim 0.7$--$1.7$ Gyr (though, as cautioned above, this low-$t_d$ dip lies in the weakly constrained short-delay region and should not be over-interpreted), and the BF$<3$ envelope shows the most pronounced striated pattern of the three cases, reflecting discrete families of trajectories with similar intermediate-delay configurations. This sharper structure is directly explained by the right panel of Figure~\ref{fig:normalized_sfrd}: the $Z < 0.1\,Z_\odot$ SFRD rises steeply from $z=0$, peaks at $z \sim 2$--$4$, and remains strongly elevated at high redshift compared to the Madau-Dickinson model. Since this SFRD places a much larger fraction of star formation at high redshift, the DTD support that best reproduces the observed mergers is more tightly concentrated at the intermediate delays that map this high-redshift star formation onto the detected population.

Despite these SFRD-dependent variations in the detailed shape of the DTD support, the overall conclusion for the $20$--$40\,M_\odot$ bin is consistent across all three cases: the data favour an extended plateau of intermediate-to-long delay times, with a broad BF$<3$ envelope and no strong preference for very short delays.

The merger rate evolution associated with these trajectories is shown in the top row of Figure~\ref{fig:Rz_all}. For the $20$--$40\,M_\odot$ bin, the best-fit merger rate rises from its local value to a clear peak at intermediate redshift and then declines, with a broad BF$<3$ envelope surrounding the best-fit curve at all redshifts, consistent with the broad DTD support identified in Section~\ref{sec:dtd_results}. The amplitude and position of this peak vary with the SFRD assumption in a physically understandable way.

Under the standard Madau-Dickinson SFRD (Figure~\ref{fig:Rz_all}, top left), the best-fit trajectory starts at a local rate of $\mathcal{R}(z=0) \approx 22\,\mathrm{Gpc^{-3}\,yr^{-1}}$, rises to a peak of $\sim$82 Gpc$^{-3}$ yr$^{-1}$ at $z \sim 0.8$, and declines to $\sim$44 Gpc$^{-3}$ yr$^{-1}$ by $z \sim 1.5$, corresponding to a peak-to-local ratio of $\sim$3.7. The BF$<3$ envelope is wide at all redshifts, with the highest-evidence trajectories (yellow-orange) tracing the central peaked behavior while lower-evidence trajectories span a broad range of amplitudes at high redshift.

Under the high-metallicity effective SFRD ($Z > 0.1\,Z_\odot$, Figure~\ref{fig:Rz_all}, top middle), which is the physically motivated assumption for this mass bin, the best-fit merger rate evolution is closely similar to the MD SFR case. The local rate is $\mathcal{R}(z=0) \approx 22$--$23\,\mathrm{Gpc^{-3}\,yr^{-1}}$, the peak reaches $\sim$92 Gpc$^{-3}$ yr$^{-1}$ at $z \sim 0.85$--$0.9$, and the curve declines to $\sim$53 Gpc$^{-3}$ yr$^{-1}$ by $z \sim 1.5$, with a comparable peak-to-local ratio of $\sim$4. The close similarity between the two cases is consistent with the close agreement between the MD SFR and the $Z > 0.1\,Z_\odot$ SFRD at low to intermediate redshifts seen in Figure~\ref{fig:normalized_sfrd}, and confirms that the $20$--$40\,M_\odot$ merger rate evolution is insensitive to the distinction between these two SFRD assumptions.

Under the low-metallicity effective SFRD ($Z < 0.1\,Z_\odot$, Figure~\ref{fig:Rz_all}, top right), which is not the physically motivated assumption for this mass bin but is shown for completeness, the merger rate evolution shows a much more pronounced peak than the previous two cases. The best-fit local rate is somewhat lower at $\mathcal{R}(z=0) \approx 18\,\mathrm{Gpc^{-3}\,yr^{-1}}$, but the curve rises steeply to a peak of $\sim$208 Gpc$^{-3}$ yr$^{-1}$ at $z \sim 1.05$ before declining to $\sim$140 Gpc$^{-3}$ yr$^{-1}$ by $z \sim 1.5$, corresponding to a much larger peak-to-local ratio of $\sim$11. This dramatically amplified intermediate-redshift feature is a direct consequence of the steeply rising low-metallicity SFRD, which places a large fraction of star formation at high redshift and, convolved with the intermediate-delay DTD support of this mass bin, maps that star formation into a strongly elevated merger rate at $z \sim 1$. The BF$<3$ envelope is correspondingly wider and reaches higher amplitudes than in the MD SFR and $Z > 0.1\,Z_\odot$ cases.

The Bayesian evidence landscape and the associated local rate for this bin are shown in the top row of Figure~\ref{fig:evidence_all}. For the $20$--$40\,M_\odot$ bin, the evidence landscape is broadly similar across all three SFRD assumptions, reflecting the insensitivity of the low-mass DTD reconstruction to the choice of star formation history. Under the MD SFR (Figure~\ref{fig:evidence_all}, top left), the best-fit trajectory has $\ln\mathcal{Z}_{\rm best} \approx -264$ at $R_0 \approx 21\,\mathrm{Gpc^{-3}\,yr^{-1}}$, with the marginalized $R_0$ distribution extending to $\sim$100 Gpc$^{-3}$ yr$^{-1}$. Under the $Z > 0.1\,Z_\odot$ SFRD (Figure~\ref{fig:evidence_all}, top middle), the evidence landscape is broadly similar with $\ln\mathcal{Z}_{\rm best} \approx -266$ at $R_0 \approx 22\,\mathrm{Gpc^{-3}\,yr^{-1}}$, consistent with the close agreement between these two SFRDs at low to intermediate redshifts. Under the $Z < 0.1\,Z_\odot$ SFRD (Figure~\ref{fig:evidence_all}, top right), the best-fit is at $\ln\mathcal{Z}_{\rm best} \approx -260.5$ and $R_0 \approx 18\,\mathrm{Gpc^{-3}\,yr^{-1}}$, with a more compact $R_0$ distribution extending to $\sim$64 Gpc$^{-3}$ yr$^{-1}$ and a more sharply striated arc structure reflecting discrete families of trajectories. In all three cases the best-fit local rate sits at the low-$R_0$ edge of a broad marginalized distribution, consistent with the broad, weakly constrained DTD of the low-mass bin.

\subsubsection{High-mass bin: $40$--$100\,M_\odot$}

For the $40$--$100\,M_\odot$ bin, the BF$<3$ DTD envelope is significantly tighter compared to the low-mass bin under all three SFRD assumptions, indicating that the high-mass data discriminate more strongly among different DTD shapes. Across all three cases, the highest-evidence trajectories show a clear, well-defined plateau of elevated probability density at intermediate delay times, with a sharper rise from short delays and a steeper decline at long delays than in the low-mass bin. As in the low-mass bin, the trajectory scatter remains largest at short delays $t_d \lesssim 1$~Gyr, so the statistically robust feature of this reconstruction is again the steep slope of the DTD across intermediate-to-long delays, which is more sharply defined here (narrower BF$<3$ envelope) than in the low-mass bin; the detailed behaviour at very short delays remains weakly constrained. This more sharply defined intermediate-delay preference is a consistent feature across all three SFRD assumptions, though the width and position of the supported plateau vary with the assumed star formation history.

Under the standard Madau-Dickinson SFRD (Figure~\ref{fig:dtd_all}, bottom left), the best-fit DTD rises steeply from $t_d \sim 0.5$ Gyr to a plateau of $p(t_d) \approx 0.4\,\mathrm{Gyr^{-1}}$ across $t_d \sim 1$--$3$ Gyr, then drops sharply by roughly two orders of magnitude at $t_d \gtrsim 3$--$4$ Gyr. The suppression at both ends of the delay time range is more pronounced than for the low-mass bin under the same SFRD, and the overall width of the BF$<3$ envelope is visibly narrower. The concentration at intermediate delays under the MD SFR already indicates a tendency toward characteristic merging timescales for this mass bin, even before invoking any metallicity-dependent star formation assumption.

Under the high-metallicity effective SFRD ($Z > 0.1\,Z_\odot$, Figure~\ref{fig:dtd_all}, bottom middle), which is not the physically motivated assumption for the $40$--$100\,M_\odot$ bin, the best-fit DTD and the BF$<3$ envelope are closely similar to the MD SFR case, with a comparable intermediate-delay plateau and a similarly narrow overall width. This similarity follows from the close agreement between the MD SFR and the $Z > 0.1\,Z_\odot$ SFRD at the low-to-intermediate redshifts that dominate the likelihood (Figure~\ref{fig:normalized_sfrd}, left panel). Despite not being the physically favoured assumption for this mass bin, the BF$<3$ envelope remains noticeably tighter than the corresponding low-mass panel, confirming that the high-mass data are more discriminating regardless of the assumed SFRD.

Under the physically motivated low-metallicity effective SFRD ($Z < 0.1\,Z_\odot$, Figure~\ref{fig:dtd_all}, bottom right), the best-fit DTD again shows a sharply structured shape, with a dip at short-to-intermediate delays $t_d \sim 0.7$--$1.7$ Gyr (which, as cautioned above, lies in the weakly constrained short-delay region) followed by a well-defined plateau at $t_d \sim 2.3$--$4$ Gyr and a steep decline at longer delays. This structured shape is directly understandable from the right panel of Figure~\ref{fig:normalized_sfrd}: the $Z < 0.1\,Z_\odot$ SFRD rises steeply from $z=0$, peaks broadly at $z \sim 2$--$4$, and remains strongly elevated at high redshifts up to $z \sim 7$--$8$, so that the DTD support is concentrated at the intermediate delays that efficiently map this high-redshift star formation onto the redshifts sampled by the current GW catalog, while retaining the suppression at very short delays that is characteristic of the high-mass bin.

Despite these SFRD-dependent variations in the detailed shape of the supported delay time range, the overall conclusion for the $40$--$100\,M_\odot$ bin is consistent across all three cases: the data show a clear, comparatively well-defined preference for intermediate delay times $t_d \sim 2$--$4$ Gyr. The tighter BF$<3$ envelopes compared to the low-mass bin, and the consistent suppression of very short delay times, together suggest that higher-mass BBHs tend to merge on somewhat more characteristic timescales than their lower-mass counterparts.

\begin{figure*}[ht]
    \centering
    \includegraphics[width=0.32\textwidth]{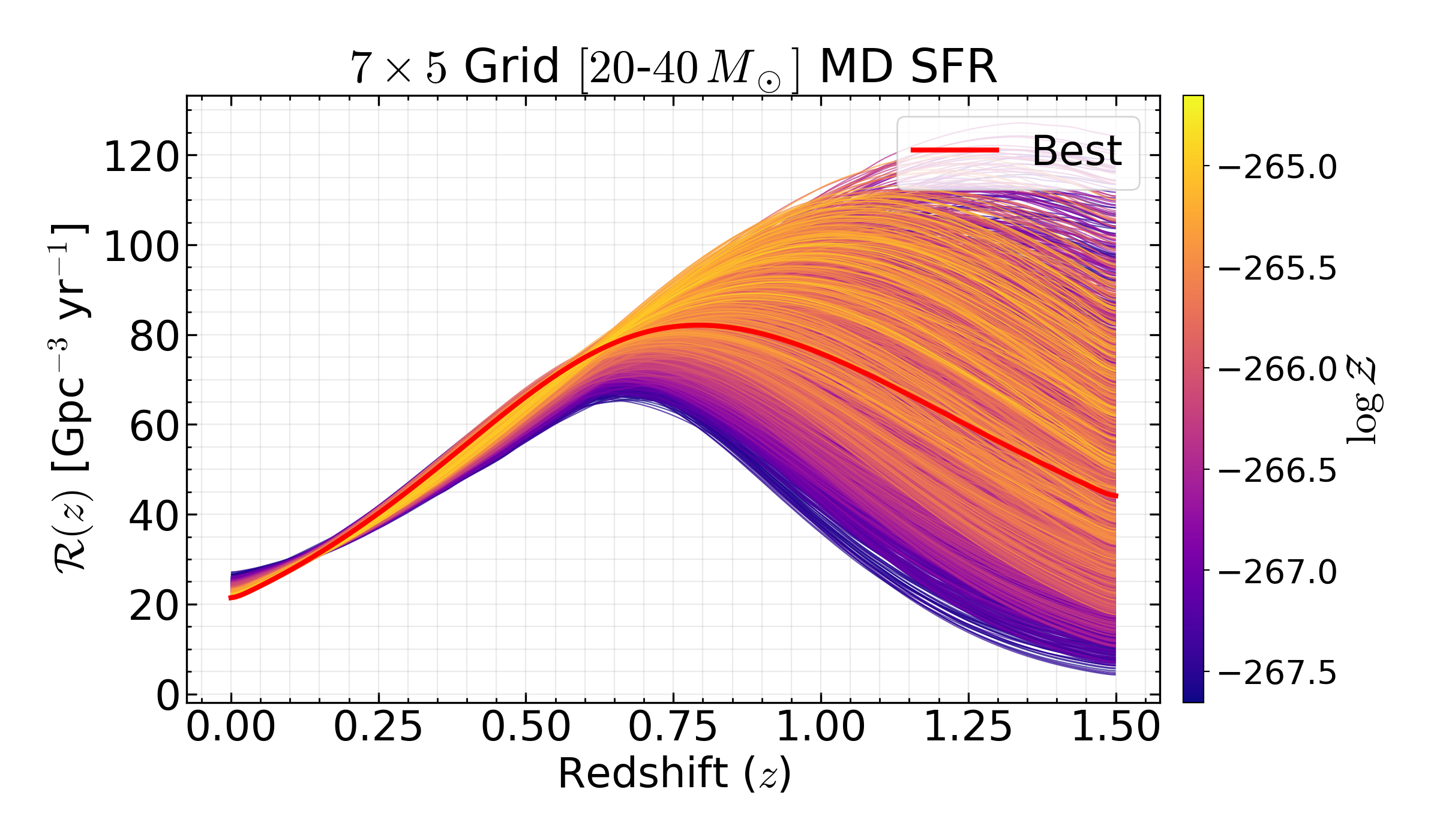}
    \includegraphics[width=0.32\textwidth]{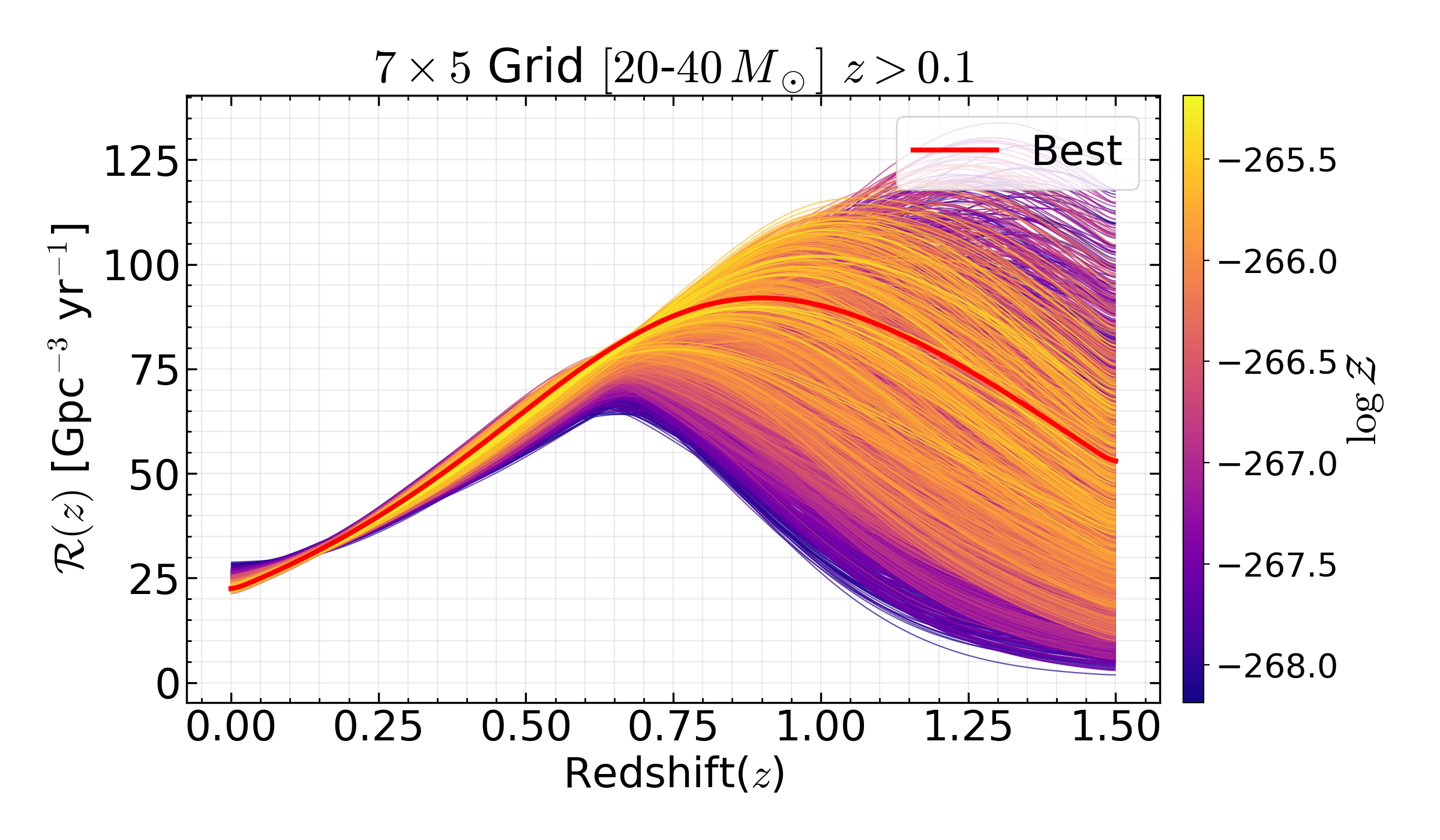}
    \includegraphics[width=0.32\textwidth]{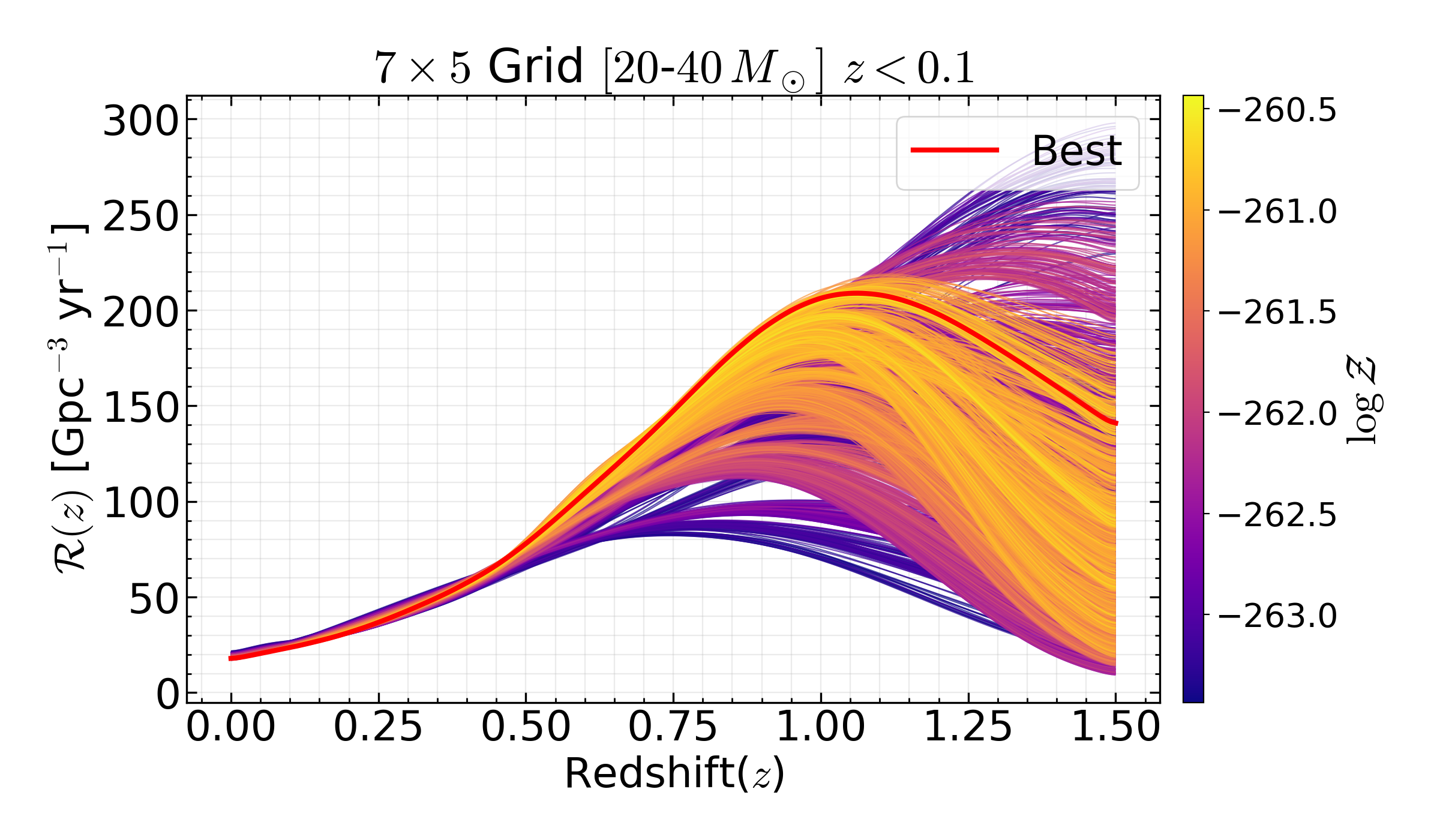}\\[6pt]
    \includegraphics[width=0.32\textwidth]{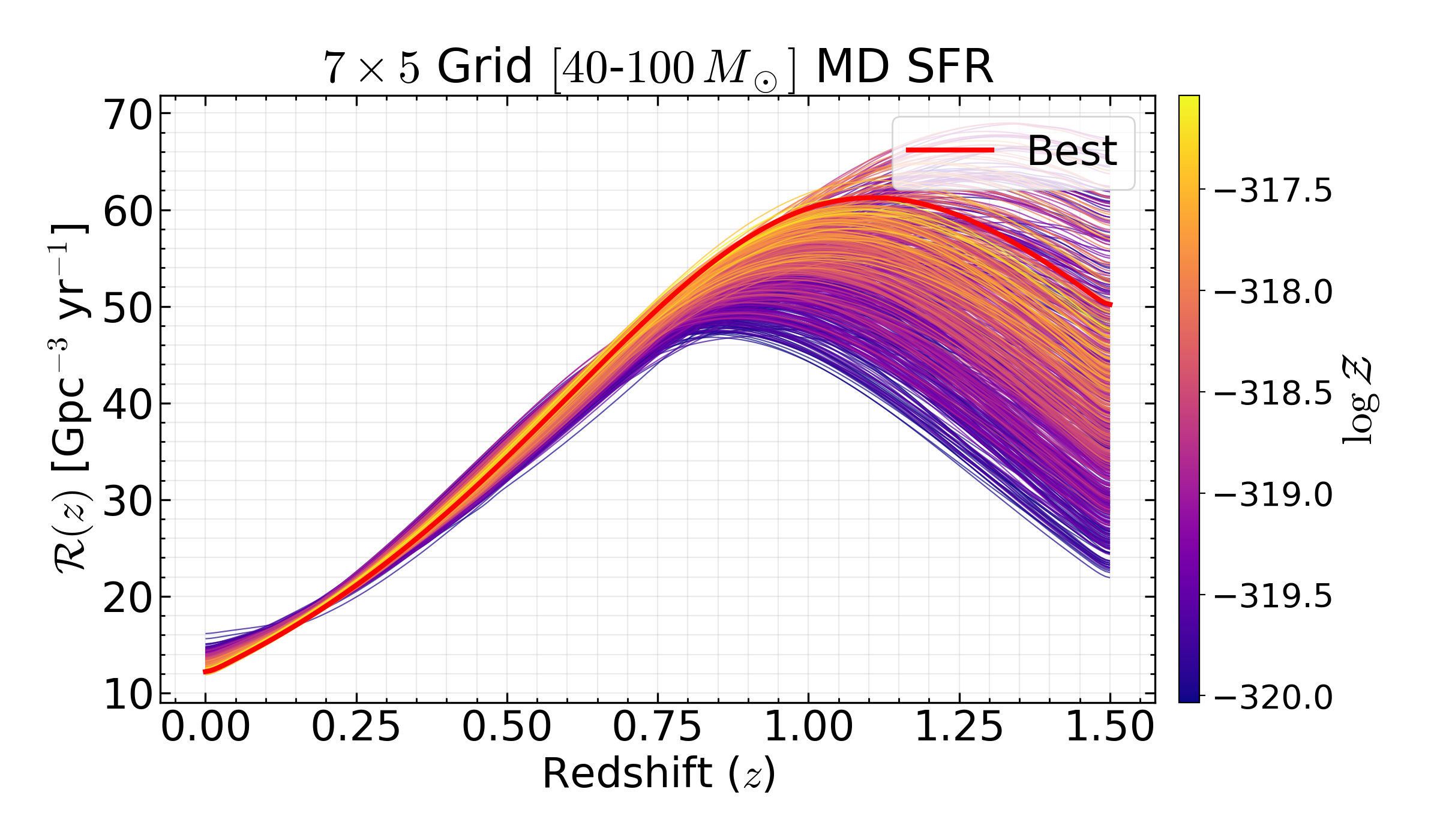}
    \includegraphics[width=0.32\textwidth]{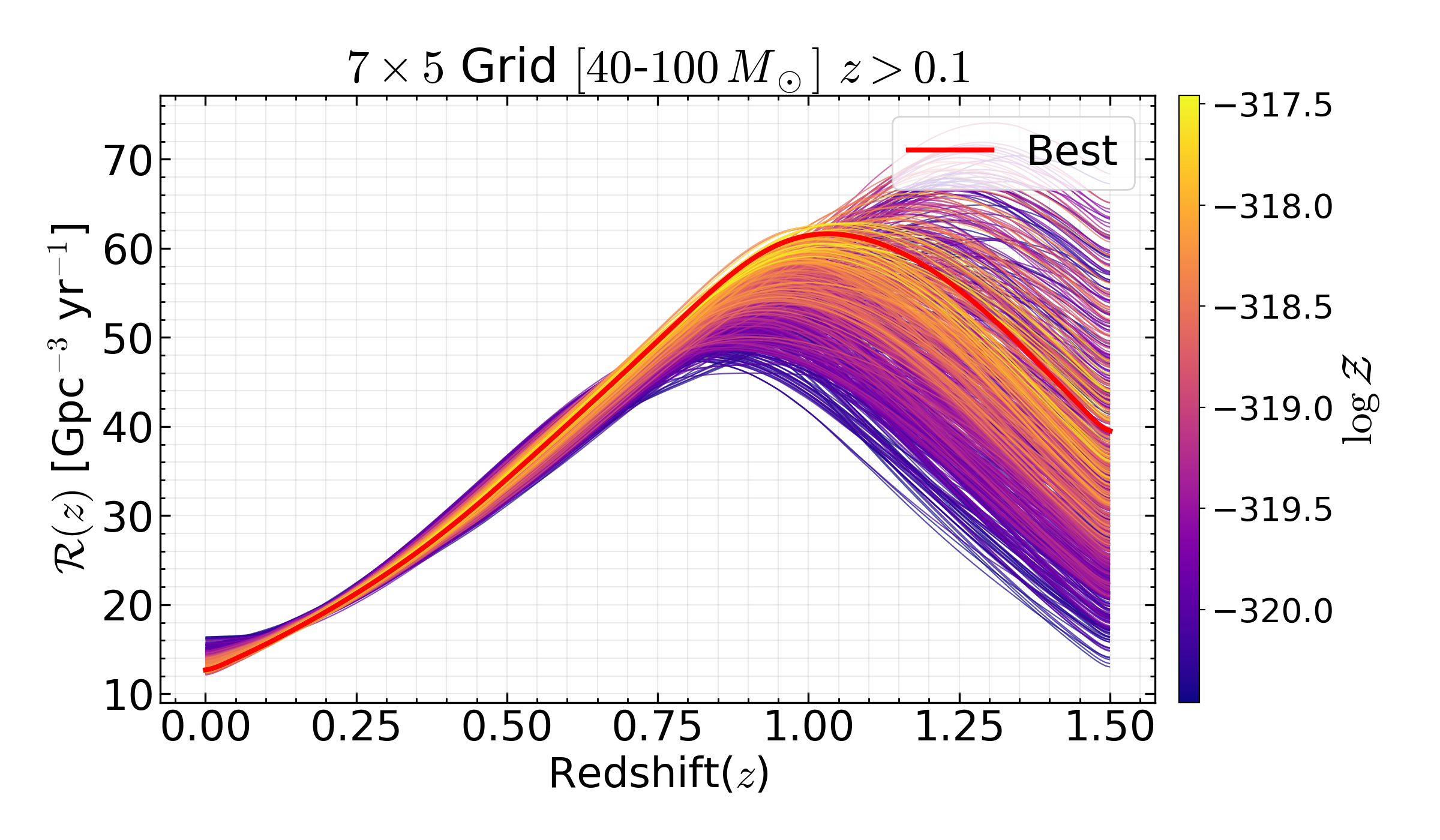}
    \includegraphics[width=0.32\textwidth]{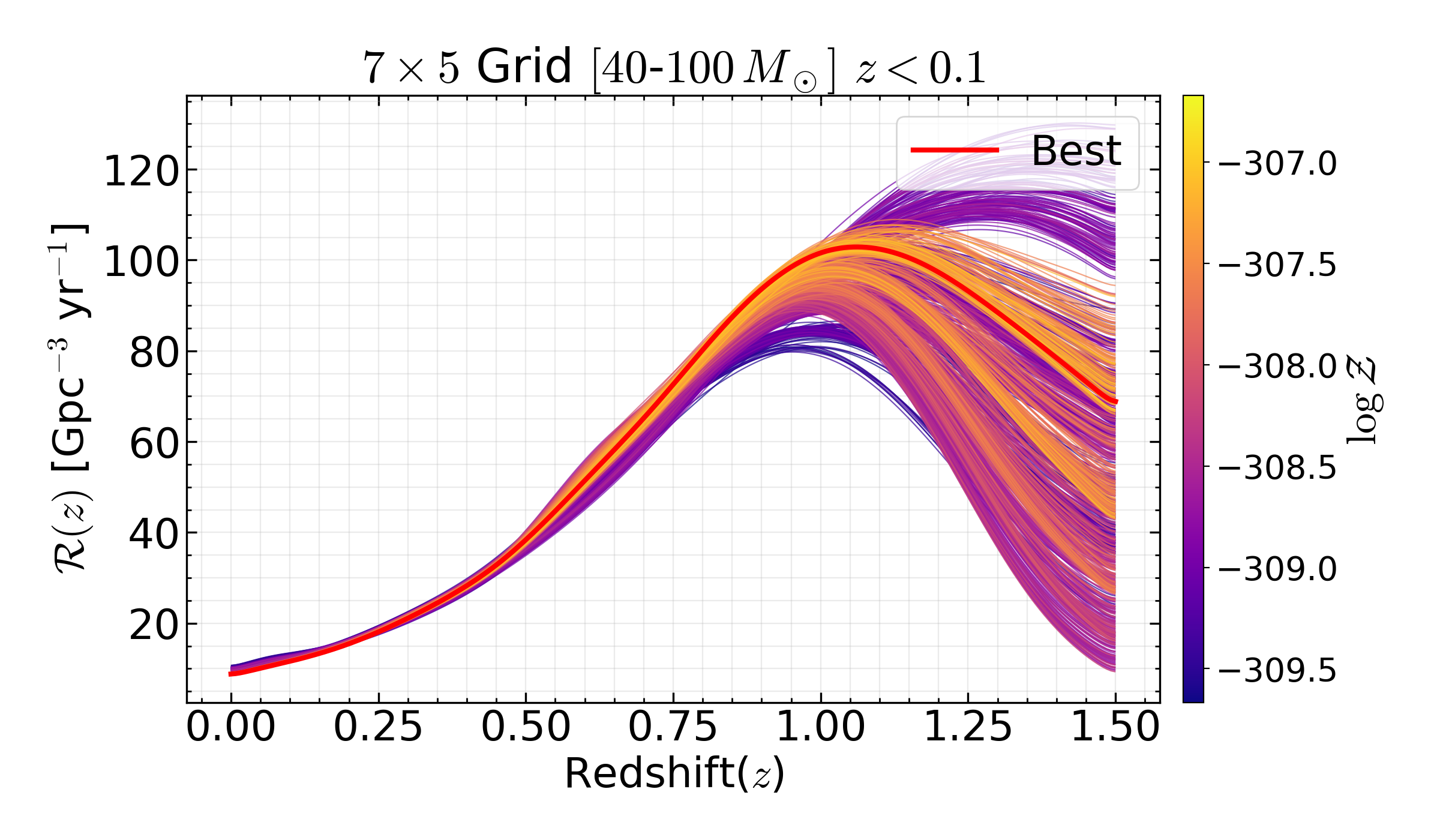}
    \caption{Physical merger rate $\mathcal{R}(z)$ [Gpc$^{-3}$ yr$^{-1}$] as a function of redshift for all trajectories within Bayes factor $\mathrm{BF} < 3$ ($\Delta\ln\mathcal{Z} < \ln 3$ relative to the highest-evidence trajectory), for the $7\times5$ grid. Top row: $20$--$40\,M_\odot$ bin. Bottom row: $40$--$100\,M_\odot$ bin. Left to right: standard Madau-Dickinson SFRD, high-metallicity effective SFRD ($Z > 0.1\,Z_\odot$), and low-metallicity effective SFRD ($Z < 0.1\,Z_\odot$). Lines are coloured by $\ln\mathcal{Z}$ from purple to yellow; the red curve marks the highest-evidence trajectory. In both mass bins the best-fit merger rate rises from its local value to a clear peak at intermediate redshift before declining; the low-metallicity SFRD produces the highest and most sharply rising peak in both bins, while the high-mass bin consistently exhibits a lower local rate and a peak shifted to slightly higher redshift than the low-mass bin.}
    \label{fig:Rz_all}
\end{figure*}

The associated merger rate evolution is shown in the bottom row of Figure~\ref{fig:Rz_all}. For the $40$--$100\,M_\odot$ bin, the merger rate evolution is qualitatively similar to the low-mass bin in that the best-fit rate rises to a clear intermediate-redshift peak, but it differs in two consistent respects across all three SFRD assumptions: the local rate $R_0$ is lower, and the peak is shifted to slightly higher redshift and is surrounded by a tighter BF$<3$ envelope. The position and amplitude of this peak vary with the assumed SFRD in a physically understandable way.

Under the standard Madau-Dickinson SFRD (Figure~\ref{fig:Rz_all}, bottom left), the best-fit trajectory starts at a local rate of $\mathcal{R}(z=0) \approx 12\,\mathrm{Gpc^{-3}\,yr^{-1}}$ and rises to a peak of $\sim$61 Gpc$^{-3}$ yr$^{-1}$ at $z \sim 1.05$--$1.1$, declining to $\sim$50 Gpc$^{-3}$ yr$^{-1}$ by $z \sim 1.5$, corresponding to a peak-to-local ratio of $\sim$5. The BF$<3$ envelope is significantly tighter than for the low-mass bin, with the highest-evidence trajectories strongly concentrated around the rising portion of the merger rate.

Under the high-metallicity effective SFRD ($Z > 0.1\,Z_\odot$, Figure~\ref{fig:Rz_all}, bottom middle), which is not the physically motivated assumption for this mass bin, the best-fit merger rate evolution is closely similar to the MD SFR case. The best-fit local rate is $\mathcal{R}(z=0) \approx 13\,\mathrm{Gpc^{-3}\,yr^{-1}}$, the peak reaches $\sim$62 Gpc$^{-3}$ yr$^{-1}$ at $z \sim 1.0$, and the envelope width and peak position are broadly consistent with the MD SFR case. This consistency is physically understandable given the close agreement between the MD SFR and the $Z > 0.1\,Z_\odot$ SFRD at the low-to-intermediate redshifts that dominate the likelihood, and mirrors the same behaviour seen for the low-mass bin.

Under the physically motivated low-metallicity effective SFRD ($Z < 0.1\,Z_\odot$, Figure~\ref{fig:Rz_all}, bottom right), the best-fit merger rate shows the most pronounced peak of the three high-mass cases. The best-fit local rate is the lowest of all six panels at $\mathcal{R}(z=0) \approx 9\,\mathrm{Gpc^{-3}\,yr^{-1}}$, and the curve rises to a peak of $\sim$103 Gpc$^{-3}$ yr$^{-1}$ at $z \sim 1.05$ before declining to $\sim$69 Gpc$^{-3}$ yr$^{-1}$ by $z \sim 1.5$, corresponding to a large peak-to-local ratio of $\sim$11. As in the low-mass bin, this amplified intermediate-redshift peak reflects the steeply rising low-metallicity SFRD, which, convolved with the intermediate-delay DTD support of this mass bin, shifts the merger rate strongly toward intermediate redshift relative to the local rate.

The contrast between the two mass bins is therefore one of degree rather than kind: in both bins the best-fit merger rate peaks at intermediate redshift, but the high-mass bin consistently exhibits a lower local rate ($R_0 \approx 9$--$13$ versus $\approx 18$--$23\,\mathrm{Gpc^{-3}\,yr^{-1}}$), a peak shifted to slightly higher redshift ($z \sim 1.0$--$1.05$ versus $z \sim 0.8$--$1.05$), and a tighter BF$<3$ envelope.

\begin{figure*}[ht]
    \centering
    \includegraphics[width=0.32\textwidth]{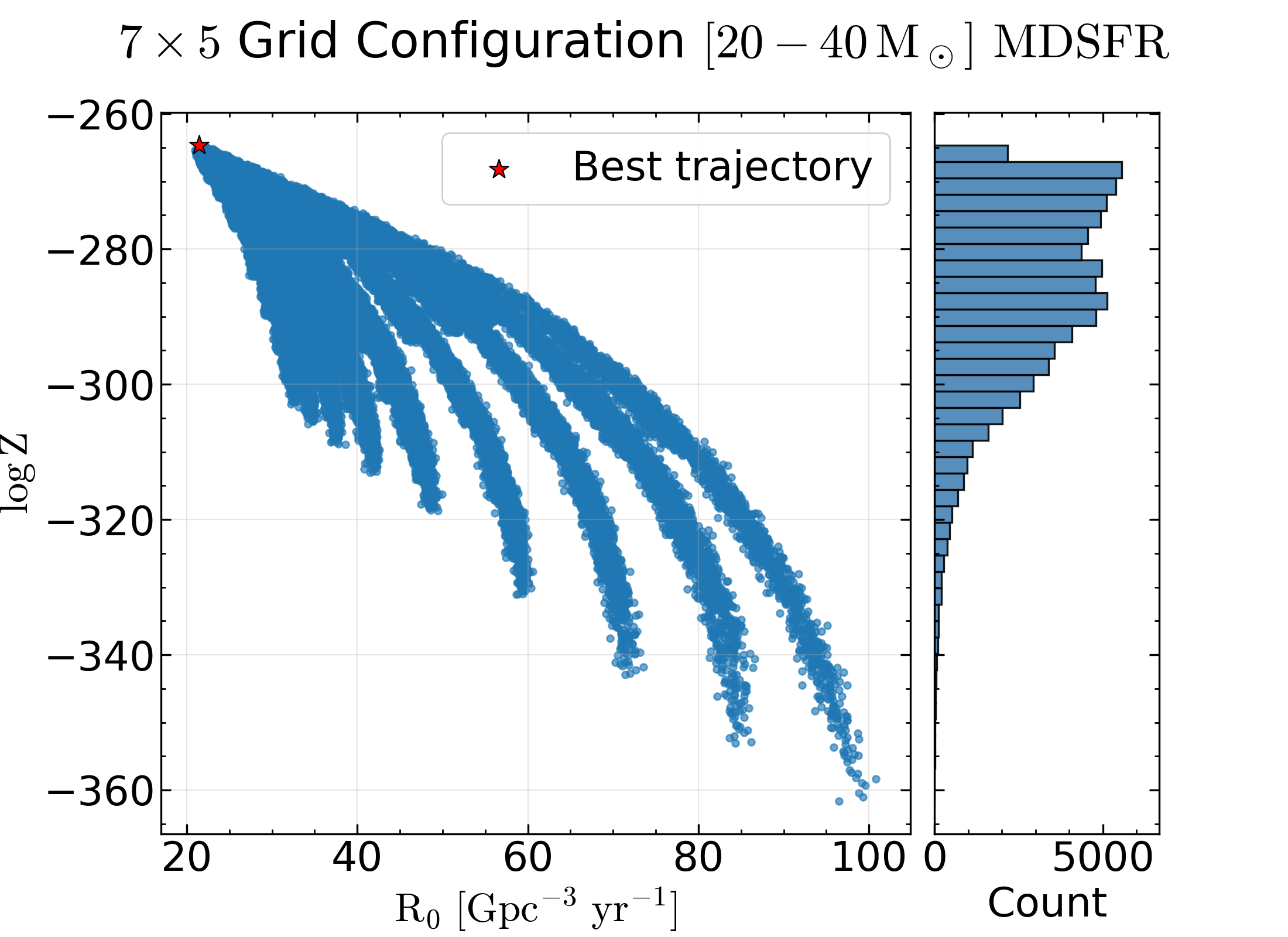}
    \includegraphics[width=0.32\textwidth]{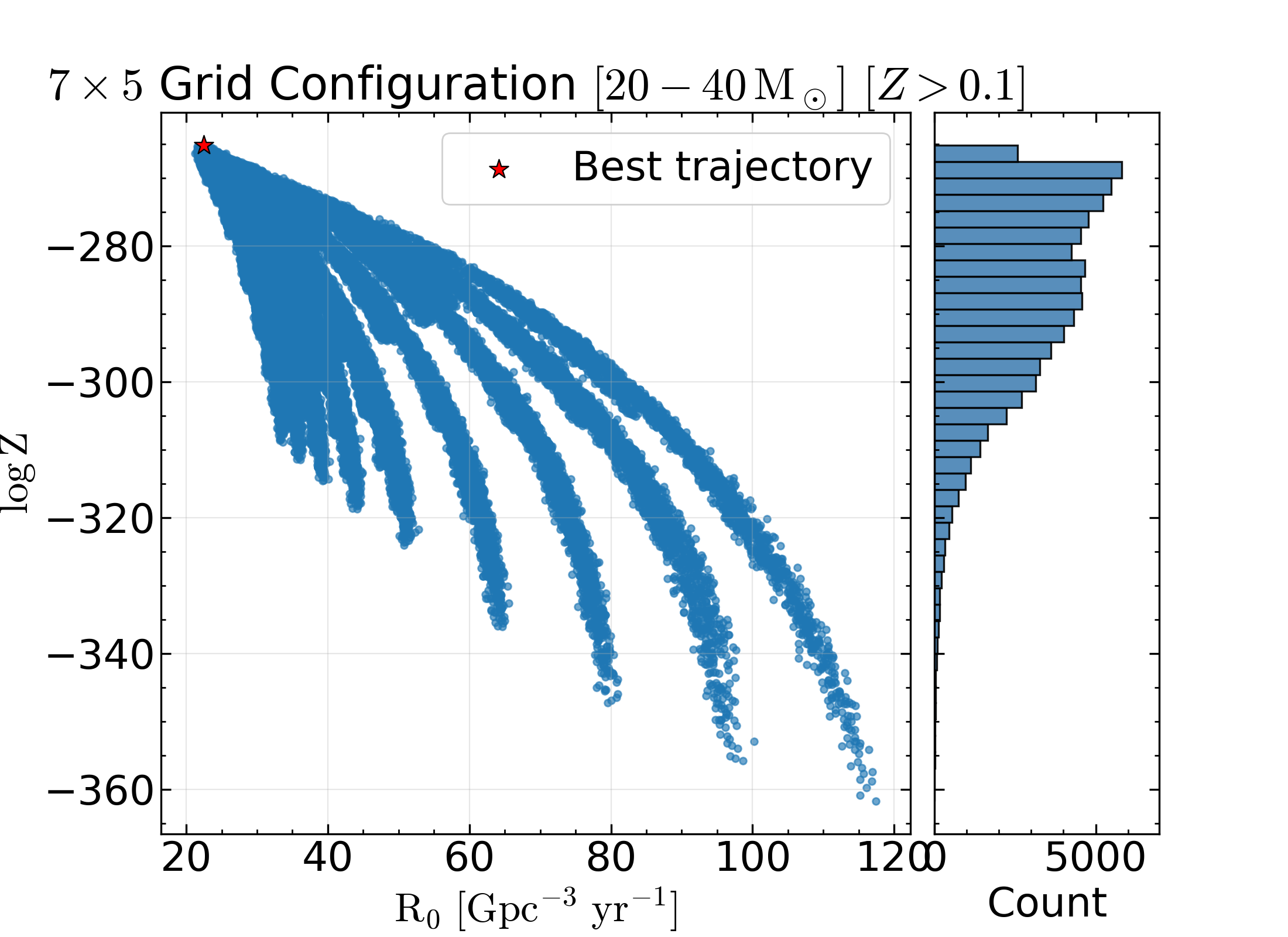}
    \includegraphics[width=0.32\textwidth]{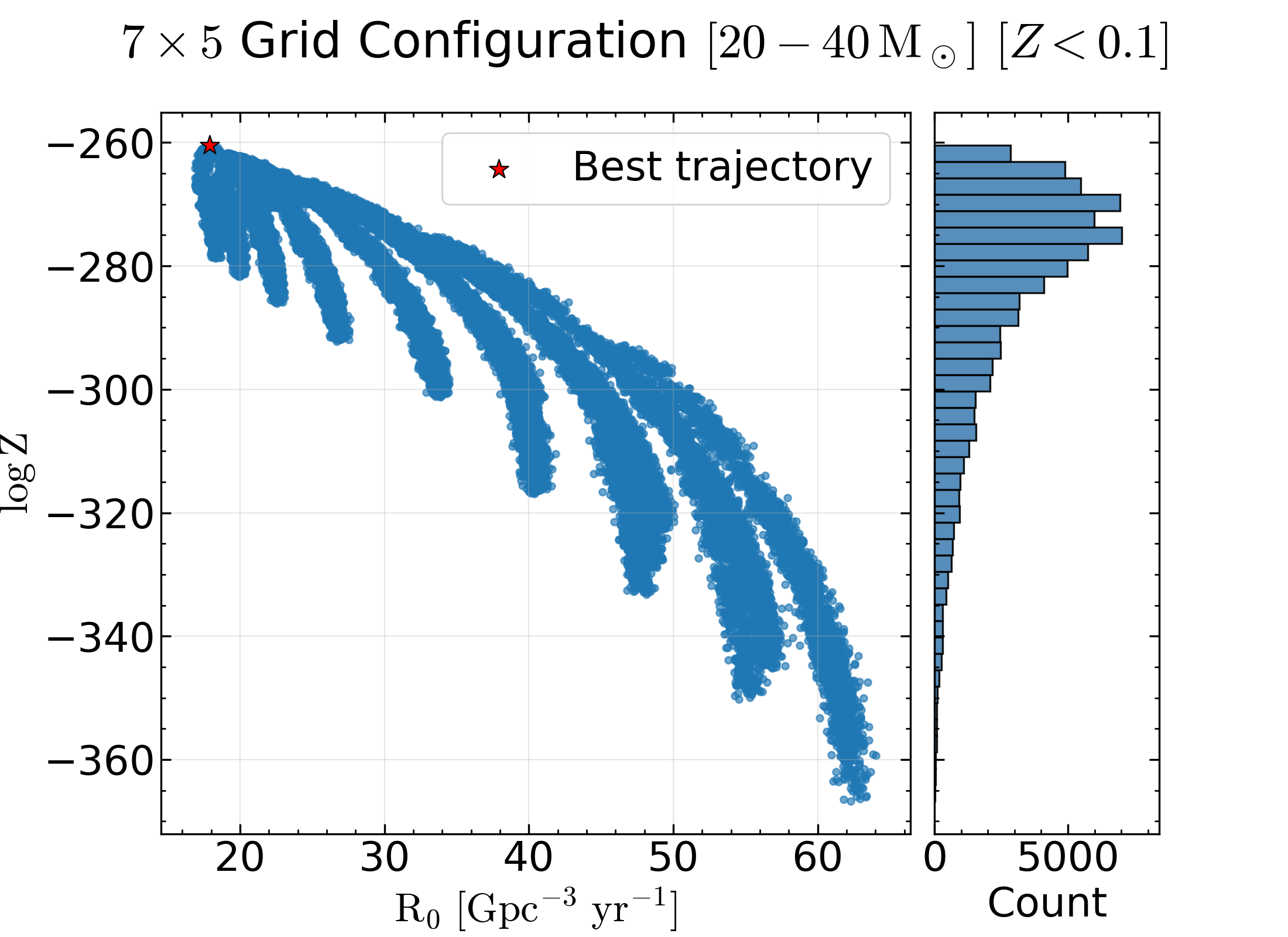}\\[6pt]
    \includegraphics[width=0.32\textwidth]{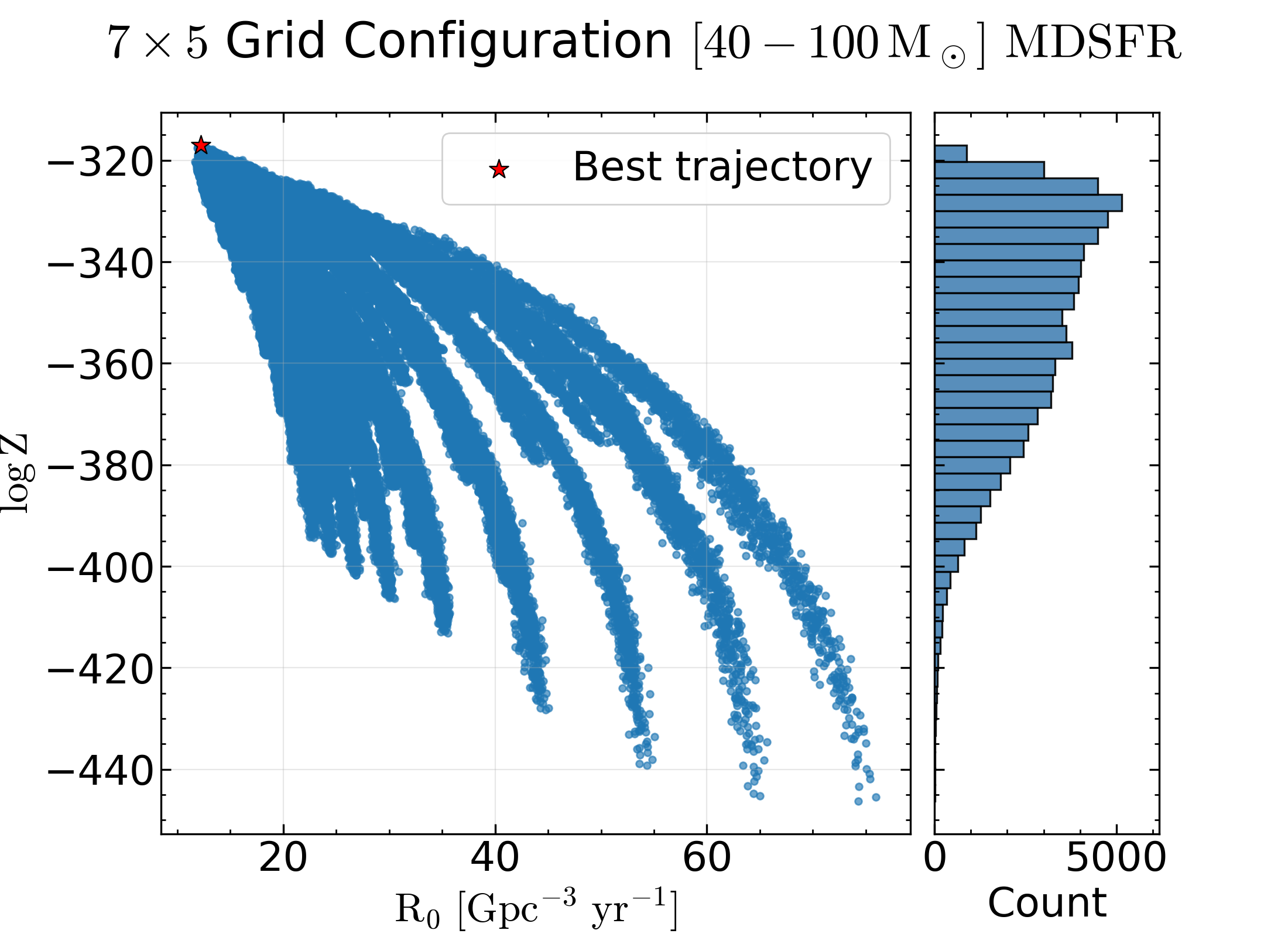}
    \includegraphics[width=0.32\textwidth]{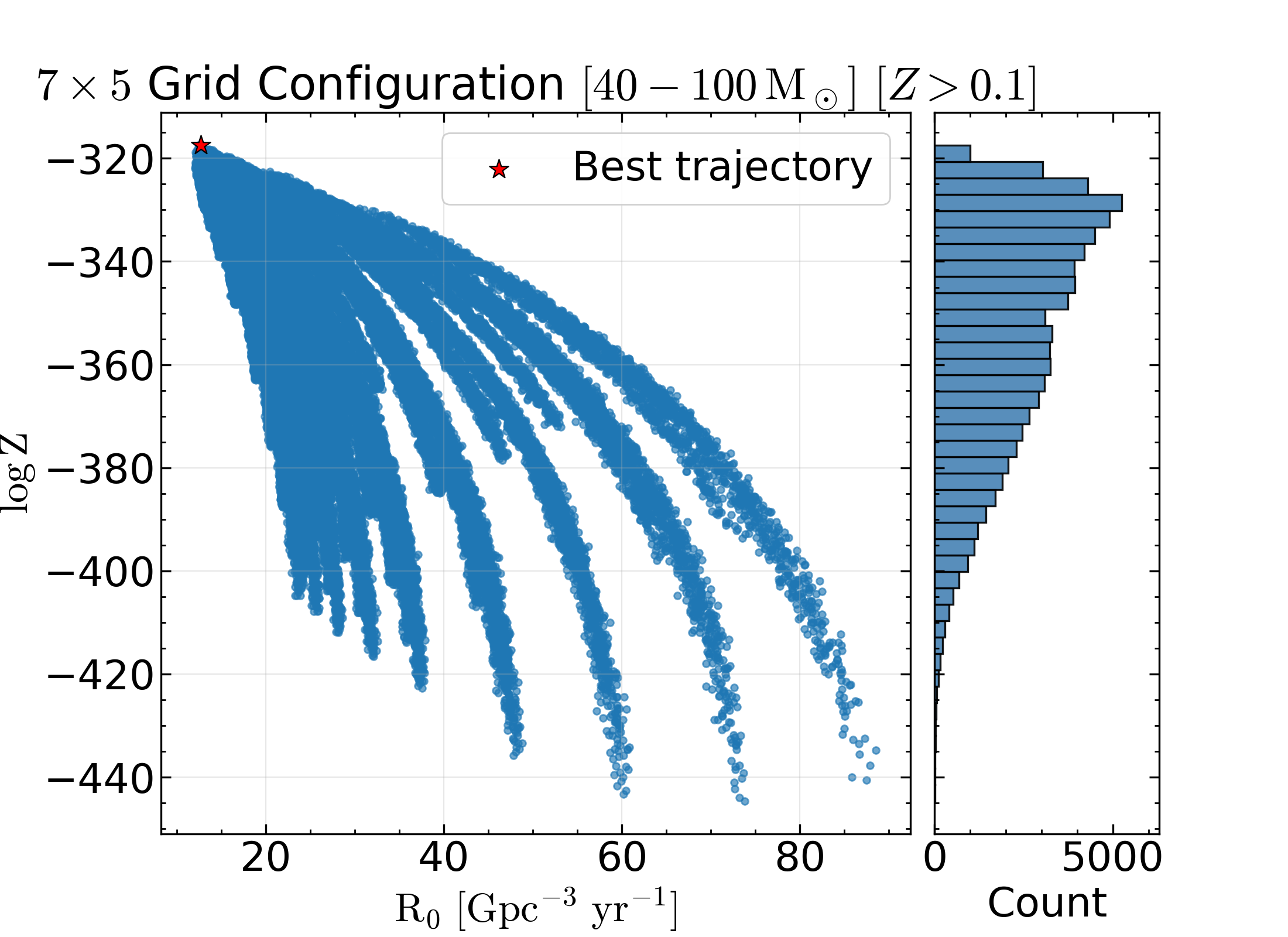}
    \includegraphics[width=0.32\textwidth]{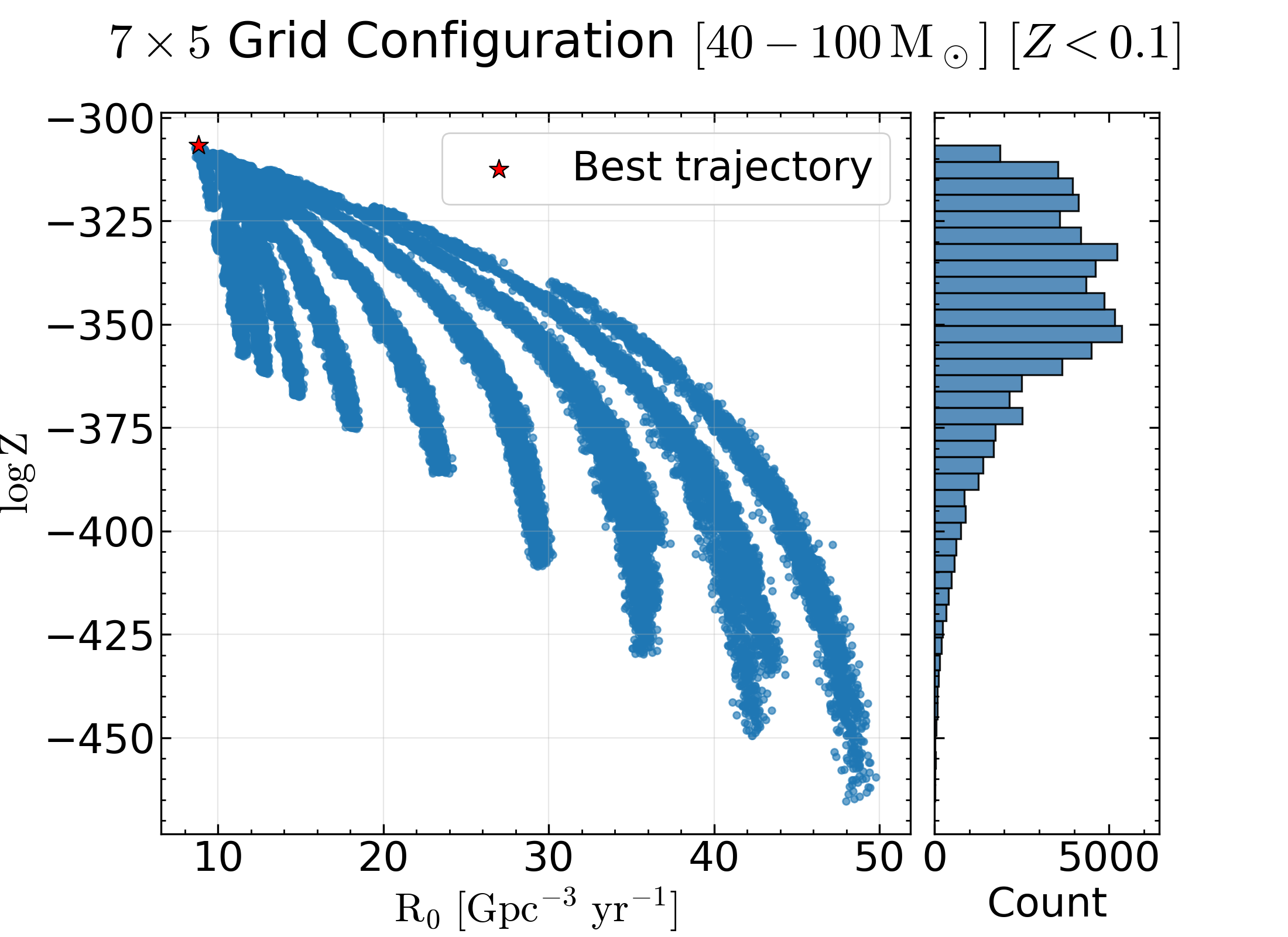}
    \caption{Distribution of Bayesian log-evidence $\ln\mathcal{Z}$ versus local merger rate $R_0$ for all trajectories of the $7\times5$ grid. Top row: $20$--$40\,M_\odot$ bin. Bottom row: $40$--$100\,M_\odot$ bin. Left to right: standard Madau-Dickinson SFRD, high-metallicity effective SFRD ($Z > 0.1\,Z_\odot$), and low-metallicity effective SFRD ($Z < 0.1\,Z_\odot$). The right panel of each subfigure shows the marginalized $R_0$ histogram, and the red star marks the single highest-evidence trajectory. In all panels the highest-evidence trajectories cluster at the lowest $R_0$ values, with $\ln\mathcal{Z}$ declining into discrete striated families toward higher $R_0$. The high-mass bin shows best-fit local rates concentrated at lower $R_0$ and a larger evidence dynamic range than the low-mass bin.}
    \label{fig:evidence_all}
\end{figure*}

The evidence landscape and local rate for this bin are shown in the bottom row of Figure~\ref{fig:evidence_all}. For the $40$--$100\,M_\odot$ bin, the evidence landscape is qualitatively different from the low-mass case across all three SFRD assumptions: the best-fit local rates are concentrated at systematically lower $R_0$, the dynamic range of $\ln\mathcal{Z}$ is larger, and the striated family structure is more pronounced, all consistent with the tighter DTD constraints identified in Section~\ref{sec:dtd_results}. Under the MD SFR (Figure~\ref{fig:evidence_all}, bottom left), the best-fit trajectory has $\ln\mathcal{Z}_{\rm best} \approx -317$ at $R_0 \approx 12\,\mathrm{Gpc^{-3}\,yr^{-1}}$, with the marginalized $R_0$ distribution extending to $\sim$76 Gpc$^{-3}$ yr$^{-1}$. Under the $Z > 0.1\,Z_\odot$ SFRD (Figure~\ref{fig:evidence_all}, bottom middle), the best-fit is at $\ln\mathcal{Z}_{\rm best} \approx -318$ and $R_0 \approx 12.5$--$13\,\mathrm{Gpc^{-3}\,yr^{-1}}$, nearly identical to the MD SFR case, consistent with the similar merger rate evolution seen for these two SFRDs above. Under the physically motivated $Z < 0.1\,Z_\odot$ SFRD (Figure~\ref{fig:evidence_all}, bottom right), the best-fit is at $\ln\mathcal{Z}_{\rm best} \approx -306.5$ and the lowest local rate of all six panels, $R_0 \approx 9\,\mathrm{Gpc^{-3}\,yr^{-1}}$, with a compact $R_0$ distribution extending to $\sim$50 Gpc$^{-3}$ yr$^{-1}$ and the most pronounced striated arc structure, each band corresponding to a family of trajectories sharing similar intermediate-delay node configurations that dominate the likelihood under this SFRD. In all three cases the best-fit local rate of the high-mass bin is lower than that of the low-mass bin, standing in clear contrast to the broader, higher-$R_0$ distributions of the low-mass bin.

In all six panels, the evidence landscape shows a characteristic structure in the $\ln\mathcal{Z}$-$R_0$ plane: the highest-evidence trajectories cluster at the lowest values of $R_0$, and $\ln\mathcal{Z}$ declines steeply and breaks into discrete striated families as $R_0$ increases, where the model increasingly over-predicts the observed event counts. The dynamic range of $\ln\mathcal{Z}$ across all trajectories quantifies how strongly the data discriminate among different DTD shapes. For the $20$--$40\,M_\odot$ bin this spread is large, extending over $\sim$100 log-scale from the best-fit value down to the worst trajectories, while for the $40$--$100\,M_\odot$ bin it is larger still, extending over $\sim$120--150 log-units, confirming quantitatively that the high-mass data carry more discriminating power over the DTD shape, consistent with the tighter BF$<3$ envelopes seen in Figures~\ref{fig:dtd_all} and~\ref{fig:Rz_all}.

\section{Conclusions}
\label{sec:conclusions}

We have presented a non-parametric, grid-based reconstruction of the mass-dependent binary black hole delay time distribution using the cumulative GWTC-4.0 catalog, under three star formation rate density assumptions: the metallicity-averaged Madau-Dickinson model and the high- and low-metallicity effective SFRDs of \citet{Chruslinska2019MNRAS}. By enumerating all causality-allowed trajectories in a delay time probability grid and evaluating their Bayesian evidence, we reconstruct the DTD, the associated merger rate evolution, and the local rate $R_0$ independently for two primary-mass bins, $20$-$40\,M_\odot$ and $40$-$100 \, M_\odot$. Our main findings, consistent across all three SFRD assumptions, are as follows.

\begin{enumerate}

    \item \textbf{Mass-dependent DTD shapes.} The two mass bins show quantitatively different DTD shapes across all three SFRD assumptions. The $20$--$40\,M_\odot$ bin supports a broad plateau of intermediate-to-long delay times within a wide BF$<3$ envelope, while the $40$--$100\,M_\odot$ bin shows a tighter, more sharply defined concentration of high-evidence trajectories at intermediate delay times $t_d \sim 2$--$4$ Gyr, with a clearer suppression at  very long delays. This difference between the two bins persists regardless of the assumed SFRD, suggesting that it reflects a genuine difference in the delay time preferences of the two populations rather than an artifact of the assumed star formation history. We emphasize that the statistically robust feature of both reconstructions is the slope of the DTD at intermediate-to-long delays, where the BF$<3$ envelope is narrow, whereas the short-delay behavior is weakly constrained by the current data.

    \item \textbf{Intermediate-redshift merger rate peak in both mass bins.} The best-fit merger rate evolution $\mathcal{R}(z)$ rises from its local value to a clear peak at intermediate redshift in both mass bins and across all three SFRD assumptions. The two bins differ in degree rather than kind: the low-mass bin peaks at $z \sim 0.8$--$1.05$ from a local rate of $R_0 \approx 18$--$23\,\mathrm{Gpc^{-3}\,yr^{-1}}$, while the high-mass bin peaks at slightly higher redshift $z \sim 1.0$--$1.05$ from a lower local rate of $R_0 \approx 9$--$13\,\mathrm{Gpc^{-3}\,yr^{-1}}$ within a tighter BF$<3$ envelope. In both bins the peak-to-local ratio is largest under the physically motivated $Z < 0.1\,Z_\odot$ case, reflecting the steeper rise of the low-metallicity SFRD with redshift.

    \item \textbf{Insensitivity of the low-mass results to the SFRD.} For the $20$--$40\,M_\odot$ bin, the DTD shape, merger rate evolution, and best-fit $R_0$ are closely consistent between the MD SFR and the physically motivated $Z > 0.1\,Z_\odot$ cases, which is expected given the close agreement between these two SFRDs at the redshifts that dominate the likelihood. The $Z < 0.1\,Z_\odot$ case produces a more sharply structured DTD and a much higher merger rate peak, but the overall character of the low-mass results remains broad and weakly constrained across all three assumptions.

    \item \textbf{Sensitivity of the high-mass results to the SFRD.} For the $40$--$100\,M_\odot$ bin, the MD SFR and $Z > 0.1\,Z_\odot$ cases show closely similar DTD shapes and merger rate evolution, while the physically motivated $Z < 0.1\,Z_\odot$ case shows a more sharply structured DTD and a substantially more pronounced peak in $\mathcal{R}(z)$. These differences are physically understandable from the shape of the low-metallicity SFRD and do not alter the qualitative conclusion that the high-mass bin prefers a tighter, more sharply defined range of intermediate delay times than the low-mass bin.

    \item \textbf{Local merger rates.} The high-mass bin consistently yields lower local merger rates $R_0 \approx 9$--$13\,\mathrm{Gpc^{-3}\,yr^{-1}}$ compared to the low-mass bin ($R_0 \approx 18$--$23\,\mathrm{Gpc^{-3}\,yr^{-1}}$) across all three SFRD assumptions. This difference reflects the combined effects of fewer intrinsic high-mass systems and the stronger shift of the high-mass merger rate toward intermediate redshift, which moves $\mathcal{R}(z)$ away from $z=0$ and reduces the local rate relative to the low-mass bin. In both bins the lowest local rate occurs under the $Z < 0.1\,Z_\odot$ assumption, which simultaneously produces the highest intermediate-redshift peak.

\end{enumerate}

Overall, our results indicate that a single power-law DTD may not simultaneously describe both mass bins, pointing toward possible mass-dependent binary evolution pathways that merit further investigation. Because the most stringent current limitation is the weak constraint on the short-delay portion of the DTD, and hence on the local rate, larger GW catalogs from future observing runs will be essential to sharpen these constraints and to test the inferred mass dependence with greater statistical significance.

\appendix
\section{Robustness to Grid Resolution: the $6\times5$ Configuration}
\label{app:6x5}

The results presented in Section~\ref{sec:result} use a $7\times5$ logarithmically spaced grid in the $(t_d, \log p_t)$ plane. To verify that our conclusions are not artifacts of the particular grid resolution, we repeat the full analysis using an alternative $6\times5$ grid, with delay-time nodes at $\{0.50, 0.94, 1.78, 3.37, 6.36, 12.0\}$ Gyr and five free $\log p_t$ levels per node, enumerating all $5^6 = 15{,}625$ causally allowed trajectories. This grid differs from the fiducial $7\times5$ configuration by removing the longest-delay node at $13.0$ Gyr, providing a coarser sampling of the delay-time axis. We present results for both mass bins and all three SFRD assumptions, and find that every qualitative and quantitative conclusion of the main analysis is reproduced.

\begin{figure*}[ht]
    \centering
    \includegraphics[width=0.32\textwidth]{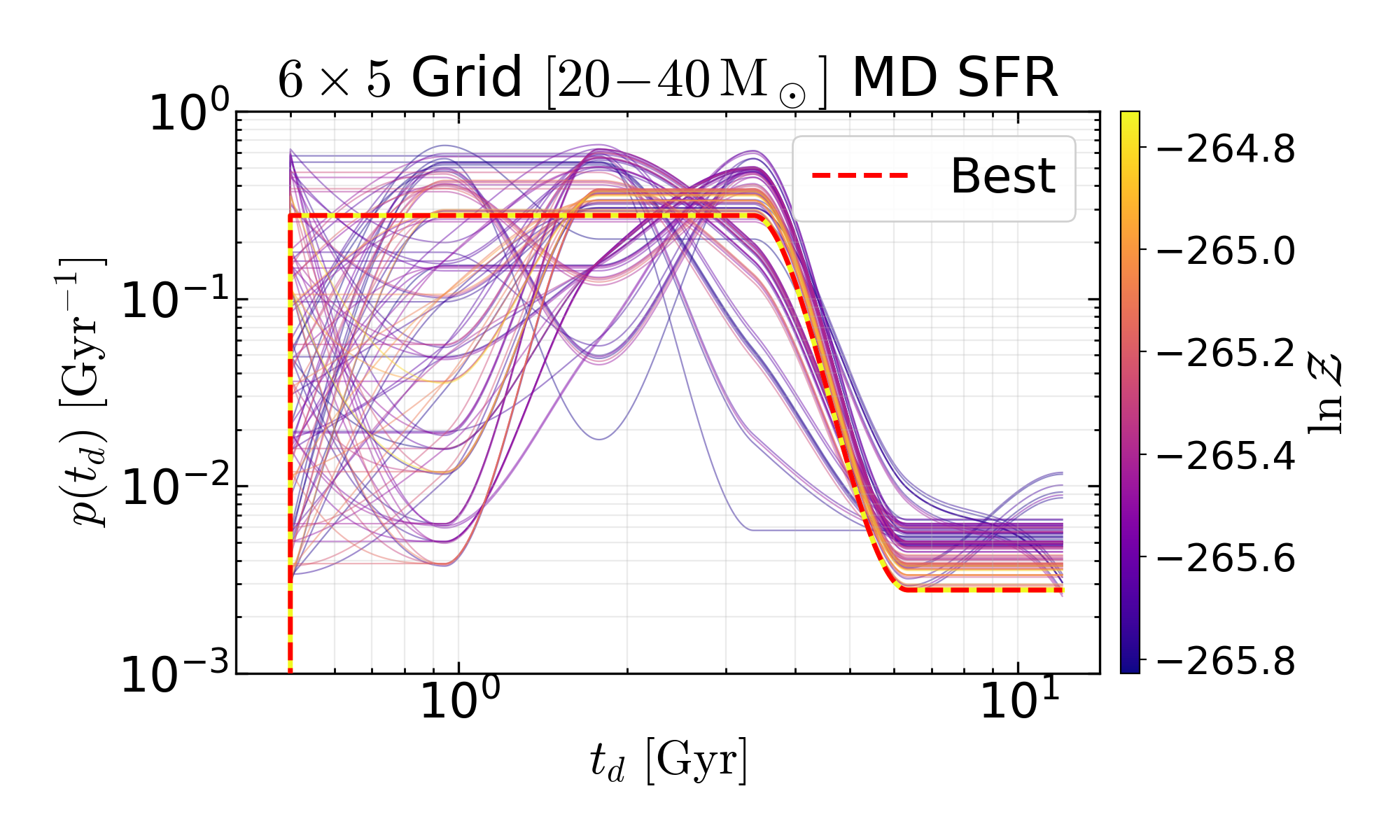}
    \includegraphics[width=0.32\textwidth]{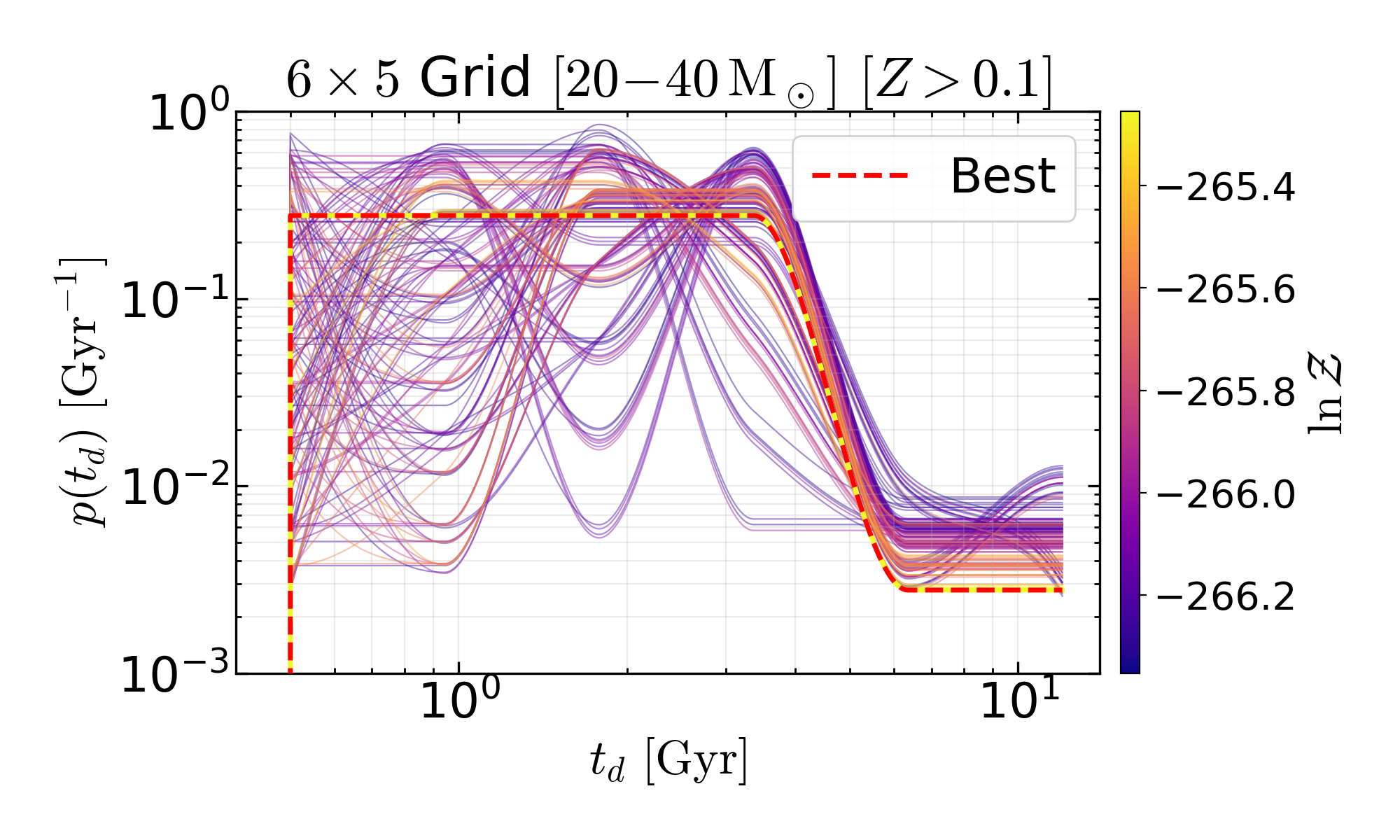}
    \includegraphics[width=0.32\textwidth]{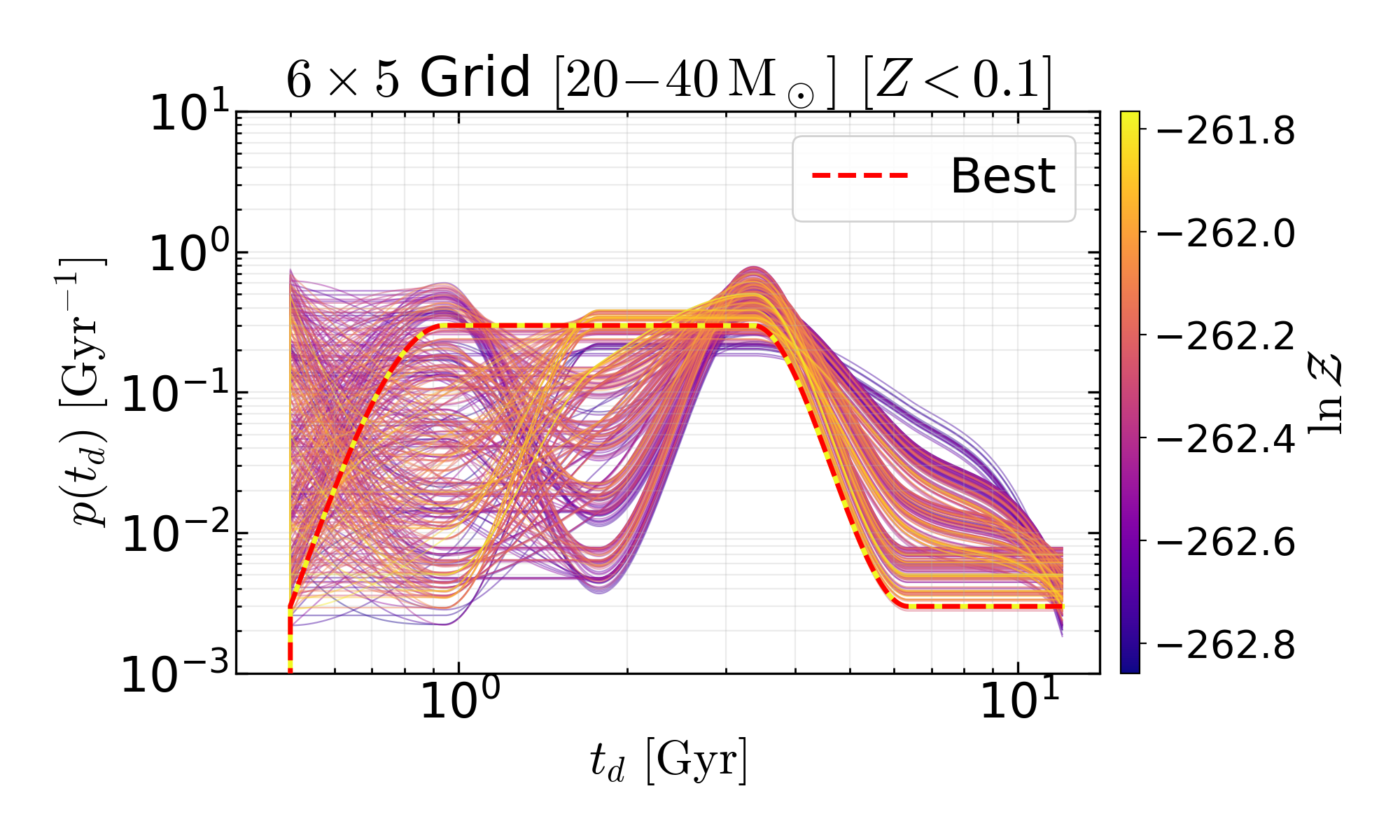}\\[6pt]
    \includegraphics[width=0.32\textwidth]{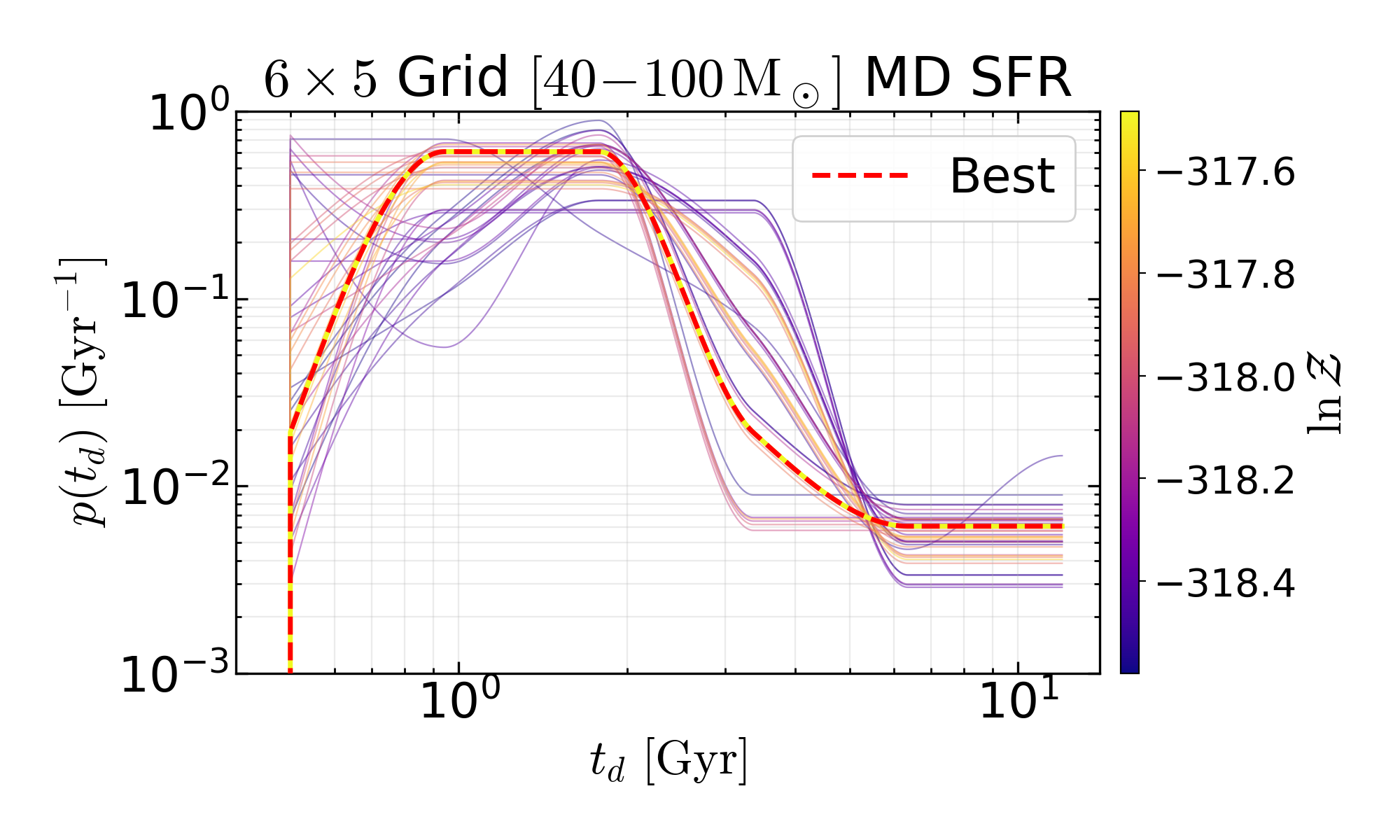}
    \includegraphics[width=0.32\textwidth]{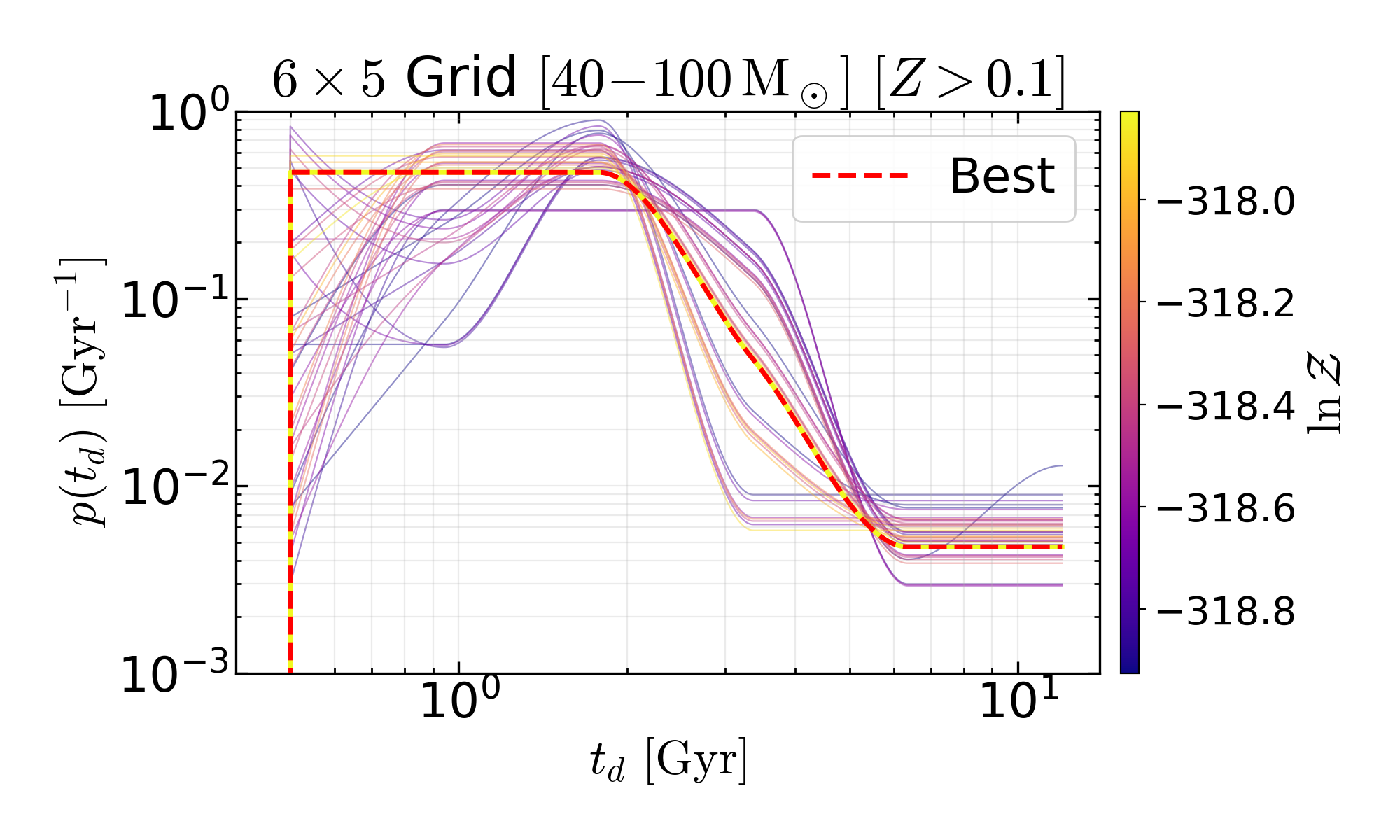}
    \includegraphics[width=0.32\textwidth]{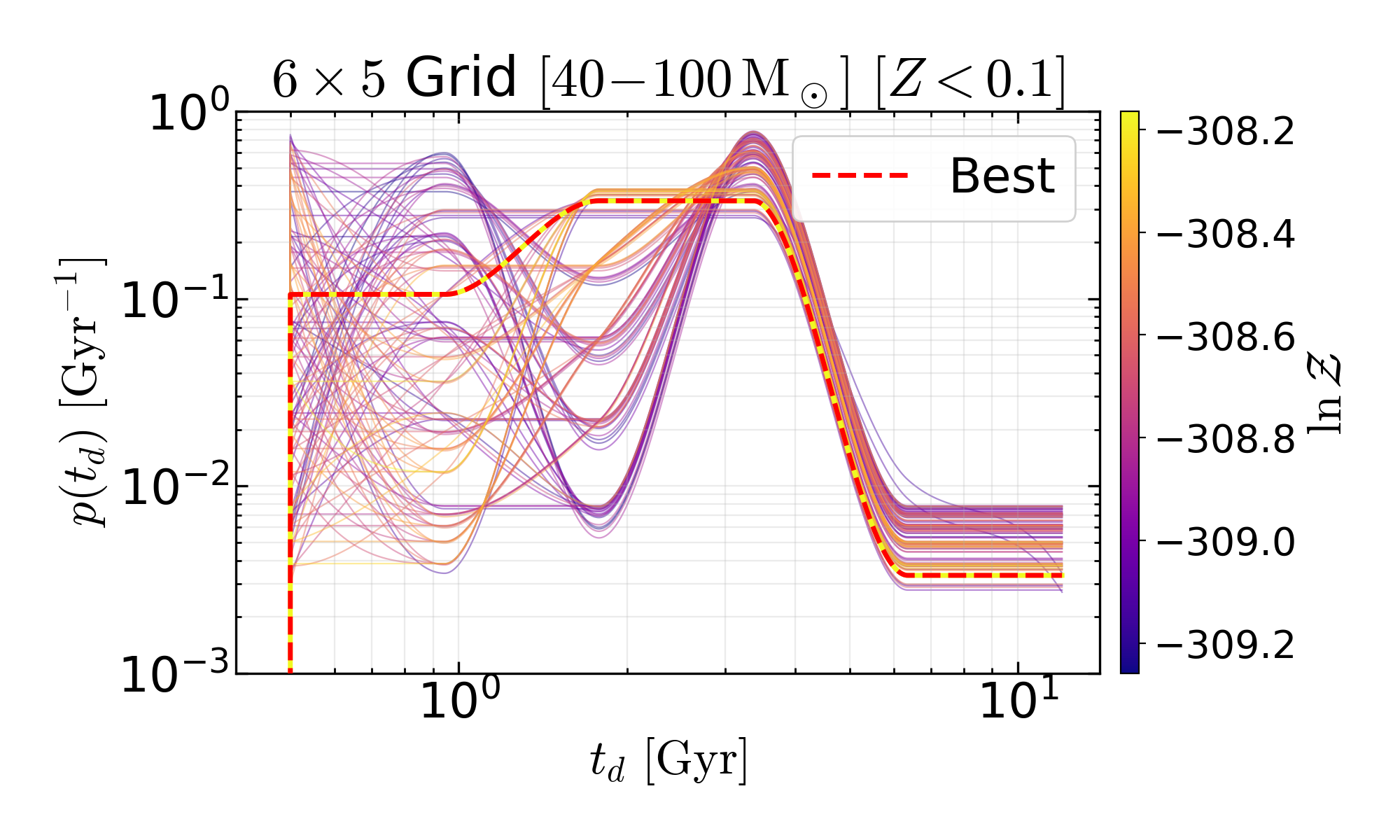}
    \caption{Reconstructed DTD trajectories $p(t_d)$ [Gyr$^{-1}$] within $\mathrm{BF} < 3$ of the best-fit solution, for the $6\times5$ grid with nodes at $\{0.50, 0.94, 1.78, 3.37, 6.36, 12.0\}$ Gyr. Top row: $20$--$40\,M_\odot$. Bottom row: $40$--$100\,M_\odot$. Left to right: Madau-Dickinson SFRD, high-metallicity effective SFRD ($Z > 0.1\,Z_\odot$), and low-metallicity effective SFRD ($Z < 0.1\,Z_\odot$). Lines are coloured by $\ln\mathcal{Z}$; the dashed red curve with yellow outline marks the highest-evidence trajectory. As in the $7\times5$ case, the best-fit trajectories favour an extended intermediate-delay plateau followed by a decline at the longest delays, with the low-metallicity cases showing a bimodal short-delay shelf, intermediate dip, and intermediate-delay plateau.}
    \label{fig:dtd_6x5}
\end{figure*}

Figure~\ref{fig:dtd_6x5} shows the DTD trajectories within $\mathrm{BF} < 3$ of the best-fit solution for the $6\times5$ grid. The reconstructed delay-time distributions reproduce the same qualitative behaviour found for the $7\times5$ grid in Section~\ref{sec:dtd_results}. For the $20$--$40\,M_\odot$ bin, the best-fit DTD is a broad intermediate-delay plateau at $p(t_d) \approx 0.3\,\mathrm{Gyr^{-1}}$ extending across $t_d \sim 1$--$4$ Gyr within a wide BF$<3$ envelope, under both the MD SFR and the physically motivated $Z > 0.1\,Z_\odot$ SFRD, which produce closely similar reconstructions. Under the $Z < 0.1\,Z_\odot$ SFRD the best-fit DTD develops the same bimodal structure seen in the $7\times5$ case, with a low-amplitude shelf at short delays, a dip near $t_d \sim 1$--$1.5$ Gyr, and a well-defined plateau at $t_d \sim 2$--$4$ Gyr, accompanied by a more sharply striated envelope.

For the $40$--$100\,M_\odot$ bin, the best-fit DTD again shows a more sharply defined intermediate-delay concentration and a tighter BF$<3$ envelope than the low-mass bin across all three SFRD assumptions. The MD SFR and $Z > 0.1\,Z_\odot$ cases yield a steep rise from short delays to a plateau at $p(t_d) \approx 0.4$--$0.5\,\mathrm{Gyr^{-1}}$ across $t_d \sim 1$--$2$ Gyr, followed by a steep decline. Under the physically motivated $Z < 0.1\,Z_\odot$ SFRD the best-fit DTD shows the characteristic dip-then-plateau structure, with a short-delay shelf near $p(t_d) \approx 0.1\,\mathrm{Gyr^{-1}}$, a rise to a plateau at $t_d \sim 2$--$3$ Gyr, and a steep cut-off at longer delays. The agreement between the two grid resolutions confirms that the inferred delay-time preferences are driven by the data and the SFRD assumption rather than by the placement of the grid nodes.

\begin{figure*}[ht]
    \centering
    \includegraphics[width=0.32\textwidth]{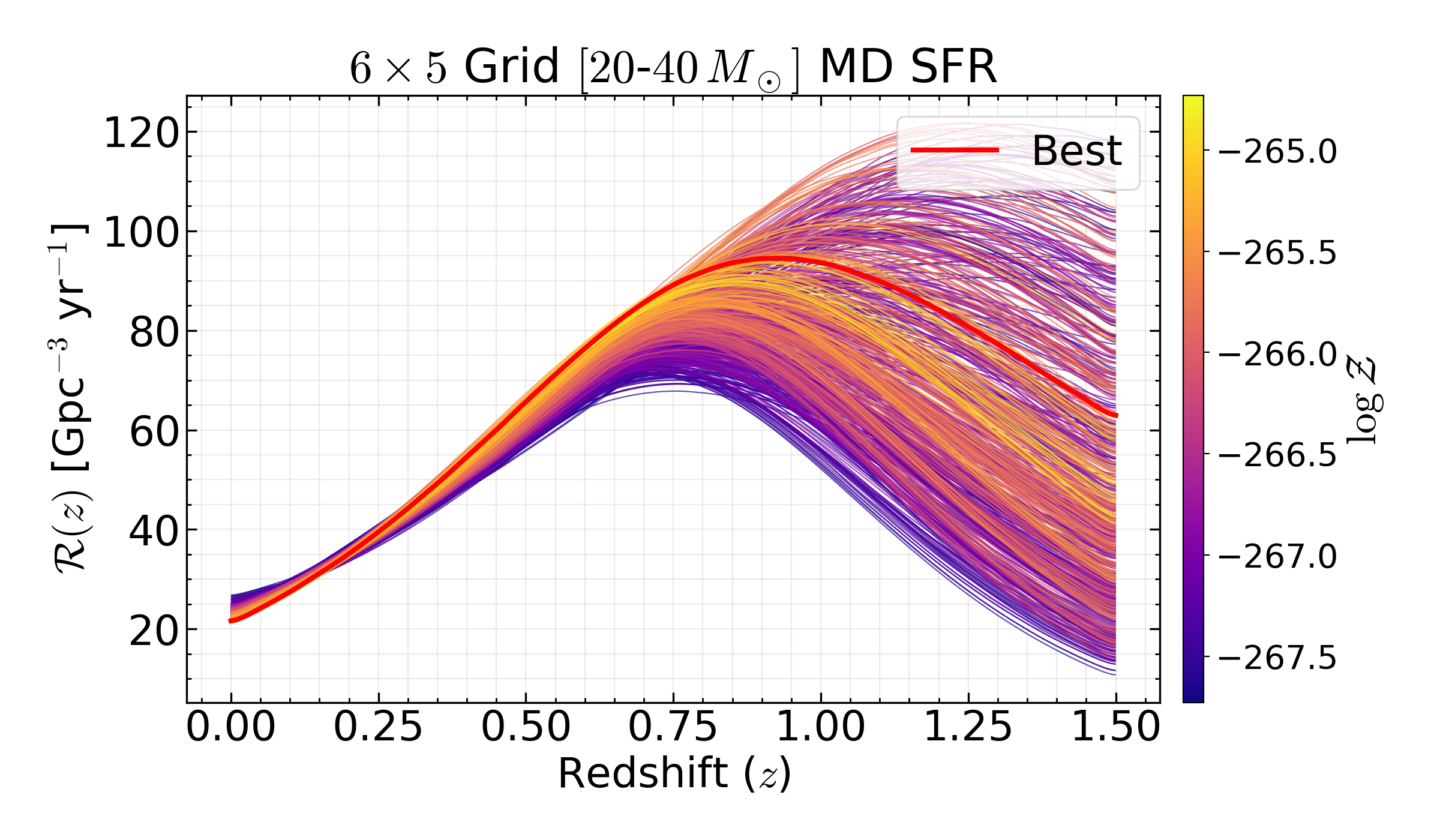}
    \includegraphics[width=0.32\textwidth]{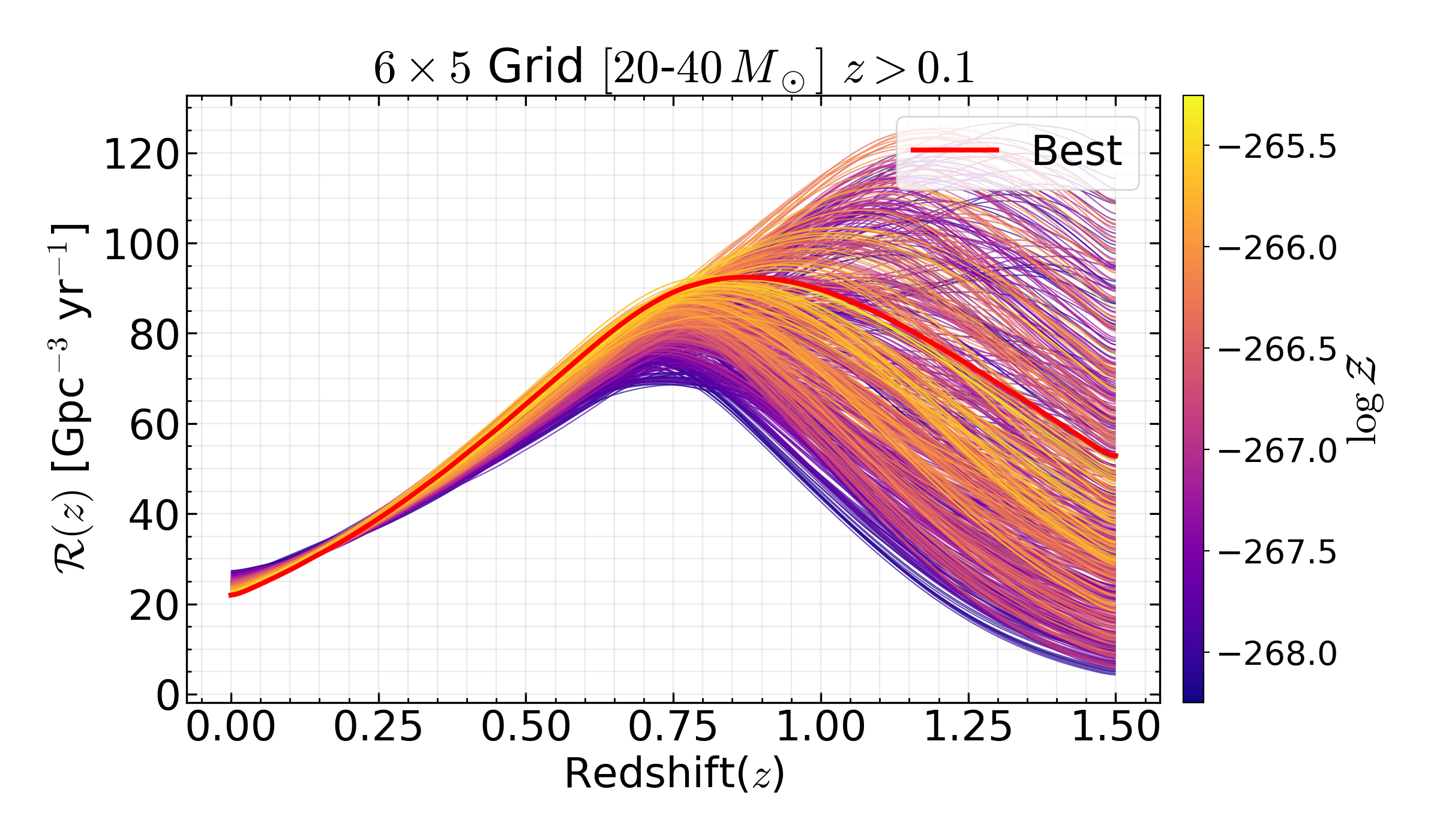}
    \includegraphics[width=0.32\textwidth]{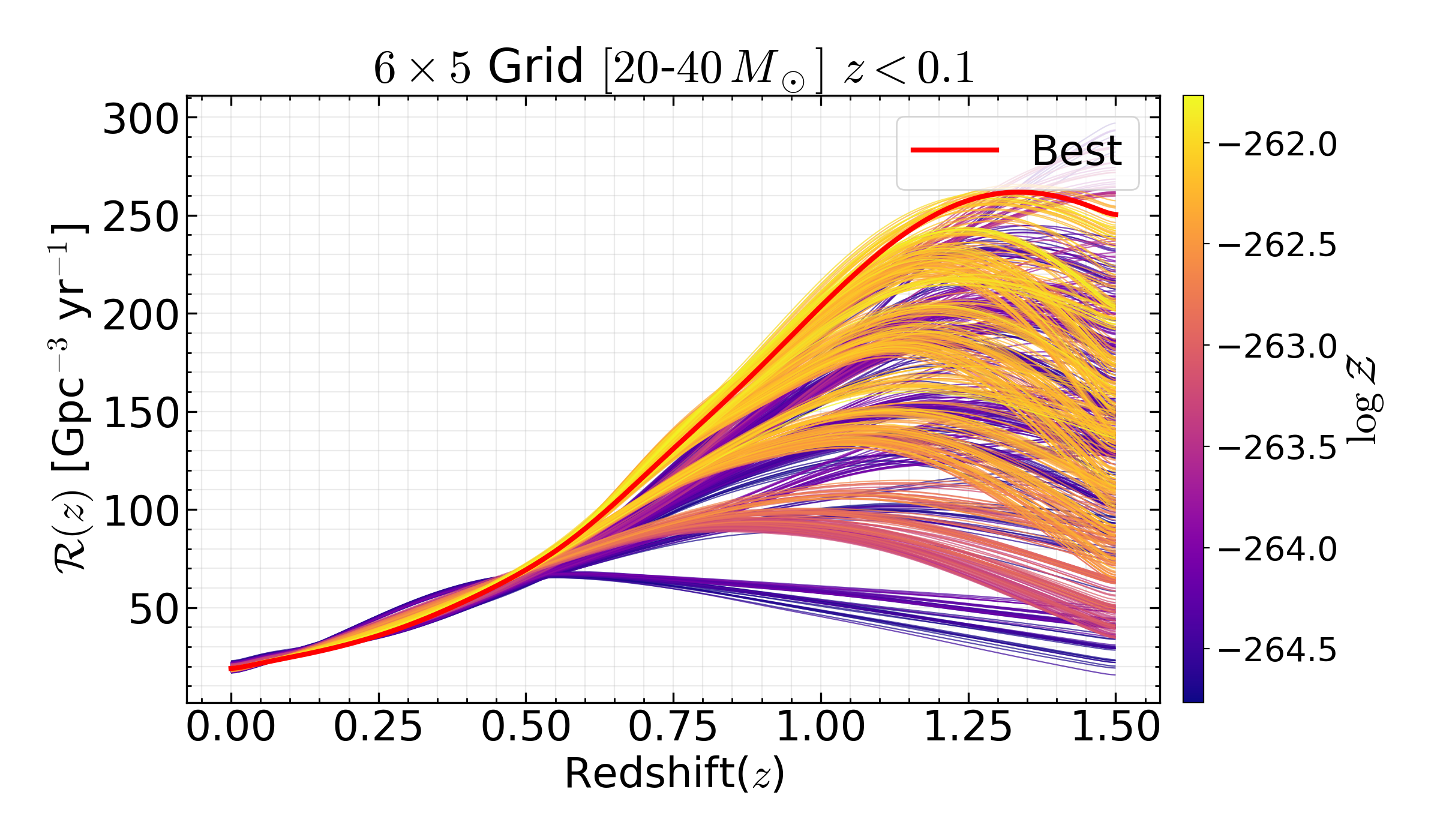}\\[6pt]
    \includegraphics[width=0.32\textwidth]{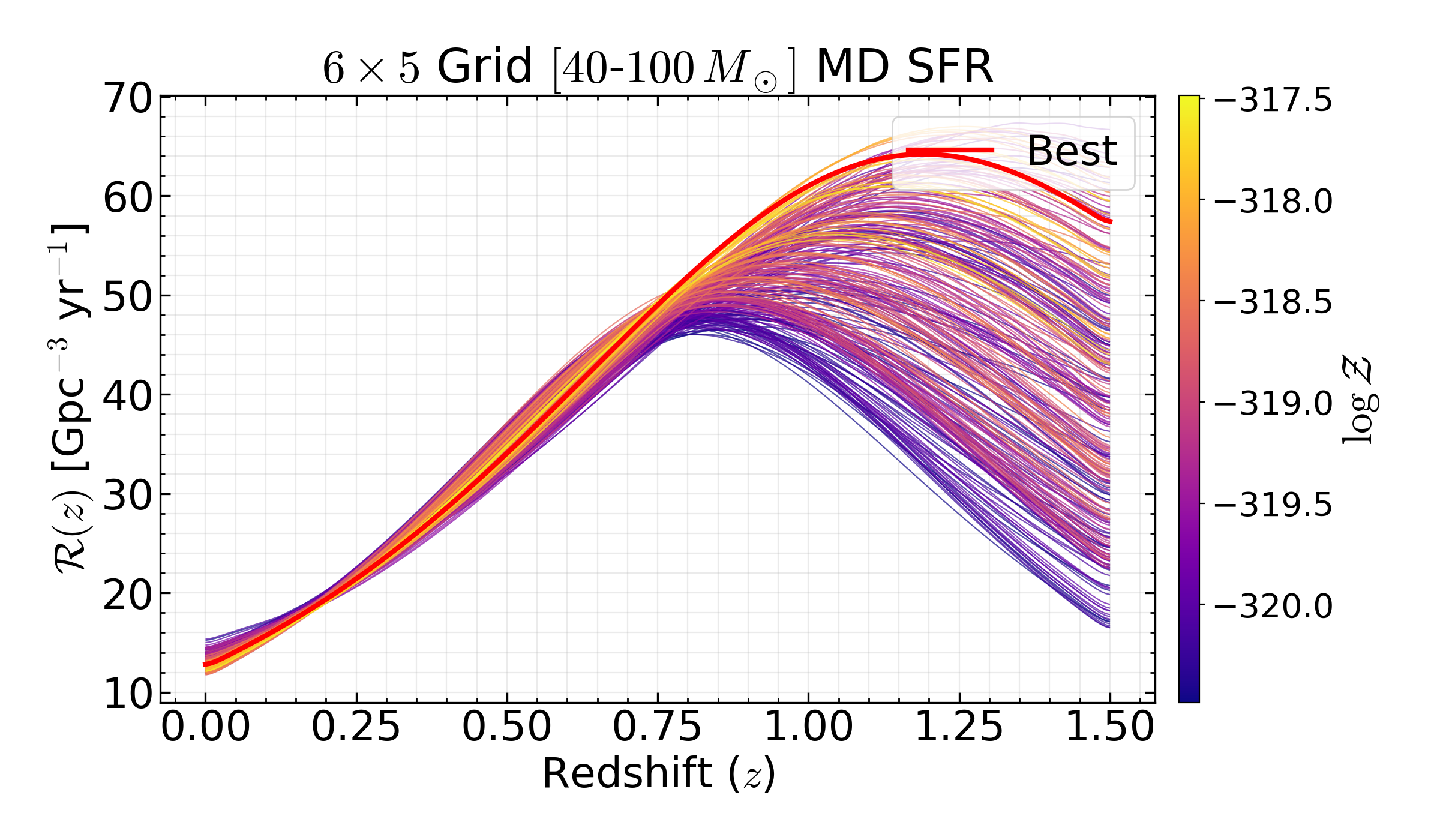}
    \includegraphics[width=0.32\textwidth]{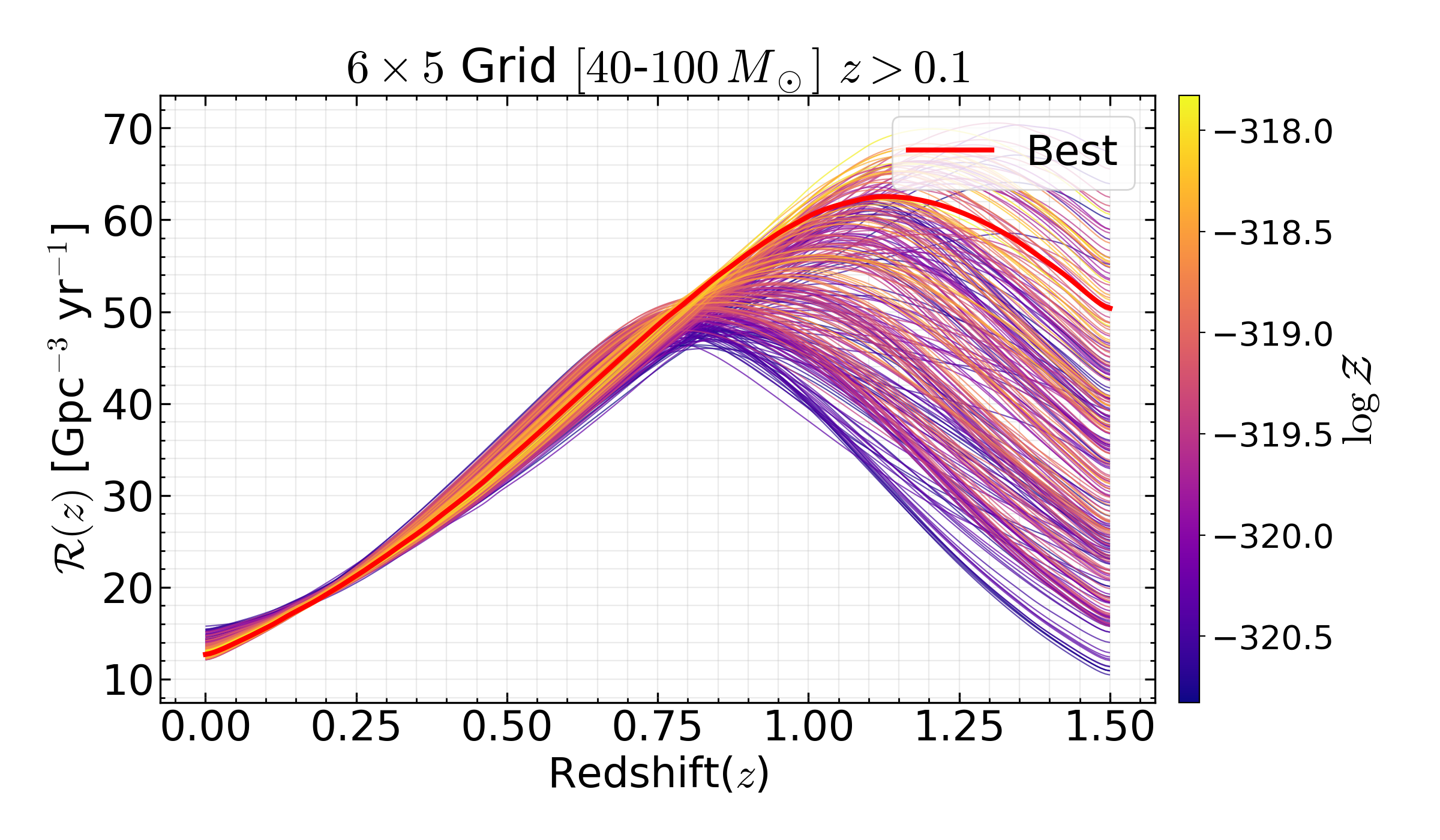}
    \includegraphics[width=0.32\textwidth]{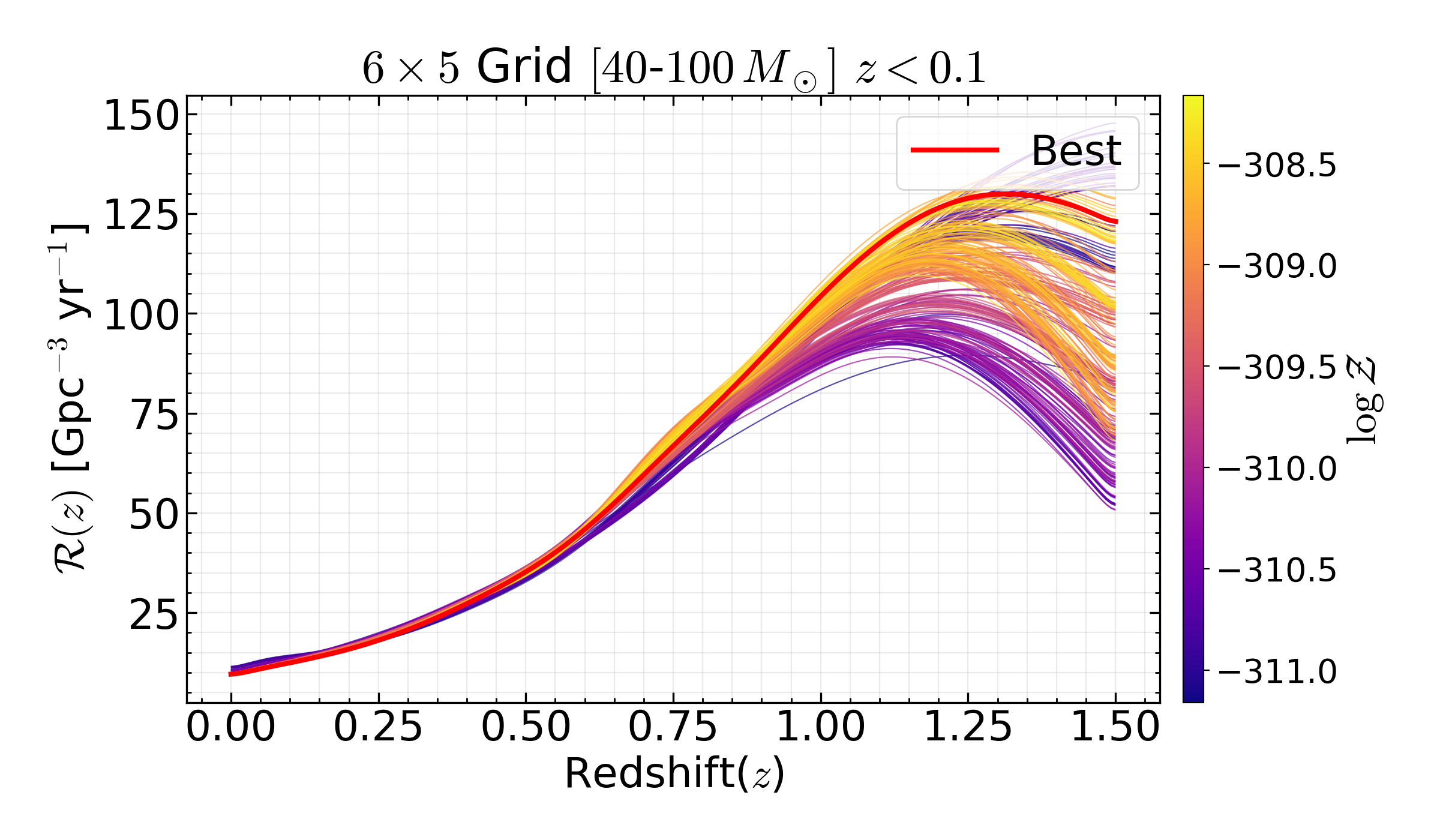}
    \caption{Physical merger rate $\mathcal{R}(z)$ [Gpc$^{-3}$ yr$^{-1}$] for all trajectories within $\mathrm{BF} < 3$ of the best-fit solution, for the $6\times5$ grid. Rows and columns as in Figure~\ref{fig:dtd_6x5}; the red curve marks the highest-evidence trajectory. As in the $7\times5$ case, the best-fit merger rate peaks at intermediate redshift in both mass bins, with the high-mass bin showing a lower local rate, a peak shifted to slightly higher redshift, and a tighter envelope, and with the low-metallicity SFRD producing the highest and most sharply rising peak.}
    \label{fig:Rz_6x5}
\end{figure*}

Figure~\ref{fig:Rz_6x5} shows the corresponding merger rate evolution $\mathcal{R}(z)$. The $6\times5$ grid reproduces the central result of the main analysis: in both mass bins the best-fit merger rate rises from its local value to a clear intermediate-redshift peak before declining. For the $20$--$40\,M_\odot$ bin, the best-fit trajectory rises from a local rate of $\mathcal{R}(z=0) \approx 21\,\mathrm{Gpc^{-3}\,yr^{-1}}$ to a peak of $\sim$94 Gpc$^{-3}$ yr$^{-1}$ at $z \sim 0.9$ under the MD SFR, and from $\mathcal{R}(z=0) \approx 22\,\mathrm{Gpc^{-3}\,yr^{-1}}$ to $\sim$92 Gpc$^{-3}$ yr$^{-1}$ at $z \sim 0.85$ under the closely similar $Z > 0.1\,Z_\odot$ SFRD. Under the $Z < 0.1\,Z_\odot$ SFRD the best-fit rate rises far more steeply, from a lower local value of $\mathcal{R}(z=0) \approx 19\,\mathrm{Gpc^{-3}\,yr^{-1}}$ to a peak of $\sim$260 Gpc$^{-3}$ yr$^{-1}$ at $z \sim 1.3$, reflecting the steeper redshift dependence of the low-metallicity SFRD. These values are consistent with the $7\times5$ results to within the read-off precision.

For the $40$--$100\,M_\odot$ bin, the best-fit merger rate peaks at slightly higher redshift and from a lower local rate than the low-mass bin, again within a tighter envelope. Under the MD SFR the best-fit rises from $\mathcal{R}(z=0) \approx 13\,\mathrm{Gpc^{-3}\,yr^{-1}}$ to $\sim$64 Gpc$^{-3}$ yr$^{-1}$ at $z \sim 1.15$; under the $Z > 0.1\,Z_\odot$ SFRD it rises from $\mathcal{R}(z=0) \approx 12.5\,\mathrm{Gpc^{-3}\,yr^{-1}}$ to $\sim$63 Gpc$^{-3}$ yr$^{-1}$ at $z \sim 1.1$; and under the physically motivated $Z < 0.1\,Z_\odot$ SFRD it rises from the lowest local rate of all six panels, $\mathcal{R}(z=0) \approx 9.5\,\mathrm{Gpc^{-3}\,yr^{-1}}$, to a peak of $\sim$130 Gpc$^{-3}$ yr$^{-1}$ at $z \sim 1.25$. As in the $7\times5$ analysis, the contrast between the two mass bins is one of degree rather than kind, and the peak-to-local ratio is largest under the low-metallicity assumption in both bins.

\begin{figure*}[ht]
    \centering
    \includegraphics[width=0.32\textwidth]{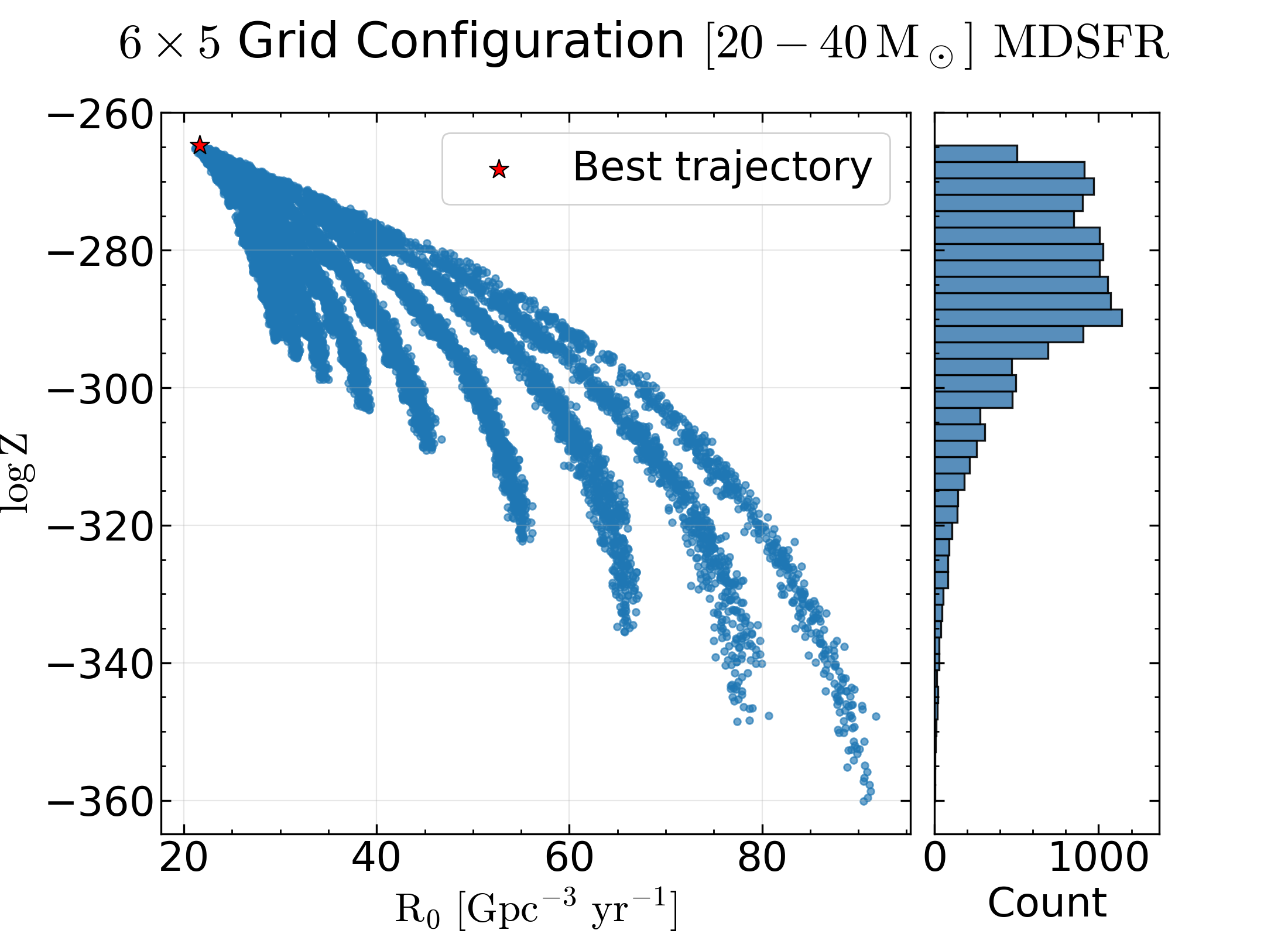}
    \includegraphics[width=0.32\textwidth]{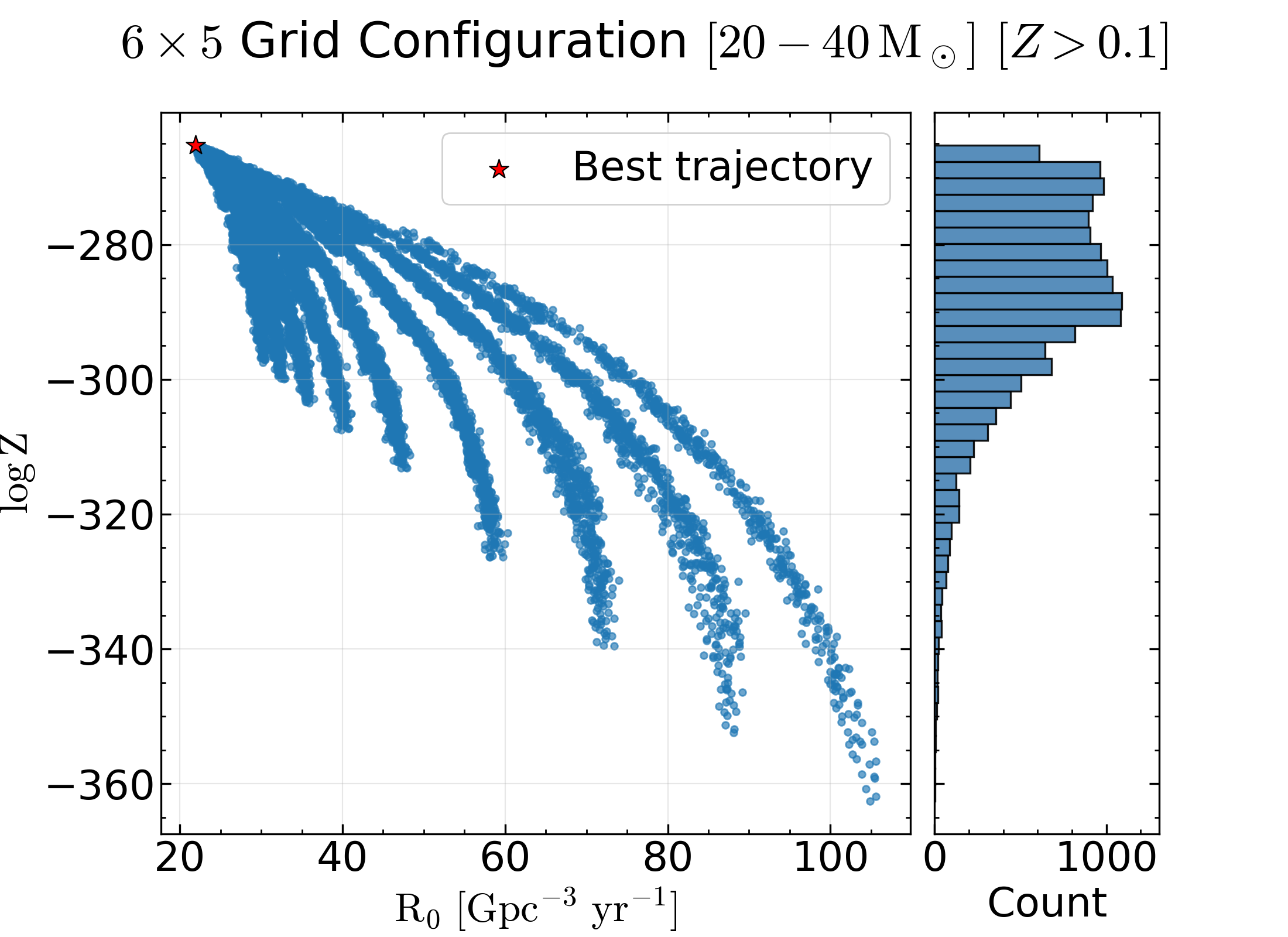}
    \includegraphics[width=0.32\textwidth]{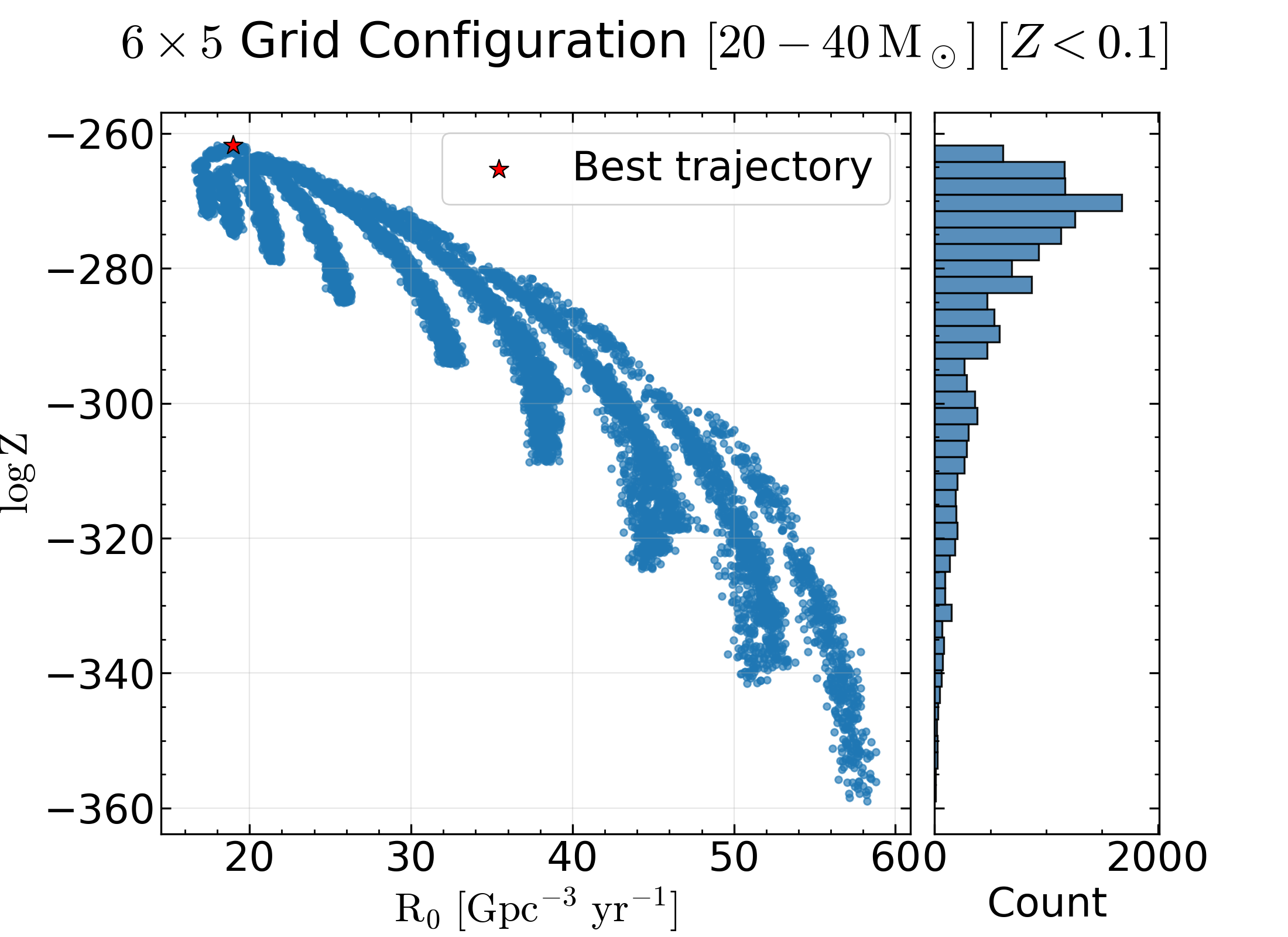}\\[6pt]
    \includegraphics[width=0.32\textwidth]{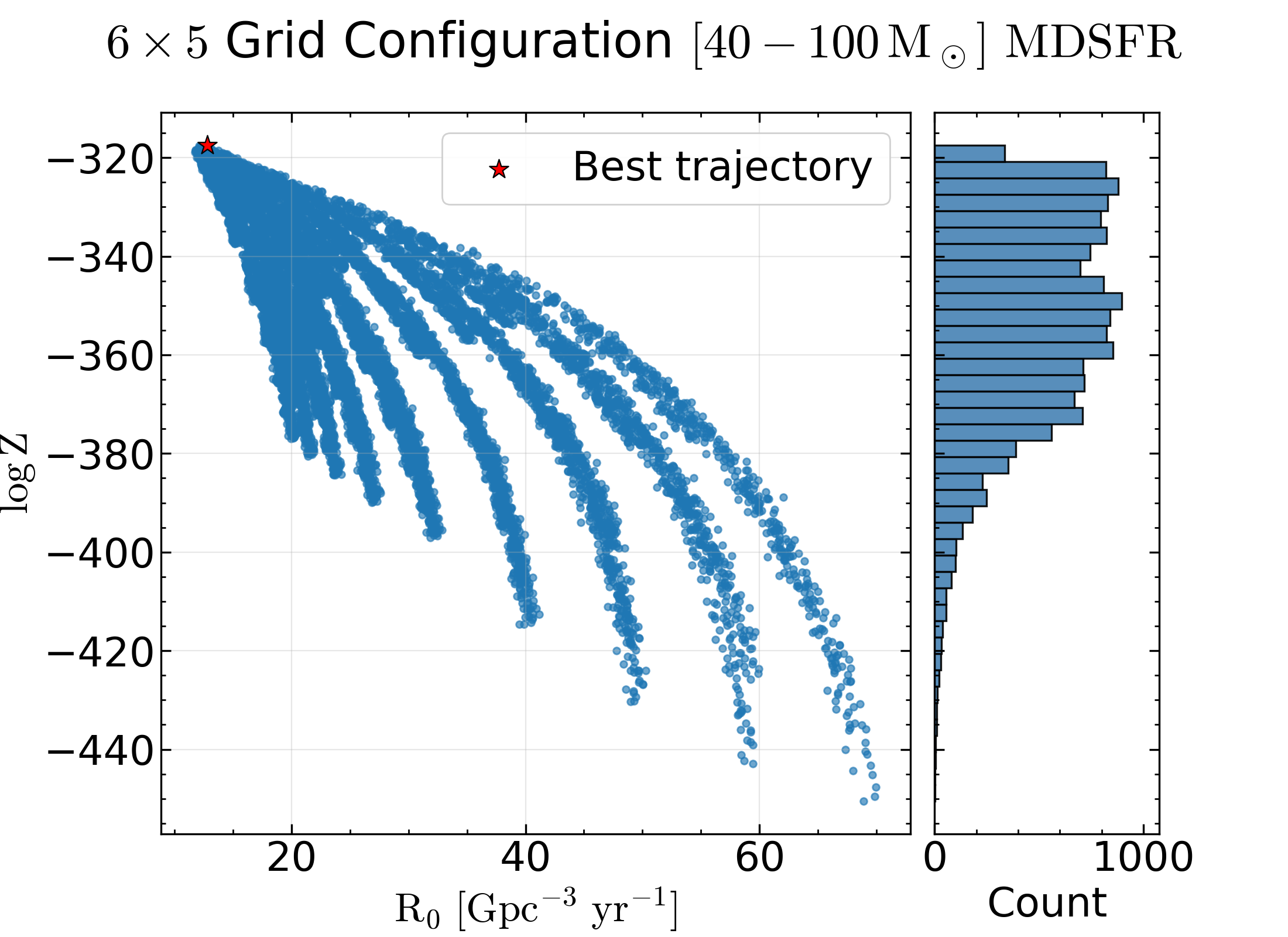}
    \includegraphics[width=0.32\textwidth]{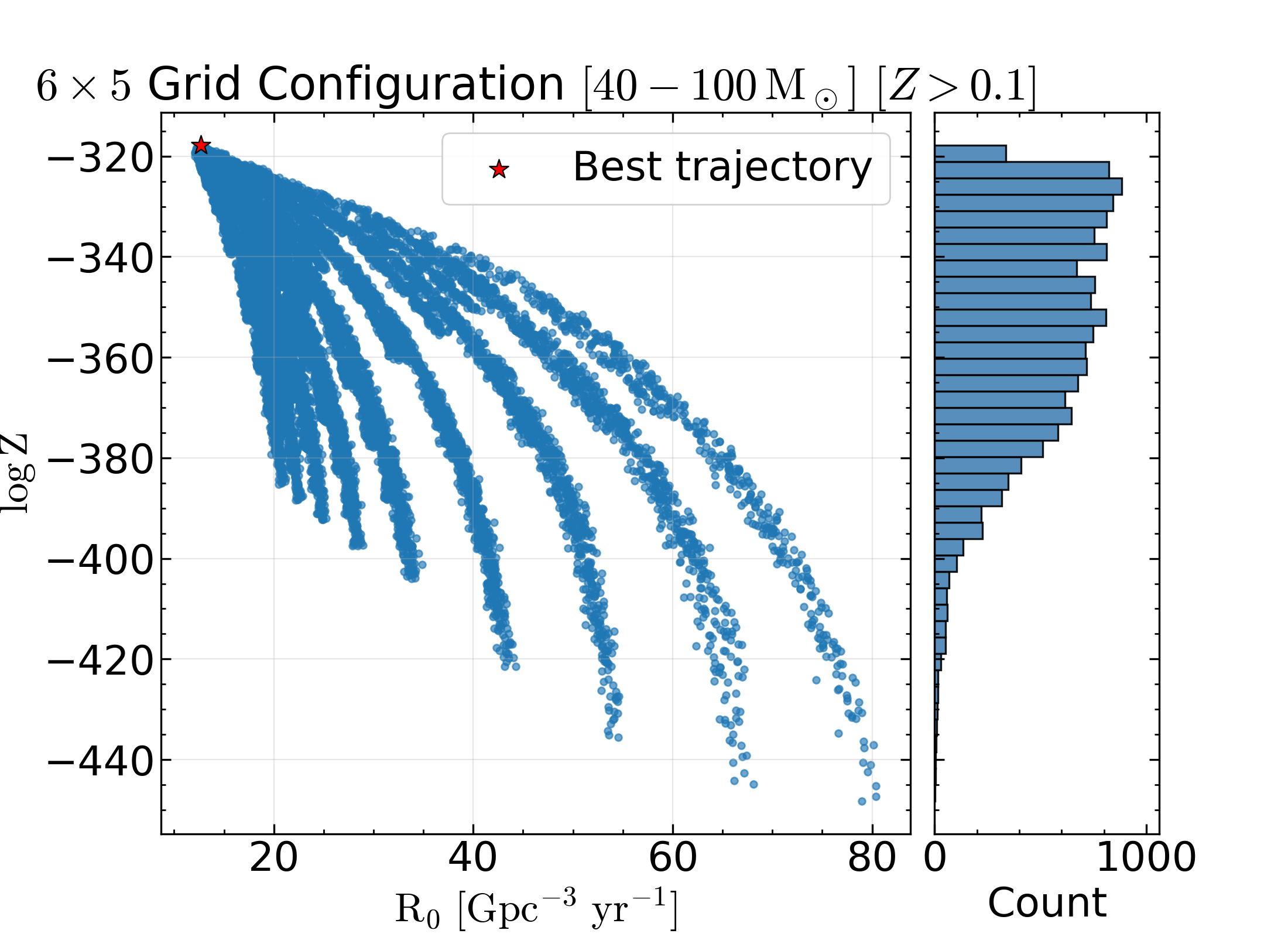}
    \includegraphics[width=0.32\textwidth]{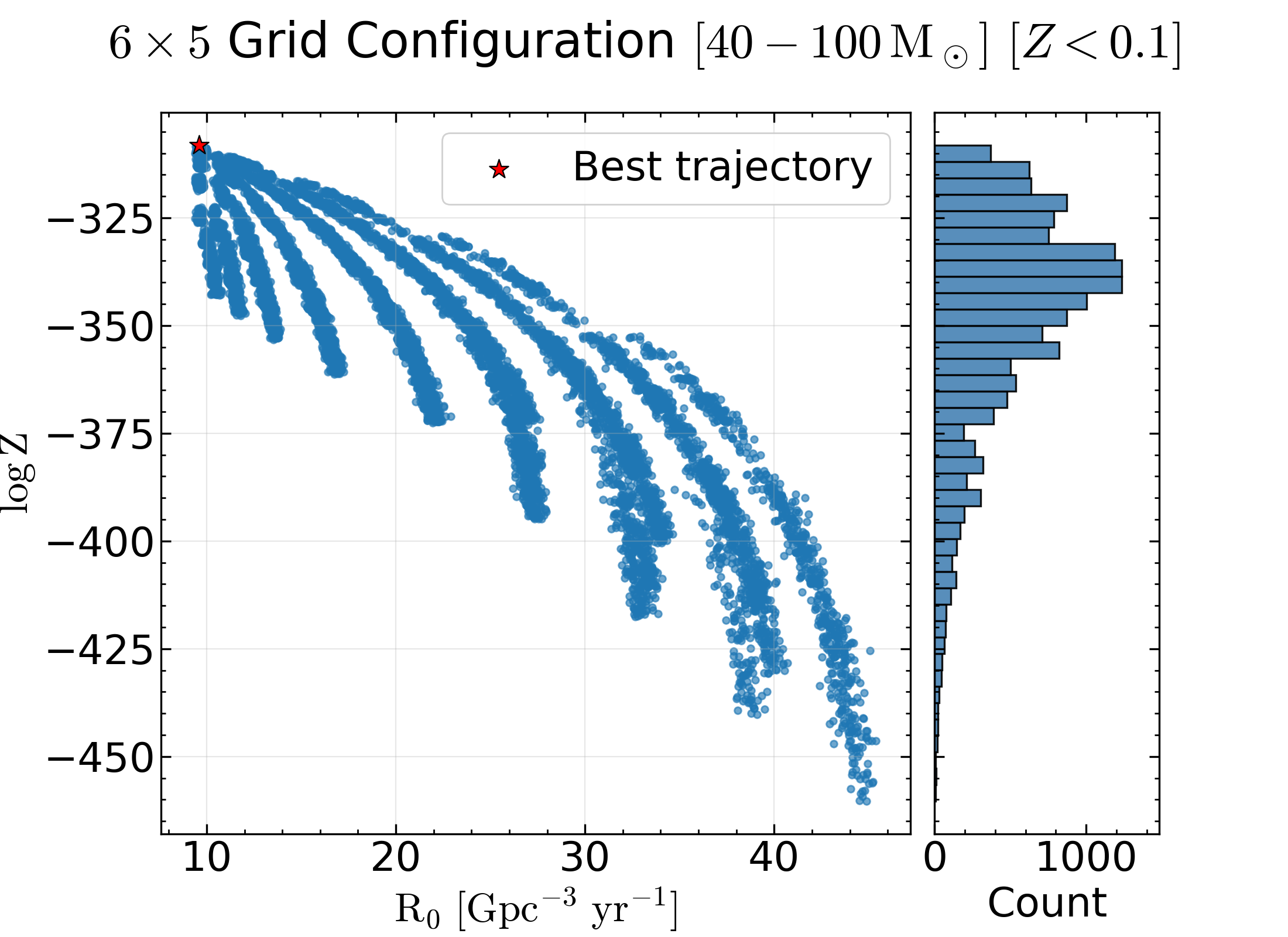}
    \caption{Bayesian log-evidence $\ln\mathcal{Z}$ versus local merger rate $R_0$ for all trajectories of the $6\times5$ grid. Rows and columns as in Figure~\ref{fig:dtd_6x5}; the right panel of each subfigure shows the marginalized $R_0$ histogram and the red star marks the highest-evidence trajectory. As in the $7\times5$ case, the best-fit trajectories cluster at the lowest $R_0$ with $\ln\mathcal{Z}$ declining into discrete striated families toward higher $R_0$, and the high-mass bin shows lower best-fit $R_0$ and a larger evidence dynamic range than the low-mass bin.}
    \label{fig:evidence_6x5}
\end{figure*}

Figure~\ref{fig:evidence_6x5} shows the $\ln\mathcal{Z}$--$R_0$ distribution for the $6\times5$ grid. The same evidence structure found in Section~\ref{sec:result} is reproduced: the highest-evidence trajectories cluster at the lowest values of $R_0$, and $\ln\mathcal{Z}$ declines into discrete striated families toward higher $R_0$. For the $20$--$40\,M_\odot$ bin the best-fit trajectory has $\ln\mathcal{Z}_{\rm best} \approx -264$ at $R_0 \approx 21\,\mathrm{Gpc^{-3}\,yr^{-1}}$ under the MD SFR, $\ln\mathcal{Z}_{\rm best} \approx -265$ at $R_0 \approx 22\,\mathrm{Gpc^{-3}\,yr^{-1}}$ under the $Z > 0.1\,Z_\odot$ SFRD, and $\ln\mathcal{Z}_{\rm best} \approx -261$ at $R_0 \approx 19\,\mathrm{Gpc^{-3}\,yr^{-1}}$ under the $Z < 0.1\,Z_\odot$ SFRD, with the marginalized $R_0$ distributions extending to $\sim$60--90 Gpc$^{-3}$ yr$^{-1}$. For the $40$--$100\,M_\odot$ bin the best-fit local rates are systematically lower and the evidence dynamic range larger: $\ln\mathcal{Z}_{\rm best} \approx -318$ at $R_0 \approx 13\,\mathrm{Gpc^{-3}\,yr^{-1}}$ under the MD SFR, $\ln\mathcal{Z}_{\rm best} \approx -319$ at $R_0 \approx 12.5\,\mathrm{Gpc^{-3}\,yr^{-1}}$ under the $Z > 0.1\,Z_\odot$ SFRD, and $\ln\mathcal{Z}_{\rm best} \approx -308$ at $R_0 \approx 9.5\,\mathrm{Gpc^{-3}\,yr^{-1}}$ under the $Z < 0.1\,Z_\odot$ SFRD. The dynamic range of $\ln\mathcal{Z}$ spans $\sim$100 log-units for the low-mass bin and $\sim$120--140 log-units for the high-mass bin, mirroring the $7\times5$ result that the high-mass data carry greater discriminating power over the DTD shape.

\section{Data Availability}
The gravitational-wave catalogs used in this study are publicly available on Zenodo as part of the LIGO-Virgo-KAGRA data releases:
GWTC-4 (\href{https://zenodo.org/records/16053484}{link}),
GWTC-2.1 (\href{https://zenodo.org/records/6513631}{link}),
and GWTC-3 (\href{https://zenodo.org/records/5546663}{link}).

\section*{Acknowledgments}
The authors express their gratitude to Sergio Andr\'es Vallejo-Pe\~na for reviewing the manuscript and providing useful comments as a part of the LIGO publication policy. This work is part of the \texttt{⟨data|theory⟩ Universe Lab}, supported by TIFR and the Department of Atomic Energy, Government of India. The authors express gratitude to the system administrator of the computer cluster of \texttt{⟨data|theory⟩ Universe Lab}. This research is supported by the Prime Minister Early Career Research Award, Anusandhan National Research Foundation, Government of India. Special thanks to the LIGO Virgo KAGRA Scientific Collaboration for providing noise curves. LIGO, funded by the U.S. National Science Foundation (NSF), and Virgo, supported by the French CNRS, Italian INFN, and Dutch Nikhef, along with contributions from Polish and Hungarian institutes. This collaborative effort is backed by the NSF's LIGO Laboratory, a major facility fully funded by the National Science Foundation. The research leverages data and software from the Gravitational Wave Open Science Center, a service provided by LIGO Laboratory, the LIGO Scientific Collaboration, Virgo Collaboration, and KAGRA. Advanced LIGO's construction and operation receive support from STFC of the UK, Max Planck Society (MPS), and the State of Niedersachsen/Germany, with additional backing from the Australian Research Council. Virgo, affiliated with the European Gravitational Observatory (EGO), secures funding through contributions from various European institutions. Meanwhile, KAGRA's construction and operation are funded by MEXT, JSPS, NRF, MSIT, AS, and MoST. This material is based upon work supported by NSF's LIGO Laboratory which is a major facility fully funded by the National Science Foundation. We acknowledge the use of the following packages in this work: Numpy \citep{van2011numpy}, Scipy \citep{jones2001scipy}, Matplotlib \citep{hunter2007matplotlib}, Astropy \citep{robitaille2013astropy}, Dynesty \citep{speagle2020dynesty} and emcee \citep{foreman2013emcee}.

\bibliography{references}
\end{document}